\documentclass[aps,prb,preprint,superscriptaddress]{revtex4-2}
\usepackage{amsmath}
\usepackage{graphicx}
\usepackage{xcolor}

\usepackage{xr}
\graphicspath{{./Figures/}}

\newcommand{\highlight}[1]{\textcolor{black}{#1}}

\begin{document}

\title{Skyrmion motion in a synthetic antiferromagnet driven by asymmetric spin wave emission}

\author{Christopher E. A. Barker}
\email[]{christopher.barker@npl.co.uk}
\affiliation{National Physical Laboratory, Hampton Road, Teddington, TW11 0LW, United Kingdom}

\author{Charles Parton-Barr}
\affiliation{School of Physics and Astronomy, University of Leeds, Leeds, LS2 9JT, United Kingdom}

\author{Christopher H. Marrows}
\email[]{c.h.marrows@leeds.ac.uk}
\affiliation{School of Physics and Astronomy, University of Leeds, Leeds, LS2 9JT, United Kingdom}

\author{Olga Kazakova}
\affiliation{National Physical Laboratory, Hampton Road, Teddington, TW11 0LW, United Kingdom}
\affiliation{Department of Electrical and Electronic Engineering, University of Manchester, Manchester, M13 9PL, United Kingdom}

\author{Craig Barton}
\email[]{craig.barton@npl.co.uk}
\affiliation{National Physical Laboratory, Hampton Road, Teddington, TW11 0LW, United Kingdom}

\date{\today}

\begin{abstract}

Skyrmions have been proposed as new information carriers in racetrack memory devices. To realise such devices, a small size; high speed of propagation; and minimal skyrmion Hall angle are required. Synthetic antiferromagnets (SAFs) present the ideal materials system to realise these aims. In this work, we use micromagnetic simulations to propose a new method for manipulating them using exclusively global magnetic fields. An out-of-plane microwave field induces oscillations in the skyrmions radius which in turn emits spin waves. When a static in-plane field is added, this breaks the symmetry of the skyrmions and causes asymmetric spin wave emission. This in turn drives motion of the skyrmions, with the fastest velocities observed at the frequency of the intrinsic out-of-phase breathing mode of the pair of skyrmions. This behaviour is investigated over a range of experimentally realistic antiferromagnetic interlayer exchange coupling strengths, and the results compared to previous works. Through this we demonstrate the true effect of varying the exchange coupling strength, and gain greater insight into the mechanism of skyrmion motion. \highlight{In an attempt to more accurately reproduce experimental SAF samples, we investigate the effect of changing the ratio between the layer magnetisations in the SAF, and demonstrate that maximum velocity can be achieved when there is 100~\% compensation. We also investigate the differences between excitation mechanism, using both electric fields and spin transfer torques to excite skyrmion motion.} These results will help to inform the design of future novel computing architectures based on the dynamics of skyrmions in synthetic antiferromagnets.
\end{abstract}

\maketitle

\section{Introduction}
	Magnetic skyrmions---a type of topological spin texture stabilised by the Dzyaloshinksii-Moriya interaction~\cite{Nagaosa2013, EverschorSitte2018, Marrows2021}---are an area of great recent interest for their potential applications in magnetic memory~\cite{Finocchio2016, Tomasello2014}, future computation~\cite{Song2020,Marrows2024} and quantum operations~\cite{petrovic2024, banerjee2024}. In all of these applications, the ability to controllably move and manipulate skyrmions is key. They can be moved with electrical currents~\cite{Fert2013,Woo2016,Juge2019}, magnetic field gradients~\cite{Zhang2018, Wang_2017}, and microwave magnetic fields~\cite{Moon2016, Wang2015} among other methods. However, in ferromagnetic systems, skyrmions exhibit the skyrmion Hall effect---where they travel at an angle to the driving force~\cite{EverschorSitte2014}---the magnitude of which depends on the velocity, skyrmion size and local pinning landscape~\cite{Jiang2017,Litzius2017,Juge2019, Zeissler2020}. This is a significant roadblock to the implementation of skyrmion based devices as skyrmions will tend to drift to the edges of devices and annihilate~\cite{Zhang2016}.\\\\
	Synthetic antiferromagnets---where two individually ferromagnetic layers are sandwiched around a nonmagnetic spacer layer and coupled antiferromagnetically~\cite{Duine2018}---have been predicted to be an advantageous platform for skyrmions~\cite{Barker2016, Zhang2016}. This is because of their proposed smaller size, faster velocity, and increased resistance to stray fields compared to skyrmions observed in ferromagnetic systems~\cite{Buttner2018}. Skyrmions with radii as small as $\sim$50~nm have been experimentally in synthetic antiferromagnets~\cite{Legrand2020, Juge2022}. Skyrmions and domain walls have been shown to move at substantially faster velocities than their ferromagnetic equivalents of up to 900~m/s~\cite{Pham2024} in the case of skyrmions, or 750~m/s in the case of domain walls~\cite{Yang2015}. Domain walls have also been shown to depin at current densities as low as $\mathrm{1 \times 10^{11}\ A/m^2}$~\cite{Lepadatu2017, Barker2023_JPhysD}. Due to the challenges in detection, much of the body of research on skyrmions in synthetic antiferromagnets remains theoretical~\cite{Wang2023}, and experimental measurements of skyrmion motion are extremely limited~\cite{Juge2022, Pham2024, Dohi2019}.\\\\
	Skyrmions have been shown to exhibit microwave resonances when excited by magnetic and electric fields~\cite{Lonsky2020_APLMat}. This was originally predicted based on numerical modelling of skyrmions in bulk systems~\cite{Mochizuki2012}, and was first demonstrated experimentally in Cu$_2$OSeO$_3$~\cite{Onose2012}. The breathing modes of skyrmions were first simulated in thin films by Kim \textit{et al.}~\cite{Kim2014}, and have recently been demonstrated experimentally in ferromagnetic thin films~\cite{Satywali2021, Srivastava2023}. In synthetic antiferromagnets, micromagnetic simulations have been used to demonstrate both the breathing modes~\cite{Lonsky2020_PRB, Barker2023_JAP} and corresponding rotational modes~\cite{Xing2018}. \highlight{The breathing and rotational modes of skyrmions have generated interest for magnonics, where, by exciting at the resonant frequency of these modes, they have been used as spin-wave emitters~\cite{Chen2021, Diaz2020, Tang2023}.} Building on this, by exciting these breathing modes in the presence of an in-plane magnetic field, it has been demonstrated that skyrmions in ferromagnetic systems will emit spin waves asymmetrically. This stimulates skyrmion motion with a peak in velocity as a function of frequency that is coupled to the breathing mode resonant frequency~\cite{Wang2015, Yuan2019}. \highlight{In the presence of the in-plane field, there will be a net momentum transfer from the skyrmion to the spin waves, which will cause the skyrmion to move through the conservation of linear momentum~\cite{Yuan2019, Liu2025}.} This has also been demonstrated in synthetic antiferromagnets where microwave electric fields were used to excite straight-line skyrmion motion~\cite{Qiu2021}. In that work, a dependence of the velocity on the strength of the antiferromagnetic interlayer exchange coupling was demonstrated, however the link between this and the skyrmion breathing modes was not discussed~\cite{Qiu2021}. This microwave field driven approach to skyrmion motion has the advantage that---unlike the conventional method of using an electric current---it does not require a conductive sample, and so can be used to manipulate skyrmions in insulating materials.\\\\
	In this work, we use microwave magnetic fields in micromagnetic simulations to drive skyrmion breathing modes in a synthetic antiferromagnet, and combine these with an in-plane, static field to excite skyrmion motion in a wiggle-like fashion. We investigate this motion as a function of frequency, and relate it to the intrinsic breathing modes of these skyrmions. We use this understanding of the coupling between velocity and breathing mode frequency to understand the true relationship between the skyrmion velocity and the antiferromagnetic interlayer exchange coupling strength. Finally, we use the additional knowledge of this mechanism to discuss the differences between the previous electric-field driven skyrmion motion~\cite{Qiu2021} and our work. The additional understanding of the relationship between the skyrmion breathing modes and the velocity can be used in applications to modulate the speed of the skyrmion using only the microwave frequency, and has particular relevance to the emerging field of skyrmion magnonics~\cite{GOBEL20211}. The emission of spin waves by a domain wall driven by microwave magnetic fields has also very recently been demonstrated~\cite{Liu2025}. This, together with the similarly recent imaging of spin waves in a synthetic antiferromagnet~\cite{Girardi2024}, paves the way to the experimental realisation of our method.

\section{Methods}\label{sec:Methods}

	The synthetic antiferromagnetic multilayer presented in this paper was studied using the micromagnetic solver MUMAX3~\cite{Vansteenkiste2014}. MUMAX3 is a finite difference numerical solver for the Landau-Lifshitz-Gilbert (LLG) equation that calculates the spatial and time dependence of the magnetisation $\mathbf{m}$ in a simulation~\cite{Vansteenkiste2014}. The LLG equation is given by
	\begin{equation}
		\dfrac{d\mathbf{m}}{dt} = - \lvert \gamma_0 \rvert \mathbf{H}_{\mathrm{eff}} + \alpha \mathbf{m} \times \dfrac{d\mathbf{m}}{dt},
		\label{eq:LLG} 
	\end{equation}
	 where $\mathbf{m}$ is the magnetisation unit vector, $\alpha$ is the Gilbert damping parameter, and $\mathbf{H}_{\mathrm{eff}}$ is the effective magnetic field. The effective magnetic field is proportional to the differential of the total micromagnetic energy $U$, given by
	 \begin{equation}
	 	\mathbf{H}_{\mathrm{eff}} = -\dfrac{1}{\mu_0 M_{\mathrm{s}}} \dfrac{\delta U}{\delta \mathbf{m}}.
	 \end{equation}
	 Here $\mu_0$ is the permeability of free space which is $4\pi \times 10^{-7} \ \mathrm{N\cdot A^{-2}}$, and $M_{\mathrm{s}}$ is the saturation magnetisation of the system. The micromagnetic energy $U$ includes the usual energy terms that describe the interaction with an external magnetic field, the effects of the demagnetisation field, the Heisenberg exchange. In these simulations we consider a single anisotropy energy term perpendicular to the plane of the field. This is given by
	 \begin{equation}
	 	U_{\mathrm{Anis}} = -K_{\mathrm{u}}\int (\mathbf{u}\cdot \mathbf{m})^2 dV,
	 \end{equation}
	 where $K_{\mathrm{u}}$ is the anisotropy constant, and $\mathbf{u}$ is a unit vector denoting the direction of the ansiotropy, in this case along the vector perpendicular to the film plane. In addition we consider an interfacial DMI term of the form
	 \begin{equation}
	 	U_{\mathrm{iDMI}} = D\int [m_z (\nabla\cdot\mathbf{m}) - (\mathbf{m}\cdot \nabla)m_z] dV,
	 \end{equation}
	 where D is the DMI constant. Finally, we add a custom energy term to the simulation to model the indirect interlayer exchange coupling, given by
	 \begin{equation}
	 	U_{\mathrm{RKKY}} = -\sigma_{\mathrm{RKKY}}\int (\mathbf{m_1}\cdot\mathbf{m_2}) dV, 
	 \end{equation}
	 where the subscript of $\mathbf{m}$ denotes the layer number, and $\sigma_{\mathrm{RKKY}}$ is the coupling energy. The system is antiferromagnetic for $\sigma_{\mathrm{RKKY}} < 0$.\\\\
	 The simulation is set up with exchange stiffness $A = 15\times 10^{-12}$ J/m, saturation magnetisation $M_{\mathrm{s}} =  580 \times 10^{3}$ A/m, uniaxial anisotropy constant $K = 7 \times 10^5\ \mathrm{J/m^3}$, interfacial DMI $D = 3 \times 10^{-3}\ \mathrm{J/m^2}$ and Gilbert damping parameter $\alpha = 0.01$. These values represent typical generic parameters of metallic multilayers with DMI that are usually chosen for synthetic antiferromagnets hosting skyrmions~\cite{Jiang_2017}. They are also chosen specifically to mirror those used in ref.~\cite{Qiu2021} in order to better compare results. The system was initialised with an in plane \textit{x-y} cellsize of 1.5625~nm and an out-of-plane \textit{z} cellsize of 0.4~nm. The configuration of cells was $256 \times 256 \times 4$ cells, for a total layer width and length of 400~nm and a total stack thickness of 1.6~nm. Only two of the layers were magnetic, with the remaining two layers not contributing to the simulation other than to mirror the spatial separation of the magnetic layers as in a real system.\\\\
	 To measure the frequency response of the system, it was excited at $t = 0$ with a sinc field along the \textit{z}-axis with a cutoff frequency of $f$ = 100~GHz of the form
	 \begin{equation}
	 	H_z = H_0 + H_1 \dfrac{\mathrm{sin}(2\pi f t)}{2\pi f t},
	 \end{equation} 
	 where $H_0$ is an optional static field, $H_1$ is the magnitude of the sinc field, 0.5~mT unless otherwise specified. 
	 This was used as it is a top-hat function in Fourier space, and so excites all modes equally up to its cutoff frequency. Simulations were then run for 10-20~ns, and the magnetisations of each layer were recorded. The change in magnetisation, based on the spatial average of the magnetisation in each layer at time $t$ $\langle m_z(t) \rangle$ is calculated as  
	 \begin{equation}
	 	\delta m_z = \langle m_z(t) \rangle - \langle m_z(0) \rangle . 
	 \end{equation}
	 The frequency response of the system is then calculated using the power spectral density (PSD), which is given by the Fourier transform of $\delta m_z$ multiplied by its complex conjugate
	 \begin{equation}
	 	\mathrm{PSD} = \Big\lvert \int_{0}^{t_0} \delta m_z e^{2\pi f t} dt \Big\rvert^2. 
	 \end{equation}
	 
	 To excite the skyrmion motion---unless otherwise specified---a static in-plane field of 0.5~T is used, together with a static out-of-plane field of 50~mT and an out-of-plane microwave excitation field of 20~mT. \highlight{The static in-plane field is chosen to be 0.5~T to be consistent with previous work studying skyrmion motion driven by excitation with an electric field~\cite{Qiu2021}. Likewise the choice of the 50~mT static out-of-plane field is in line with previous work on skyrmion breathing driven by a magnetic field~\cite{Lonsky2020_PRB, Barker2023_JAP}. As discussed in Ref.~\cite{Lonsky2020_PRB}, this small static field serves to aid in the observation of the in-phase skyrmion breathing mode, but does not otherwise alter the behaviour of the breathing modes for fields much smaller than the saturation field, as is the case for our chosen value (see Supplementary Note S1 for measurement of the saturation field.).}  The microwave excitation field is smaller than that used by other works~\cite{Moon2016} but a comparable order of magnitude. In order to track the motion of the skyrmion, for each timestep we calculate the guiding centre $\mathbf{R}_{\mathrm{Sk}}$ of each magnetic layer, given by
	 \begin{equation}
	 	\mathbf{R_{\mathrm{Sk}}} = \dfrac{\int \mathbf{r}q\ dxdy}{\int q\ dx dy},
	 \end{equation}
	 where $\mathbf{r}$ is the position vector of each cell, and $q$ is the topological charge density
	 \begin{equation}
	 	q = \mathbf{m}\cdot \Big( \dfrac{\partial\mathbf{m}}{\partial x} \times \dfrac{\partial\mathbf{m}}{\partial y} \Big),
	 \end{equation}
	 such that the topological charge of the skyrmion would be $Q = (1/4\pi)\int q dxdy$, which will be equal to 1 and -1 for the two skyrmions. We also note here that throughout this paper, quantities of the skyrmions are often plotted separately for the skyrmions in each layer, and also split further for the behaviour of that quantity along the \textit{x}- and \textit{y}-axis. Throughout this paper we will denote the skyrmion in the bottom layer (layer 1) of the SAF with the colour blue, and the skyrmion in the top layer (layer 2) with the colour black. Also see Fig.~\ref{fig:stack_structure}(a) for a sketch of the simulation stack. We further subdivide this into the behaviour along the \textit{x}-axis which will be represented by diamond-shaped points, and the behaviour along the \textit{y}-axis which will be denoted by circular points. We also adopt the notation $Sk_{\mathrm{n,i}}$ to label these quantities, where $n$ denotes the layer number and $i$ is either \textit{x} or \textit{y}. This notation and colour scheme is used throughout.

\section{Static and Dynamic Characterisation}
	\subsection{Effect of in-plane field on skyrmions}\label{sec:Static_Char}
	\begin{figure}[h!]
		\includegraphics[width=\linewidth]{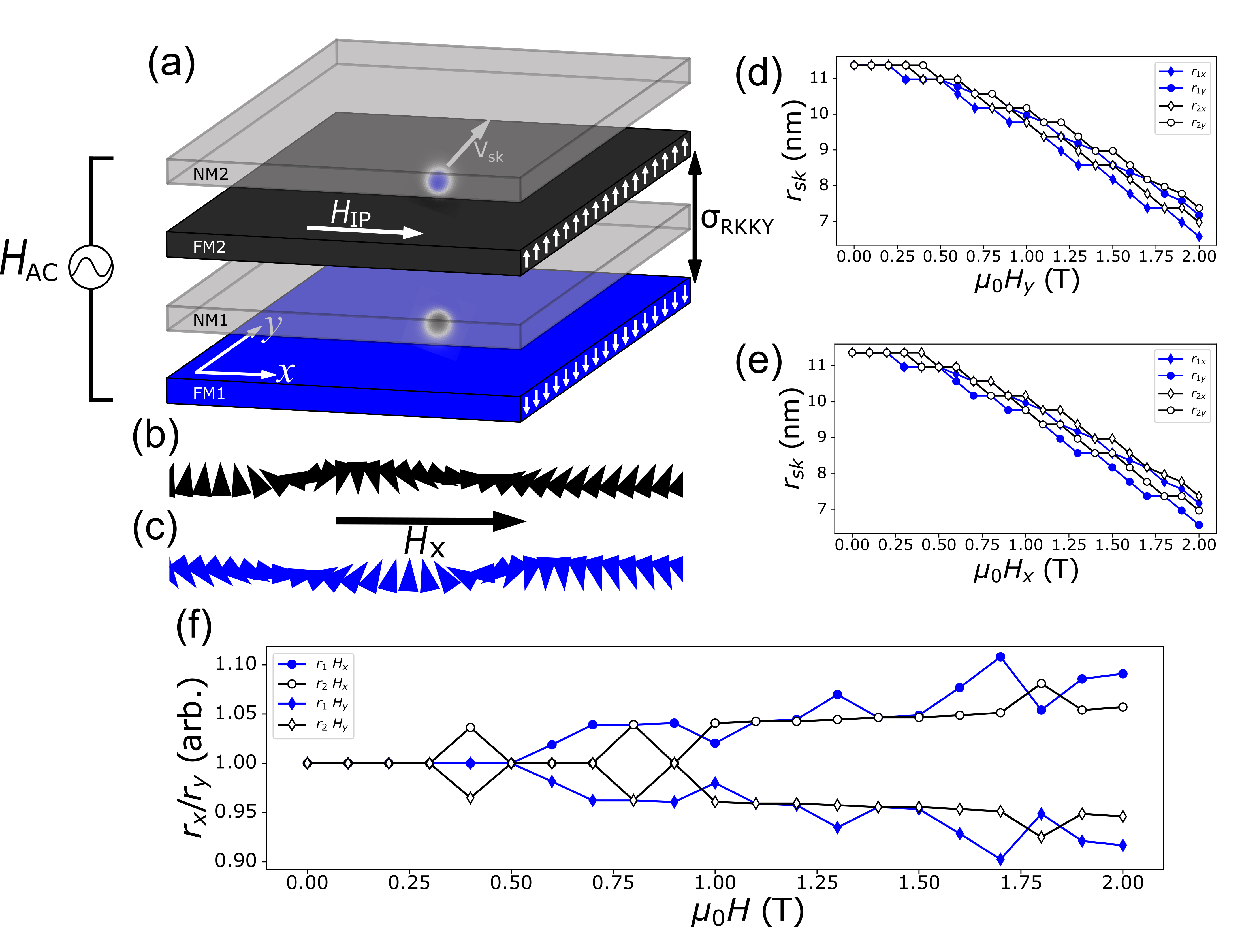}
		\caption{Illustration of simulation set-up and characterisation of skyrmion radius as a function of field. (a) Illustration of the simulated multilayer. Two ferromagnetic (FM) layers, labelled FM1 and FM2 are coupled together antiferromagnetically through a non-magnetic (NM) spacer layer. An in-plane field ($H_{\mathrm{IP}}$) causes the skyrmion to elongate, while the sinusoidal driving field causes it to emit spin waves and move perpendicular to $H_{\mathrm{IP}}$. (b) and (c) show the cross sections of the two skyrmions in layers 1 and 2 along the \textit{x}-axis in an in-plane field along the \textit{x}-axis of 1.7~T. The effect of the in plane field is shown on the radius of the two coupled skyrmions for field along the \textit{y}-axis in (d) and field along the \textit{x}-axis in (e). The ratio of the radius along the \textit{x}- and \textit{y}-axes of each skyrmion for each field orientation is shown in (f). The quanization is related to the cell size of the simulations. 
			\label{fig:stack_structure}}
	\end{figure}
	The effect of a static in-plane field on the skyrmion radius is shown in Fig.~\ref{fig:stack_structure}(b-e) for an antiferromagnetic interlayer exchange coupling strength of $-3 \times 10^{-4}\ \mathrm{J/m^2}$ in a static out-of-plane field of 50~mT. For this antiferromagnetic interlayer exchange coupling strength, the interlayer coupling is sufficiently strong to overcome the effects of the out-of-plane field and so the two skyrmions have equal radius in zero in-plane field $H_{\mathrm{IP}}$. As the field increases above $\sim$0.3~T, the skyrmions in each layer begin to both stretch and shrink. This shrinking is caused by the in-plane field pulling the cores of the two skyrmions apart laterally---as shown in Fig.~\ref{fig:stack_structure}(b,c)---which introduces a breaking of the antiferromagnetic interlayer coupling between spins inside the skyrmion. Thus, the skyrmion size shrinks to minimize this. Because of this, the stretching of the skyrmion is better demonstrated by the ratio between $r_x$ and $r_y$, as shown in Fig.~\ref{fig:stack_structure}(f). Above fields of $\sim$0.3~T this behaviour starts to change, and steadily changes from 0.5~T onwards, indicating the stretching of the two skyrmions along the direction of the field. On this basis, unless otherwise specified, an in-plane field of 0.5~T is chosen for further simulations which causes a modest deformation of the skyrmion, and also allows for better intercomparison of results with Ref.~\cite{Qiu2021}. Simulations of the full hysteresis loops are shown in supplementary note~S1. The saturation field along the x- and y-axes is 4.3~T, demonstrating that this in-plane field is only a modest $\sim$10~\% of saturation. 
	
	\subsection{Skyrmion breathing modes}\label{sec:Dynamic_Char}
	\begin{figure}
		\centering
		\includegraphics[width=\linewidth]{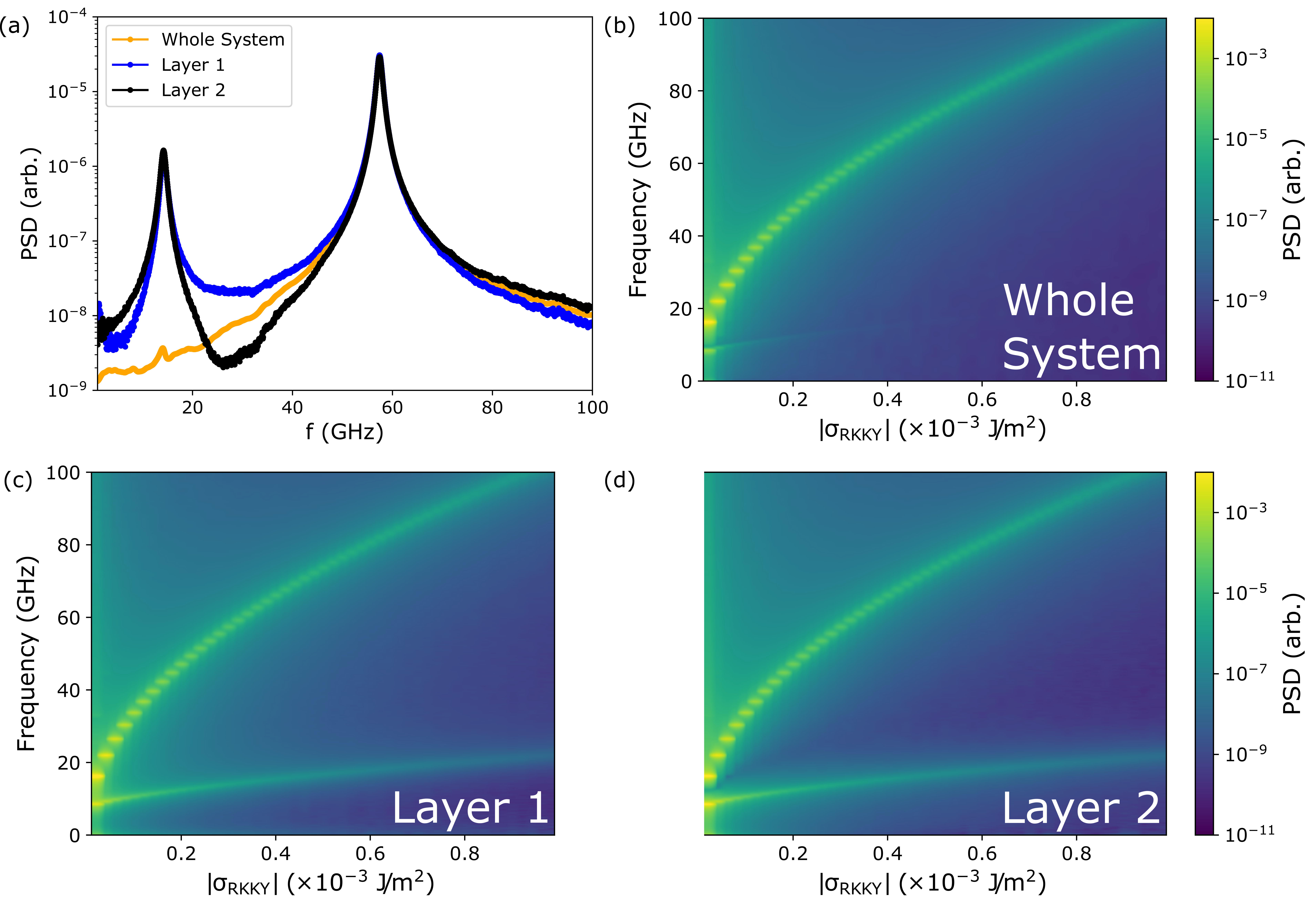}
		\caption{Skyrmion frequency response as a function of frequency and antiferromagnetic interlayer exchange coupling strength. (a) Shows the frequency response for an interlayer exchange coupling strength of $-3\times 10^{-4} \ \mathrm{J/m^2}$. Panels (b-d) show colour maps of the response as a function of interlayer exchange coupling strength, denoted as $\mathrm{\vert \sigma_{RKKY}\vert}$. (b) Shows the response of the entire simulated system, whereas (c) and (d) show the response when calculating the PSD of just the magnetisation of the cells of  layer 1 and layer 2, respectively. (b), (c) and (d) have a common colour bar.}
		\label{fig:Skyrmion_breathing}
	\end{figure}
	Before the motion of the skyrmions is induced, we first study the breathing modes of the system following the approach of Ref.~\cite{Lonsky2020_PRB}. As outlined in the methods, the system is first initialised with a N{\'e}el-type skyrmion in each layer and allowed to relax in a static out-of-plane field of 50~mT. Then a sinc pulse of amplitude 0.5~mT and frequency 100~GHz is applied at time 0 and the simulation is run for 20~ns. The results of this are shown in Fig.~\ref{fig:Skyrmion_breathing}. Fig.~\ref{fig:Skyrmion_breathing}(a) shows an illustration of the skyrmion frequency response with an interlayer exchange coupling strength of $-0.3\ \mathrm{mJ/m^2}$ for the two individual layers as well as the whole system. Because periodic boundary conditions were used for these simulations, there is no curling of the magnetisation at the edge, and so there are no modes arising from this and the frequency response seen here comes entirely from the skyrmion~\cite{Barker2023_JAP}. We see at $\sim$14~GHz there is a mode present in each of the individual layers, whereas, in the total system, there is almost no response. At the higher frequency mode at $\sim$57~GHz, there is a response in all three plots. This has been well studied~\cite{Lonsky2020_PRB, Barker2023_JAP}; the difference arises from the interplay between the two skyrmions at each mode. When excited at the lower mode, the two skyrmions breathe in phase with each other, and so at any time $t$, spins in each magnetic layer are coupled antiparallel over the whole lateral extent of the system, and so the net moment is always cancelled. Conversely at the higher frequency mode, the two skyrmions breathe in antiphase, and so at any time $t$ there is always a small net moment originating from the difference in size of these two skyrmions, and thus there is an observable peak in the frequency response of the system as a whole. Heatmaps of the frequency response of the whole system and the two individual layers are then shown in Fig.~\ref{fig:Skyrmion_breathing}(b-d). There is a considerable variation in the frequencies of the two peaks, especially that of the second as a function of the interlayer antiferromagnetic exchange strength. 
	
	\section{All magnetic field driven skyrmion motion}
	\subsection{Skyrmion motion driven by the addition of an in-plane magnetic field}\label{sec:B_Field_Motion_Intro}
	\begin{figure}
		\centering
		\includegraphics[width=0.8\linewidth]{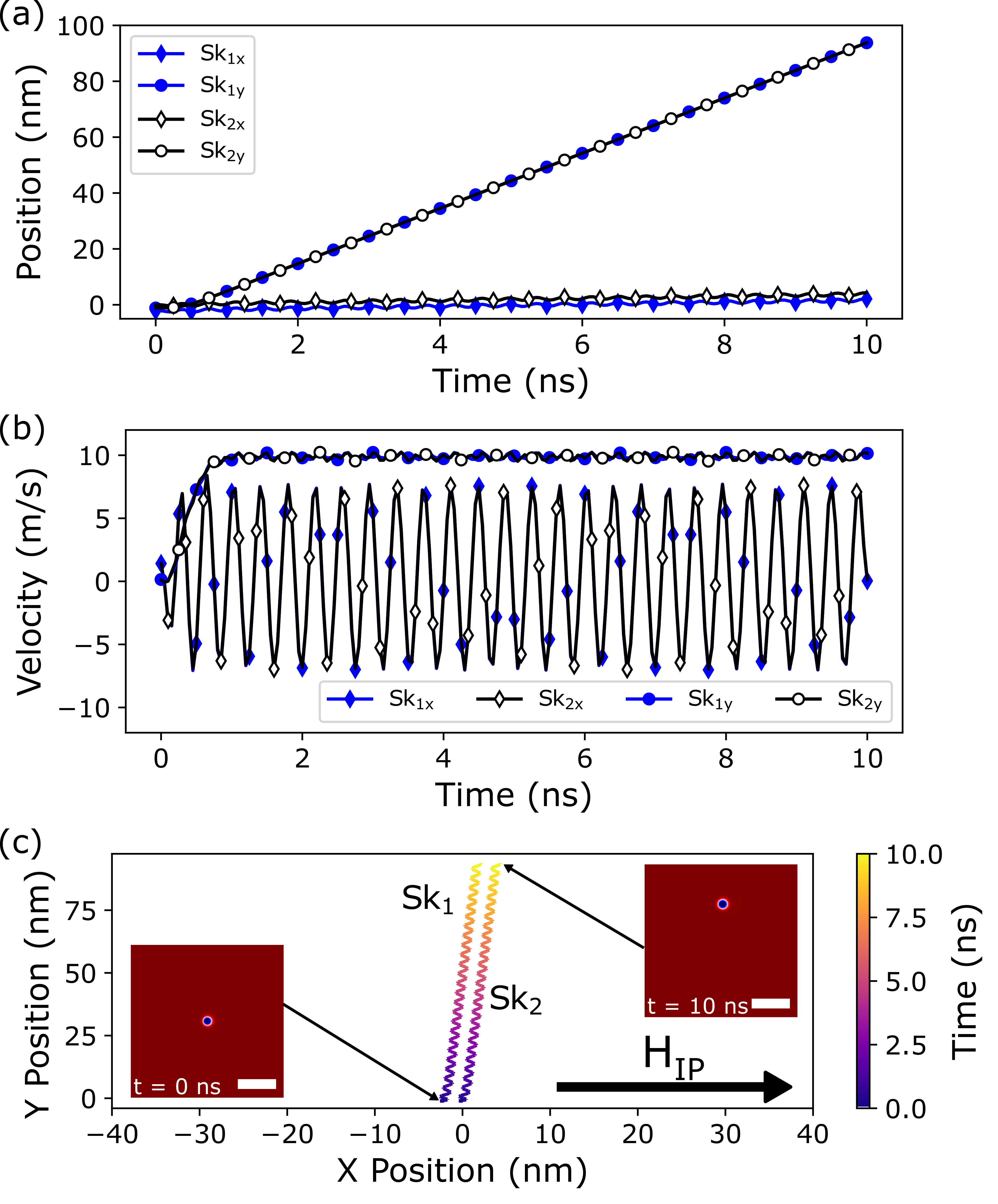}
		\caption{In-plane field-driven skyrmion motion in a field of 0.5~T along \textit{x}, driven at a frequency of 57.4~GHz. Panel (a) shows the \textit{x}- and \textit{y}-positions of the skyrmion guiding centre for both the skyrmions in layers 1 and 2 of the SAF. Panel (b) shows the calculated velocities of these guiding centre \textit{x}- and \textit{y}-positions as a function of time. Points on the lines in panels (a) and (b) are deliberately undersampled to see both the datapoints of layers 1 and 2. Panel (c) shows the trajectory of both skyrmions, with the \textit{y}-positions of each skyrmion plotted against the \textit{x}-position. The colour gradient of each line denotes the time, which is shown on the colour bar to the right of the panel. \highlight{The two insets show the position of the skyrmion in one of the layers at the start (t = 0~ns) and end (t = 10~ns) of the simulation. Scale bar is 100 nm. The arrow in the bottom right} shows the direction of the in plane static field.}
		\label{fig:Skyrmion_Motion}
	\end{figure}
	Considering the static and dynamic characterisation of skyrmions in the simulated system described in Sec.~\ref{sec:Static_Char} and Sec.~\ref{sec:Dynamic_Char}, we now move on to describing the effect of an in-plane field together with an excitation field on the motion of the SAF skyrmion. The frequencies of the two peaks of Fig.~\ref{fig:Skyrmion_breathing}(a) are found by fitting the data to a Lorentzian lineshape to be 14.3~GHz and 57.4~GHz. A sinusoidal driving field is then used at the higher frequency to drive the skyrmion together with a static out-of-plane field of 50~mT as discussed in Sec.~\ref{sec:Dynamic_Char}. This is also done with an in-plane static field of 0.5~T. The results of this are shown in Fig.~\ref{fig:Skyrmion_Motion}. Fig.~\ref{fig:Skyrmion_Motion}(a) shows the \textit{x}- and \textit{y}-positions of the guiding centres of the skyrmions in each layer as a function of time for 20~ns. We see that---while the skyrmion moves steadily along the \textit{y}-axis at a seemingly constant rate---the \textit{x}-axis motion proceeds in an oscillatory fashion with an oscillation frequency of $\sim$1~GHz, with an additional very small net movement along the \textit{x}-axis. This can also be visualised with the trajectories of the two skyrmions shown in Fig.~\ref{fig:Skyrmion_Motion}(c)\highlight{, together with the insets showing the initial and final positions of the skyrmion}. We see that the skyrmions move in a wiggle-like fashion, the first report of this kind of motion in a synthetic antiferromagnet. This is also unlike the motion observed in ref.~\cite{Qiu2021} but similar to the trajectories in similar ferromagnetic systems reported in~\cite{Yuan2019}. The position as a function of time is used to calculate the \textit{x}- and \textit{y}-velocities of each skyrmion as a function of time, which is shown in Fig.~\ref{fig:Skyrmion_Motion}(b). The \textit{y} velocities ramp up from 0 for the first $\sim$2~ns, before maintaining a steady value of $\sim$6~m/s for the rest of the simulation time. Meanwhile, the \textit{x} velocities oscillate around 0, driving the wiggle-like motion. To calculate the average velocities presented later, an average is taken of the \textit{y}-axis velocity once it reaches its steady value, here around 3~ns into the simulation time.\\\\
	
	\subsection{Skyrmion velocity coupled to out-of-phase breathing}\label{sec:B_Field_Motion_Fn_Freq}
	\begin{figure}
		\centering
		\includegraphics[width=.8\linewidth]{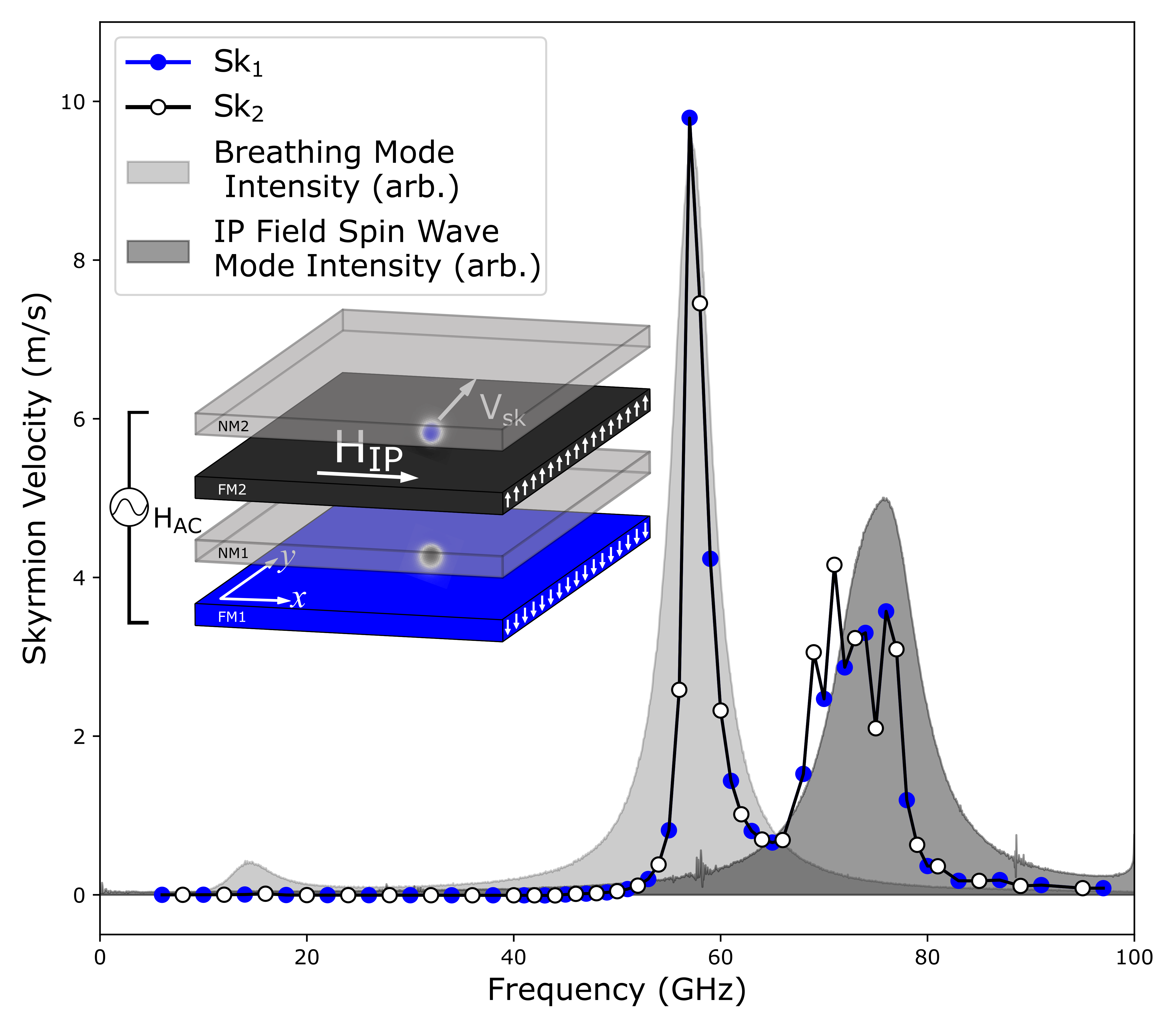}
		\caption{Velocities of the two skyrmions as a function of frequency of the driving field. Datapoints are deliberately undersampled so that both datasets are visible. The intensity of the breathing modes from Fig.~\ref{fig:Skyrmion_breathing} is overlaid with light grey shading and rescaled to match the peak amplitudes. \highlight{The power spectral density of the spin wave mode arising from the canted spins in the IP field (See supplementary note S2) is also overlaid in a dark grey shading.} Inset shows the same schematic from Fig.~\ref{fig:stack_structure} to illustrate the directions of the magnetic fields and the direction of skyrmion motion.}
		\label{fig:velocity_fn_freq}
	\end{figure}
	
	The average velocity as a function of frequency is presented in Fig.~\ref{fig:velocity_fn_freq} for an antiferromagnetic interlayer exchange coupling strength of -0.3~$\mathrm{mJ/m^2}$, over a range of frequencies from 6~GHz to 97~GHz. Added for comparison is the PSD of the breathing modes from Fig.~\ref{fig:Skyrmion_breathing}(a) rescaled to allow comparison with the calculated velocities. There is a strong overlap between the peak velocity of the skyrmion and the peak in the breathing mode intensity, suggesting a coupling between the breathing mode and the velocity. However, perhaps surprisingly there is no similar peak when driving the skyrmion motion at the lower frequency of the in-phase peak.\\ 
	Between frequencies of $\sim$70 and $\sim$80~GHz in Fig.~\ref{fig:velocity_fn_freq}(a)---above the out-of-phase breathing resonant frequency---there is a second broader peak in velocity that is seemingly not related to any skyrmion breathing mode. This is reminiscent of the higher frequency hybridisation of skyrmion breathing with spin wave modes of ferromagnetic~\cite{Kim2014} and synthetic antiferromagnetic~\cite{Lonsky2020_PRB} nanodots that has been previously observed. However, as discussed in Sec.~\ref{sec:Methods}, in this simulation periodic boundary conditions in the \textit{x-y} plane are used. This excludes such spin wave resonant modes, since they originate from spin canting at the edges of the simulation, which the application of periodic boundary conditions removes. We instead suggest that this higher frequency broad peak in velocity arises from exciting a hybrid mode between the skyrmion breathing and a spin wave mode arising from the canting of spins in the applied in-plane field. Because of the motion that would be induced by performing the simulations used to generate Fig.~\ref{fig:Skyrmion_breathing} in an in-plane field, this is difficult to show directly. However, in supplementary note~S2 we show and discuss simulation of the PSD in the presence of an in-plane field but without a skyrmion, which yields a peak of similar centre frequency and width to this second peak in velocity. \highlight{The power spectral density arising from this is shown in Figure~\ref{fig:velocity_fn_freq} as the darker gray shaded region, with a peak at higher frequencies, approximately matching the corresponding peak in velocity. This is discussed in detail in supplementary note S2, however it clearly demonstrates that the peak in velocity arises from hybridisation between the skyrmion breathing mode and the spin-wave resonance of the canted spins when an in-plane field is introduced.}\\\\

	\subsection{The effect of the spin wave emission amplitude}\label{sec:Spin_Wave_Emission_Vs_Radius}
	\begin{figure*}
		\centering
		\includegraphics[width=\linewidth]{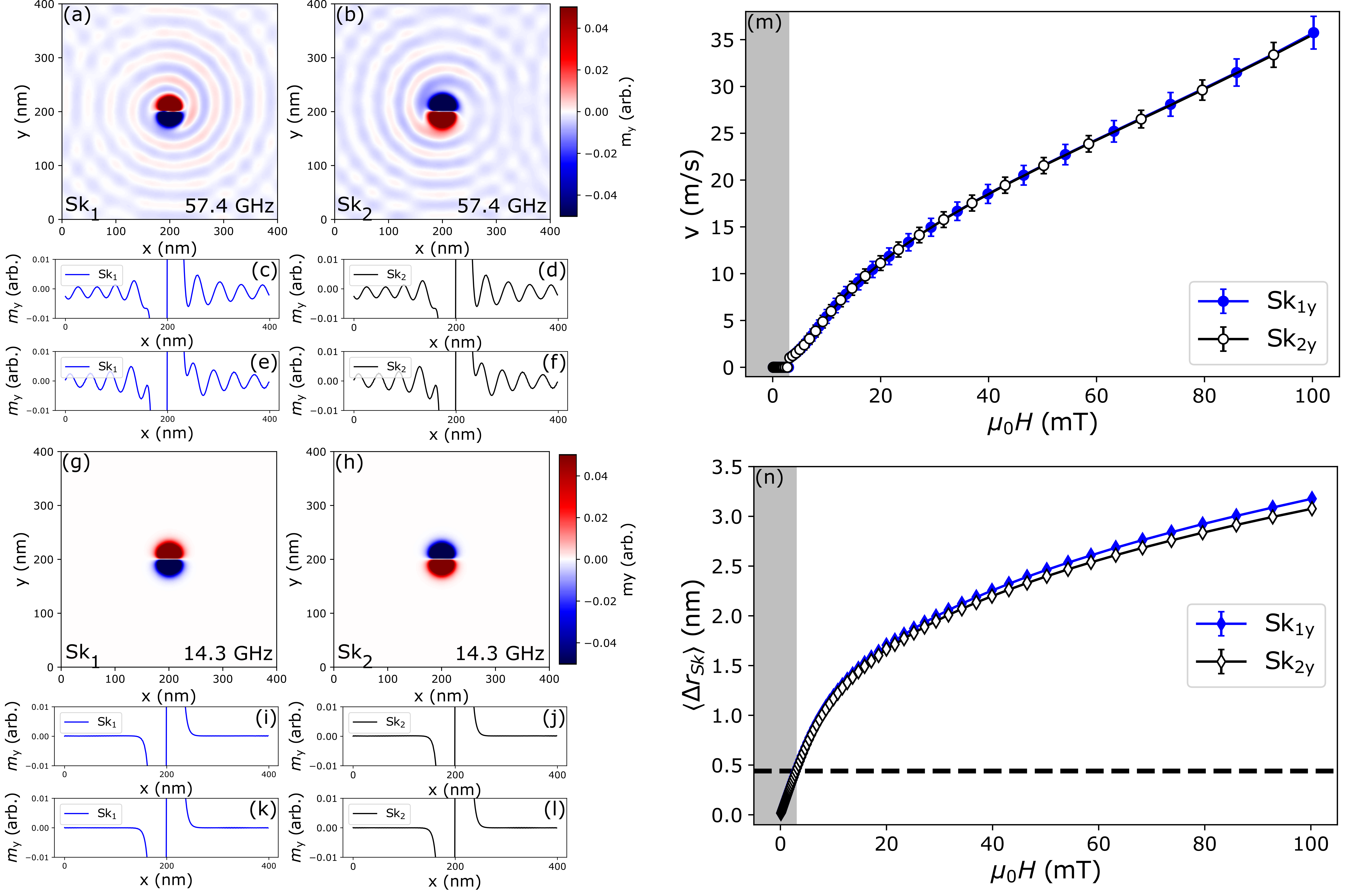}
		\caption{(a,b) Snapshots of the $y$-component of the magnetisation in layers 1 and 2 respectively at t = 0.2~ns when driven at 57.4~GHz. The colormap has been windowed to show the spin wave emission. (c,d) Show line cuts through the centre of the skyrmion from (a,b), while (e,f) show the equivalent line cuts through the centre of the skyrmion in the case where the in-plane field is removed, but still at 0.2~ns after the start of the simulation with the same out-of-plane driving field. Panels (g-l) are all the same but with a 14.3~GHz driving field. (g,h) Are snapshots of layers 1 and 2 respectively at 0.2~ns, (i,j) are line cuts through (g,h), and (k,l) are the equivalent line cuts in the absence of an in-plane field. Panel (m) shows the average skyrmion velocity when driven with a static in-plane field of 0.5~T as a function of the out-of-plane excitation field. The gray shaded region denotes the field range below which there is no motion. Datapoints are deliberately undersampled so that both datasets are visible. Panel (n) shows the average peak-to-peak variation in the skyrmion radius when breathing as a function of the amplitude of the out-of-plane field in the absence of an in-plane field. The gray shaded region is taken from panel (m) and the black dashed line is a guide to the eye to separate radii measured with excitation fields above and below this region. Note that the error bars on these series are too small to be visible.}
		\label{fig:Velocity_Amplitude_Radius_Compare}
	\end{figure*}
	
	It has been discussed that the motion of the skyrmions in these systems is driven by spin wave emission~\cite{Yuan2019,Wang2015,Qiu2021}, however, the mechanism of this spin wave emission has not been investigated in detail. This emission is represented more clearly in Figs.~\ref{fig:Velocity_Amplitude_Radius_Compare}(a,b) which show the skyrmion when driven at 57.4~GHz in a 0.5~T in-plane magnetic field at 0.2~ns after the start of the simulation; at the onset of skyrmion motion. There is some interference present because of the periodic boundary conditions. Similarly Figs.~\ref{fig:Velocity_Amplitude_Radius_Compare}(g,h) show the same situation when the system is driven with a field of 14.3~GHz, the frequency of the in-phase breathing mode. These images are complemented by supplementary videos S1 and S2 which show this spin wave emission over the first 1~ns of the simulation in the presence of an in-plane magnetic field. To further illustrate the spin-wave emission, each of these pairs of colormaps are complemented with line cuts that are take through the centre of the skyrmion, shown in Fig.~\ref{fig:Velocity_Amplitude_Radius_Compare}(c,d) for the 57.4~GHz case, and Fig.~\ref{fig:Velocity_Amplitude_Radius_Compare}(i,j) for the 14.3~GHz case. These line cuts are in turn complemented by supplementary videos S3 and S4, which again show this profile over the first 1~ns of the simulation, in an in-plane magnetic field of 0.5~T, driven by an out-of-plane oscillating field of 20~mT, at 57.4~GHz and 14.3~GHz in supplementary videos S3 and  S4, respectively. These videos show that the spin waves are emitted at the maxima of the oscillations in the skyrmion radius. This is remarkably similar to the experimental demonstration of spin-wave emission from an oscillating domain wall~\cite{Liu2025}, where the oscillating wall `knocks' the domain interior, stimulating spin waves~\cite{Liu2025}. It is clear from Fig.\ref{fig:Velocity_Amplitude_Radius_Compare}(a,b) and Fig.~\ref{fig:Velocity_Amplitude_Radius_Compare}(g,h) that the main difference between being driven at 57.4~GHz---where there is skyrmion motion---and at 14.3~GHz---where there is no skyrmion motion---is the amplitude of the emitted spin waves. It is well understood that the oscillations in radius at the out-of-phase breathing mode are considerably larger than at the in-phase breathing mode~\cite{Lonsky2020_PRB, Barker2023_JAP}. We suggest that the primary factor determining the strength of the emitted spin waves is the magnitude of the change in radius. This is in turn affected by the magnitude of the driving field. However, for the same magnetic field amplitude, this will increase when tuning onto a skyrmion breathing mode resonant frequency, and will also be considerably larger for the out-phase breathing mode frequency. \\\\
	To investigate this in more detail, we examine the amplitude of the variation in radius in the absence of an in-plane magnetic field. We do this because the deformation of the skyrmions---as shown in Fig.~\ref{fig:stack_structure}---will affect the accuracy of determining the change in radius. Fig.~\ref{fig:Velocity_Amplitude_Radius_Compare}(e,f) show snapshots of the same slice through the centre of the skyrmion at the same time as Fig.~\ref{fig:Velocity_Amplitude_Radius_Compare}(c,d) at the same driving field frequency of 57.4~GHz, but without the in-plane field. Likewise, Fig.~\ref{fig:Velocity_Amplitude_Radius_Compare}(k,l) show the same, but at a driving field frequency of 14.3~GHz, equivalent to Fig.~\ref{fig:Velocity_Amplitude_Radius_Compare}(i,j) but without the in-plane magnetic field. Both of these sets of snapshots are supported by supplementary videos S5 and S6 which show the same data over the first 1~ns of the simulation. For a detailed description of all of the supplementary videos see supplementary note~S3.\\\\
	In Fig.~\ref{fig:Velocity_Amplitude_Radius_Compare}(m,n), we show a simultaneous study of the skyrmion velocity (Fig.~\ref{fig:Velocity_Amplitude_Radius_Compare}(m)) and the magnitude of the variation in radius (Fig.~\ref{fig:Velocity_Amplitude_Radius_Compare}(n)), both as a function of the magnitude of the driving microwave magnetic field at a frequency of 57.4~GHz. For details of how the average change in radius was calculated, see supplementary note~S4. In Fig.~\ref{fig:Velocity_Amplitude_Radius_Compare}(m), at fields above $\sim$20~mT there is a clear linear relationship between the velocity and the driving field. However, below fields of $\sim$20~mT, the skyrmion velocity drops off faster towards 0~m/s, and there is no motion below driving fields of $\sim$3~mT. This region of no skyrmion motion is represented with the grey shaded box. Since this is a simulated system with no pinning, this is clearly an intrinsic effect of their being an insufficient spin-wave emission amplitude to drive motion. Likewise, examining the magnitude of the variation in radius in Fig.~\ref{fig:Velocity_Amplitude_Radius_Compare}(n), we draw the same threshold field for skyrmion motion from Fig~\ref{fig:Velocity_Amplitude_Radius_Compare}(m) also as a grey shaded box. For these magnetic parameters, we are then able to define a variation in radius of the skyrmions below which there is no motion of $\langle \Delta r_{\mathrm{Sk}}\rangle$ = 0.44~nm, or $\sim$3.5~\% of the static skyrmion radius $r_0$.\\\\
	To summarise, the motion of the skyrmions is primarily driven by the emission of spin waves. As we have shown, this spin wave emission occurs when the magnitude of the skyrmion expansion is at its widest. The results of Fig.~\ref{fig:Velocity_Amplitude_Radius_Compare} show that there is a minimum variation of skyrmion radius required to emit spin waves that are strong enough to drive motion. This radius variation will itself be dependent on the driving field; however, the relationship between radius variation and driving field will be different at the two resonant breathing modes. For this reason, we have determined empirically that it is more accurate to define a minimum variation in radius rather than a minimum driving field. This interpretation also explains the absence of skyrmion motion at the resonant frequency of the in-phase breathing mode in Fig.~\ref{fig:velocity_fn_freq}(a). The variation in skyrmion radius at this frequency is well understood to be significantly smaller for the same driving magnetic field amplitude than when driven at the out-of-phase resonant frequency~\cite{Lonsky2020_PRB, Barker2023_JAP}. Performing the same analysis on the variation in radius of the in-phase breathing mode when driven by a out-of-plane excitation field of 20~mT gives a $\langle\Delta r_{\mathrm{Sk}} \rangle$ of 0.286$\pm$0.002~nm, below this threshold for motion.\\\\
	\highlight{Furthermore, by performing these simulations, we have also developed an understanding of the relationship between skyrmion velocity and microwave power. Figure~\ref{fig:Velocity_Amplitude_Radius_Compare}(m) shows the velocity of the skyrmion as a function of the amplitude of the driving field, which is simply a representation of the microwave power driving the spin wave emission. As we have already discussed, for the lowest fields there is no observable skyrmion motion, before a non-linear ramp up in velocity between 3~mT and 20~mT, after which there is a linear relationship between skyrmion velocity and microwave power. This result will help to inform experimental study of skyrmion motion driven by this effect, where the exact relationship between microwave power and magnetic field amplitude will depend on factors such as fabrication of waveguides and losses in connectors, among others.}\\\\
	
	\subsection{The effect of antiferromagnetic interlayer exchange coupling}\label{sec:exchange_coupling_motion}
	\begin{figure}
		\centering
		\includegraphics[width=0.8\linewidth]{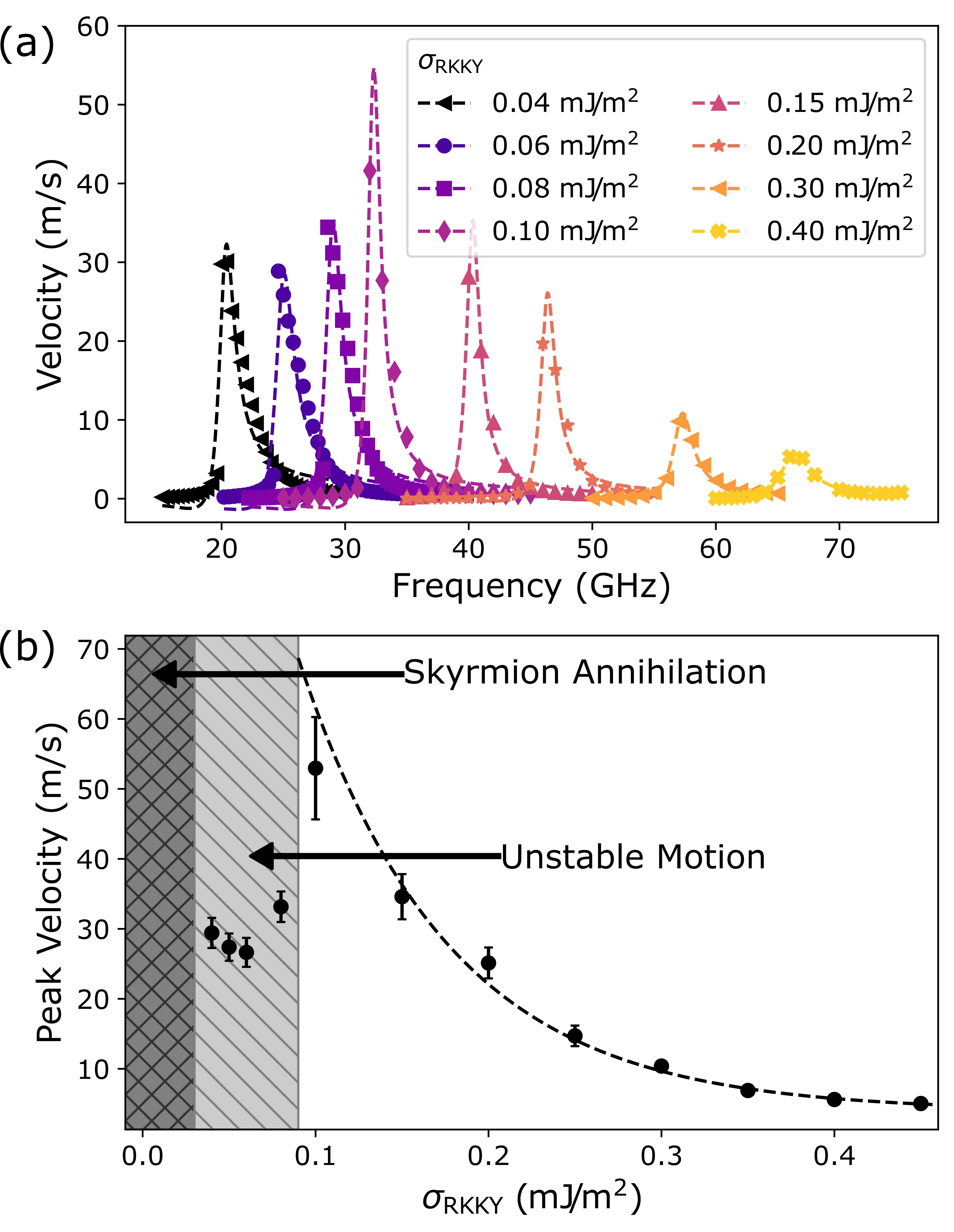}
		\caption{(a) Skyrmion velocity as a function of frequency plotted for varying antiferromagnetic exchange coupling $\sigma_{\mathrm{RKKY}}$. Dashed lines show fits to an asymmetric Lorentzian function used to extract the peak velocity. All simulations were performed with static fields of 50~mT out-of-plane and 0.5~T in-plane, along with a microwave driving field of 20~mT. (b) The extracted peak velocities as a function of the coupling strength. The regime where at least one skyrmion is annihilated when driven on resonance is cross-hatched, while the coupling regime where the skyrmions move unstably is single hatched. The dashed line shows an inverse cube law as a guide to the eye for peak velocities in the high coupling strength regime.}
		\label{fig:Vel_Fn_RKKY}
	\end{figure}
	With our new understanding of the coupling between skyrmion velocity and the frequency of the out-of-phase breathing mode, we now move to studying it as a function of the antiferromagnetic interlayer exchange coupling strength. The results of measuring the velocity as a function of frequency at different coupling strengths $\sigma_{\mathrm{RKKY}}$ are presented in Fig.~\ref{fig:Vel_Fn_RKKY}(a). The data presented here has been trimmed around the frequency of the out-of-phase breathing mode for clarity, so the broad peak in velocity associated with the in-plane field is not shown. This however follows the same behaviour---albeit shifted in frequency---as the one for $\sigma_{\mathrm{RKKY}} = 0.3\ \mathrm{mJ/m^2}$ shown in Fig.~\ref{fig:velocity_fn_freq}(a) and discussed in supplementary note~S2. The extracted values of skyrmion velocity as a function of frequency are fitted to an asymmetric Lorentzian lineshape, chosen for its empirical match to the data, which allows us to extract a value for the peak velocity. This peak velocity as a function of $\sigma_{\mathrm{RKKY}}$ is then shown in Fig.~\ref{fig:Vel_Fn_RKKY}(b).\\
	In Ref.~\cite{Qiu2021}, Qiu et al. discuss the variation of the skyrmion velocity (driven by modulations in anisotropy) as a function of $\sigma_{\mathrm{RKKY}}$, and propose this as a way of modulating the skyrmion velocity. Our results show that while this variation in velocity with $\sigma_{\mathrm{RKKY}}$ is true at a fixed frequency, it is not fundamentally related to the value of $\sigma_{\mathrm{RKKY}}$. Instead, we have clearly shown that the magnitude of the velocity is correlated to the size of the oscillations in skyrmion size. This modulation of the skyrmion size is itself a result of tuning onto the intrinsic frequency of the out-of-phase breathing mode. Thus, by examining the peak velocity value for each $\sigma_{\mathrm{RKKY}}$ we can understand the true impact of the antiferromagnetic interlayer exchange coupling strength on the skyrmion motion. We define three principal regimes of the skyrmion velocity when driven on the resonant frequency.\\
	First, we define the skyrmion annihilation regime for low values of $\sigma_{\mathrm{RKKY}}$ below $\sim 0.03\ \mathrm{mJ/m^2}$. Here the skyrmions are stable in the static case and skyrmion motion is still possible when driving with microwave magnetic fields. However, when it is driven close to the resonant frequency, the oscillations of the skyrmions at this field strength are sufficiently large to drive the annihilation of the skyrmions in at least one of the layers. Clearly the strength of the antiferromagnetic coupling is insufficient to stabilise the two skyrmions as a topological object.\\
	For intermediate values of $\sigma_{\mathrm{RKKY}}$ between $\sim 0.03\ \mathrm{mJ/m^2}$ and $<0.1\ \mathrm{mJ/m^2}$, the strength of the coupling is sufficient to prevent the annihilation of the skyrmions, however their oscillations are substantially more unstable than at higher coupling strengths, and the \textit{x}-axis trajectories (the direction orthogonal to the principal direction of motion) of the two guiding centres drift apart by up to several nanometres over the course of the simulations. We suggest that this disparity between the two skyrmions restricts their ability to move together efficiently, and we see that the peak velocity over this narrow region of $\sigma_{\mathrm{RKKY}}$ is almost completely flat. In the simulation at $\sigma_{\mathrm{RKKY}} = 0.08\ \mathrm{mJ/m^2}$ there is a small uptick in velocity as we transition to the high $\sigma_{\mathrm{RKKY}}$ regime.\\
	For values of $\sigma_{\mathrm{RKKY}}$ from $0.1\ \mathrm{mJ/m^2}$ and up, we see a very clear inverse relationship between the peak skyrmion velocity and the strength of the antiferromagnetic interlayer exchange coupling. The two skyrmions move together with regular oscillations in size for the whole simulation space. The decrease in velocity with $\sigma_{\mathrm{RKKY}}$ can be clearly understood as a consequence of the constraining effect of the antiferromagnetic coupling strength, which as it increases will reduce the magnitude of the oscillations in skyrmion radius, and in turn reduce the velocity.\\
	While experimental measurements of skyrmions and other magnetic textures are limited in synthetic antiferromagnets, several studies report observations of skyrmions and domain walls. In Table~\ref{table:coupling_strengths} we summarise the reported antiferromagnetic coupling strengths from a literature search of DMI stabilised textures in synthetic antiferromagnets. The fact that most of these reported values fall in the range of $\sigma_{\mathrm{RKKY}}$ in our study where we report skyrmion motion suggests that this method would be appropriate for driving SAF skyrmions in a future computing device. 
	\begin{table}
		\caption{Compilation of reported antiferromagnetic interlayer exchange coupling strengths from several previous studies.}
		\label{table:coupling_strengths}
		\begin{ruledtabular}
			\begin{tabular}{c c c c}
				Material & Magnetic & Coupling & Reference \\
				System &   Texture & Strength ($\mathrm{mJ/m^2}$) &   \\
				\hline
				Co/Ru/Pt/Co & Skyrmions & -0.23 & \cite{Legrand2020} \\
				Pt/Co/CoFeB/ & Skyrmions & -0.13 & \cite{Dohi2019} \\
				Ir/Co/CoFeB/W &  &  &  \\
				Co/Ni/Co/Ru/ & Domain Walls & -0.5 & \cite{Yang2015} \\
				Co/Ni/Co &  &  &  \\
				$[$CoFeB/Ru/Pt/ & Domain Walls & -0.018 & \cite{Barker2023_JPhysD} \\
				CoB/Ru/Pt$]_{\times 5}$ &  &  &  \\
			\end{tabular}
		\end{ruledtabular}
	\end{table}
	
	\section{Skyrmion Motion in Synthetic Ferrimagnets}
	
	\begin{figure}
		\centering
		\includegraphics[width=\linewidth]{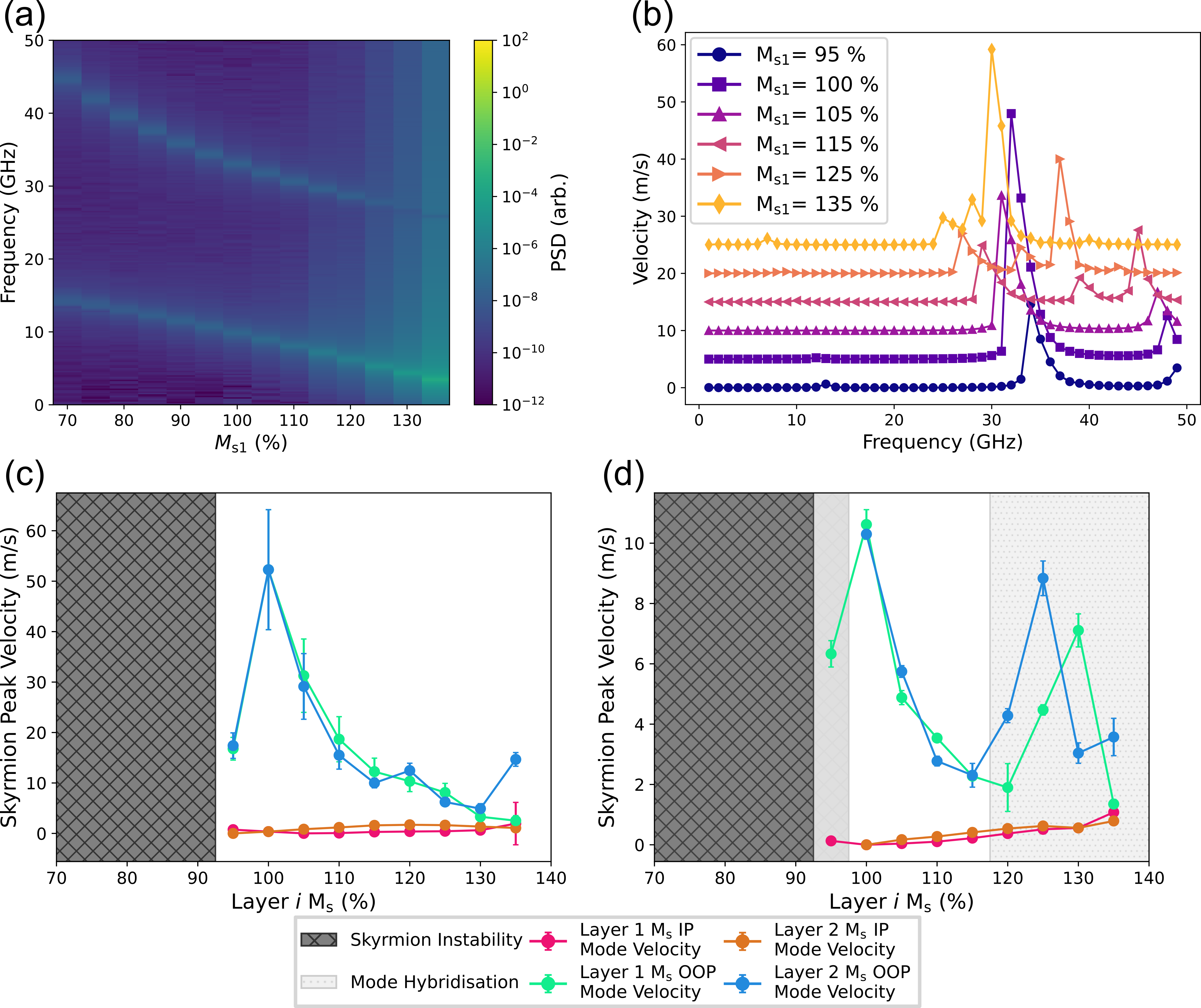}
		\caption{\highlight{Breathing modes and motion as a function of decompensation of one of the magnetic layers, expressed as a percentage of the original $M_{\mathrm{s}}$ value of 580$\times 10^3\ \mathrm{A/m^2}$. (a) The PSD of the system, varying the $M_{\mathrm{s1}}$ for an antiferromagnetic interlayer exchange coupling of $-0.1\ \mathrm{mJ/m^2}$. (b) Plots of the velocity as a function of frequency for several selected values of decompensation of layer 1, for the same antiferromagnetic interlayer exchange coupling. Note that each dataset has been deliberately offset to better see the individual trends. (c,d) Values of the peak velocities at both the IP and OOP breathing mode peaks as a function of decompensation of the $M_{\mathrm{s}}$ of both layer 1 and layer 2, for antiferromagnetic interlayer exchange coupling strengths of $-0.1\ \mathrm{mJ/m^2}$ and  $-0.3\ \mathrm{mJ/m^2}$, respectively. Note that (c,d) share a common legend, seen below both panels.}}
		\label{fig:SyFM}
	\end{figure}
	
	\highlight{In experimental SAFs, it is often difficult to grow magnetic layers with perfect compensation, especially when using layers of multiple compensations~\cite{Barker2023_JPhysD, Barker2024_PRB}. To extend our understanding of the effects presented in this paper, we consider synthetic ferrimagnets, where we introduce a decompensation of the magnetisation of one of the two layers. The measurement of the breathing modes has already been demonstrated in Ref.~\cite{Lonsky2020_PRB}, where simulation of a synthetic ferrimagnet (SFiM) was achieved by varying the $M_{\mathrm{s}}$ of one layer. Similarly, the current driven dynamics of skyrmions in angularly decompensated SAFs have been studied by varying the ration of $M_{s1}/M_{s2}$ of the two layers~\cite{haltz2023_arxiv}.  Here, we follow this method, and examine our system at two characteristic antiferromagnetic interlayer exchange coupling strengths, while varying the $M_{\mathrm{s}}$ of each layer individually by $\pm$30\% of its original value. We note that in most experimental SAFs the decompensation between the magnetisations of each layer is $\leq$10\%~\cite{Barker2023_JPhysD, Barker2024_PRB}. Thus, our chosen range is adequate to describe experiments well, while also extending our understanding to higher decompensations. Throughout we conduct every simulation by fixing the $M_{\mathrm{s}}$ of one layer to its original value, and vary the $M_{\mathrm{s}}$ of the second layer. We use the labels $M_{s1}$ and $M_{s2}$ to denote these two values, and for every simulation we consider the case of varying each of these two quantities.\\\\
	The breathing mode frequencies for an antiferromagnetic interlayer exchange coupling strength of $\sigma_{\mathrm{RKKY}} = -0.1\ \mathrm{mJ/m^2}$ and a variation of $M_{s1}$ by $\pm$30\% are shown in Figure~\ref{fig:SyFM}(a) as a heatmap. The PSD here is taken for only layer 1 of the system, in order to be able to observe the signal of the in-phase breathing mode, however the behaviour of the mode frequencies is the same for the PSD of layer 2 and the whole system. In general, the resonant frequencies of both the in-phase and the out-of-phase modes both decrease as the value of $M_{s1}$ increases. This matches the results of Ref.~\cite{Lonsky2020_PRB}, where it was shown that in compensated SAFs, the frequencies of the resonant modes decrease with $M_{\mathrm{s}}$. Likewise, our results match those considering SFiM systems~\cite{Lonsky2020_PRB}, where it was shown that the frequencies of the resonant modes continue to follow this trend even when the variation in $M_{\mathrm{s}}$ is restricted to only one of the magnetic layers.\\\\
	With this understanding of the behaviour of the two resonant modes, we move on to consider the velocity of the skyrmions in uncompensated systems. The velocity as a function of frequency for several representative values of $M_{s1}$ is shown in Figure~\ref{fig:SyFM}(b). This data is for an antiferromagnetic interlayer exchange coupling strength of $\sigma_{\mathrm{RKKY}} = -0.1\ \mathrm{mJ/m^2}$. Here, each dataset is deliberately offset along the $y$-axis to be able to see the individual trends. It is clear that the velocity at the out-of-phase peak is lower than at full compensation (see Figure~\ref{fig:Vel_Fn_RKKY} for the measurement at this value of $\sigma_{\mathrm{RKKY}}$), and this decreases as a function of the decompensation of $M_{s1}$. Conversely, at the highest decompensations, there is a clear evidence of a small peak in velocity near the in-phase breathing mode frequency. These two effects are better demonstrated in Figure~\ref{fig:SyFM}(c), which shows the measured velocity at each of these two resonant frequencies as a function of decompensation for both $M_{s1}$ and $M_{s2}$. There is a clear peak in velocity at the out-of-phase resonant frequency when $M_{s1} = M_{s2}$, with skyrmion velocity trailing off at decompensation either side of this value. For $M_{s1,2}$ compensations below 90\%, there is no velocity data. This is due to the small value of $M_{\mathrm{s}}$. For these low $M_{\mathrm{s}}$ values, the increased field amplitude used to drive motion (compared to the excitation of the breathing modes in Figure~\ref{fig:SyFM}(a)), annihilates the skyrmions.\\\\
	As expected, the behaviour is reversed when considering the velocity at the in-phase resonant frequency. The velocity is at a minimum value when $M_{s1} = M_{s2}$, and increases away from this. This can be simply understood considering the relationship between skyrmion velocity and breathing amplitude from Section~\ref{sec:Spin_Wave_Emission_Vs_Radius}. As the difference between the $M_{\mathrm{s}}$ of the layers increases, so too will the difference in the oscillation amplitudes of each skyrmion. As we have discussed this is the main factor determining the skyrmion velocity, and so naturally the velocity increases at this resonant frequency with the increase in unbalance.\\\\
	Finally, we reproduce this work for a second antiferromagnetic interlayer exchange coupling strength of $\sigma_{\mathrm{RKKY}} = -0.3\ \mathrm{mJ/m^2}$. The measurement of the breathing modes and the individual velocity as a function of frequency plots for each decompensation, analogous to Figures~\ref{fig:SyFM}(a) and \ref{fig:SyFM}(b) are not shown here, but are instead summarised in supplementary note S5. The velocity of each of the resonant modes as a function of the decompensation is shown in Figure~\ref{fig:SyFM}(d). As before, the skyrmion is annihilated when the $M_{\mathrm{s}}$ is below 95~\% of the original value, demonstrating that this is an intrinsic effect of the low magnetisation, rather than a consequence of the weaker coupling present in Figure~\ref{fig:SyFM}(c). We see similar behaviour at compensations near to $M_{s1} = M_{s2}$, with a peak in velocity at the out-of-phase resonant frequency at full compensation, and increasing velocity at the in-phase resonant frequency with increasing decompensation. However, for values of $M_{s1,2} \geq 120\%$, there is a large degree of variation in the velocity at the out-of-phase resonant frequency. This is due to hybridisation between the skyrmion resonant frequency and the spin wave mode of the canted spins discussed in Section~\ref{sec:B_Field_Motion_Fn_Freq} and Supplementary Note~S2. Here the resonant frequency of the spin wave mode arising from the canted spins decreases faster with increasing $M_{s1,2}$ than the skyrmion breathing mode frequency. Because the two modes are closer in frequency at this interlayer exchange coupling strength, at the higher decompensations, the spin wave mode overlaps with the skyrmion's out-of-phase resonant frequency, and the two effects interact to create more unpredictable spin wave emission. There is also a trace of this effect in Figures~\ref{fig:SyFM}(b,c) for the lower interlayer coupling strength, however it only occurs at the very highest decompensation, and so is less noticeable.\\\\}
	
	\section{Electric Field Driven Skyrmion Motion}
	\begin{figure}
		\centering
		\includegraphics[width=\linewidth]{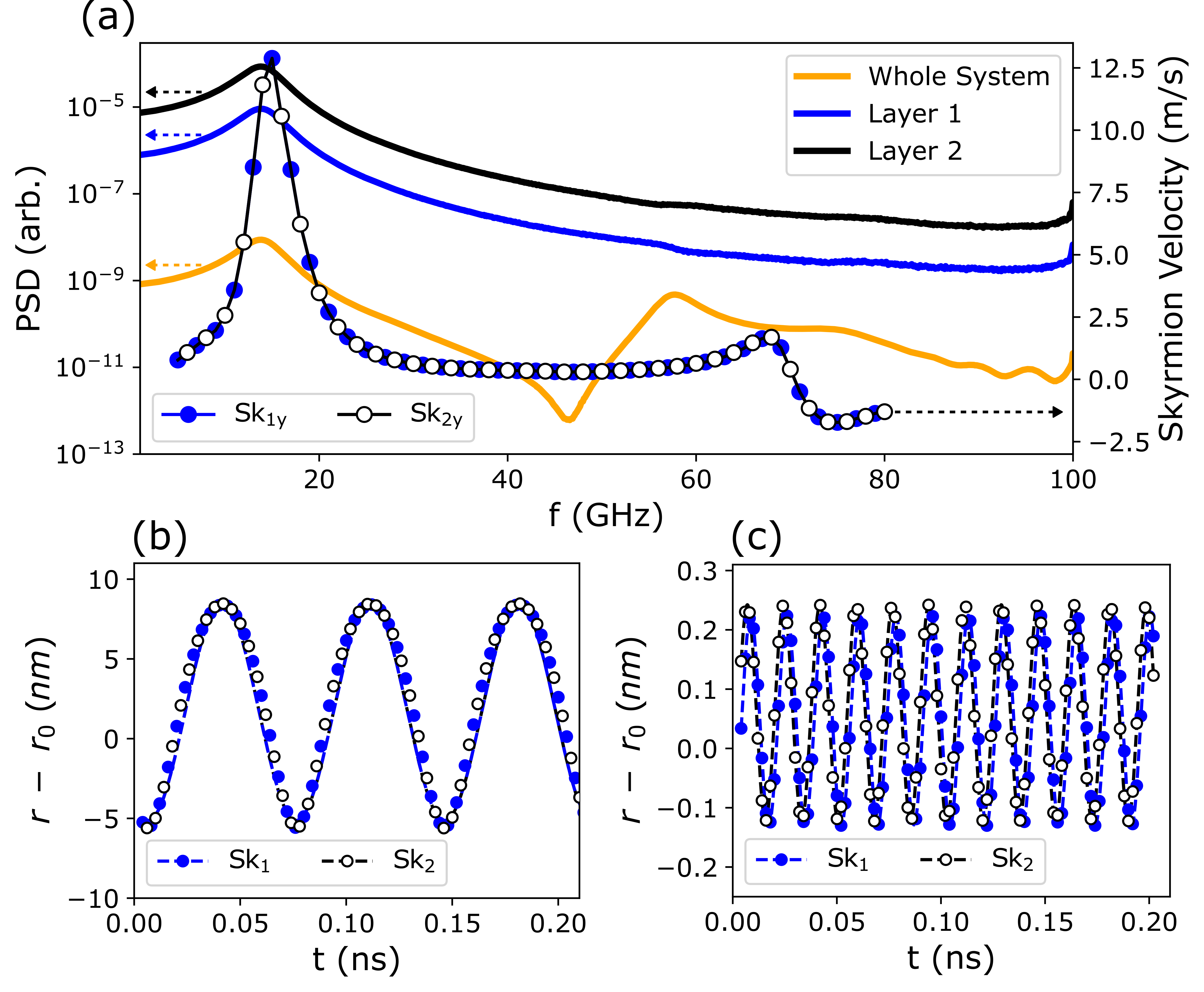}
		\caption{Examination of the effects of electric field on the skyrmion breathing and velocity at $\sigma_{\mathrm{RKKY}} = -0.3\ \mathrm{mJ/m^2}$. This is performed in an out-of-plane magnetic field of 50~mT, and in the case of the motion an in-plane magnetic field of 0.5~T. (a) The PSD of the whole system along with the two individual layers when excited with a modulation of the uniaxial anisotropy of $3\times 10^4 \ \mathrm{J/m^3}$ is plotted on the left axis as a function of frequency. On the right vertical axis, the skyrmion velocity when driven by an electric field as a function of frequency is plotted. Panels (b) and (c) show the variation in radius of the two skyrmions when driven by a sinusoidal electric field at 14.3 GHz and 57.4 GHz, respectively.}
		\label{fig:E_Field_Fig}
	\end{figure}
	Previous reports have studied skyrmion motion driven by a microwave electric field~\cite{Qiu2021}. In this work they simulated a similar system with an in-plane magnetic field and drove the motion of skyrmions by adding a microwave frequency perturbation to the anisotropy energy term. This mirrors the voltage controlled magnetic anisotropy that has been demonstrated in thin film multilayers~\cite{Mishra2024}. Ref.~\cite{Qiu2021} reported skyrmion velocities as a function of frequency, demonstrating a peak in the skyrmion velocity. However, the relationship between velocity and the resonant frequency of the skyrmion breathing modes was not identified. The frequencies at which they observed their peak velocites were close to the resonant frequency of the in-phase breathing mode that we have shown in Fig.~\ref{fig:Skyrmion_breathing}. To understand the differences between the mechanisms of magnetic-field driven skyrmion motion and the electric-field driven case previously reported~\cite{Qiu2021}, we finally perform some basic simulations using an electric field. In the same way as Ref.~\cite{Qiu2021}, the effects of an electric field are simulated by modulating the uniaxial anisotropy with an additional microwave frequency perturbation. We use the same value of $3\times 10^4 \ \mathrm{J/m^3}$ for the magnitude of this modulation of the uniaxial anisotropy in all simulations. An interlayer exchange coupling of $\sigma_{\mathrm{RKKY}} = -0.3\ \mathrm{mJ/m^2}$ was used, well into the high $\sigma_{\mathrm{RKKY}}$ regime from Sec.~\ref{sec:exchange_coupling_motion}. To match the results from Sec.~\ref{sec:B_Field_Motion_Fn_Freq}, we perform these simulations in a static out-of plane field of 50~mT as well as an in-plane field of 500~mT in the case of the simulations of motion.\\\\
	We first examine the breathing modes of the skyrmions, which is shown in Fig.~\ref{fig:E_Field_Fig}(a) for both the individual layers of the system as well as for the system as a whole. Here the features of the PSD occur at the same frequencies as in the magnetic field driven-breathing case, however the presence or lack thereof of each mode in the different layers is different. There is a strong peak at 14.3~GHz in all three traces, which in the magnetic-field driven case originates from in-phase breathing of the two skyrmions. However, the presence of the peak in the response of the whole system is less straightforward to account for. Alternatively, there is a peak in the whole system at 57.4~GHz where there is only a faint trace of in the responses of the individual layers. This behaviour is intriguing and merits further investigation.\\\\
	We also drive the motion of the skyrmions with an electric field of the same magnitude and varying frequency. The measured velocities as a function of the electric field frequency are also plotted on the right hand axis of Fig.~\ref{fig:E_Field_Fig}(a). Here we observe a strong peak in velocity that is well coupled to the lower frequency breathing mode. Unlike the magnetic field driven case, there is no observable motion at the higher frequency mode, but there is some motion around the frequency of the mode excited by the in-plane field (see supplementary note~S2). To understand this we present in Figs.~\ref{fig:E_Field_Fig}(b) and (c) simulations of the radius variation of the skyrmions when driven by a sinusoidal electric field with no in-plane magnetic field at each of the principal frequencies from Fig.~\ref{fig:E_Field_Fig}(a). At the lower 14.3~GHz peak, the radii of the two skyrmions varies massively with this electric field strength of $\langle\Delta r_{\mathrm{Sk_1}} \rangle = \langle\Delta r_{\mathrm{Sk_2}} \rangle = 13.6 \pm0.2$~nm. Our hypothesis in Sec.~\ref{sec:Spin_Wave_Emission_Vs_Radius} was that to stimulate skyrmion motion, a minimum variation in the skyrmion radius---of 0.44~nm for these micromagnetic energy terms---is required. The variation in radius at the in-phase resonant frequency is clearly large enough to drive motion. Likewise, this also explains the absence of motion at the other resonant breathing mode. The average variation in the radii of the two skyrmions when driven by an electric field at 57.4~GHz are $\langle\Delta r_{\mathrm{Sk_1}} \rangle = 0.1768 \pm 0.0001$~nm and $\langle\Delta r_{\mathrm{Sk_2}} \rangle = 0.1821 \pm 0.0001$~nm respectively, well below the threshold for motion. \highlight{This is a marked difference between the case of driving with an electric field and a magnetic field. We suggest that this arises from the difference in the mechanism of excitation. In the case of the electric field, this is simulated through a change in the anisotropy, which affects the radius of both skyrmions equally. On the other hand, when the breathing modes are excited with a magnetic field, the Zeeman coupling will have an opposite effect on each of the skyrmions due to the antiferromagnetic coupling. Thus, in the magnetic field case, at the in-phase resonant mode, the Zeeman effect will always suppress the oscillation of one of the two skyrmions and so the overall amplitude is smaller. There is no such competition in the electric field/anisotropy case, and so at this resonant mode maximum oscillations are allowed.}
	
	\section{Skyrmion Motion Driven By Perpendicular Current Densities}
	
	\begin{figure}
		\centering
		\includegraphics[width=\linewidth]{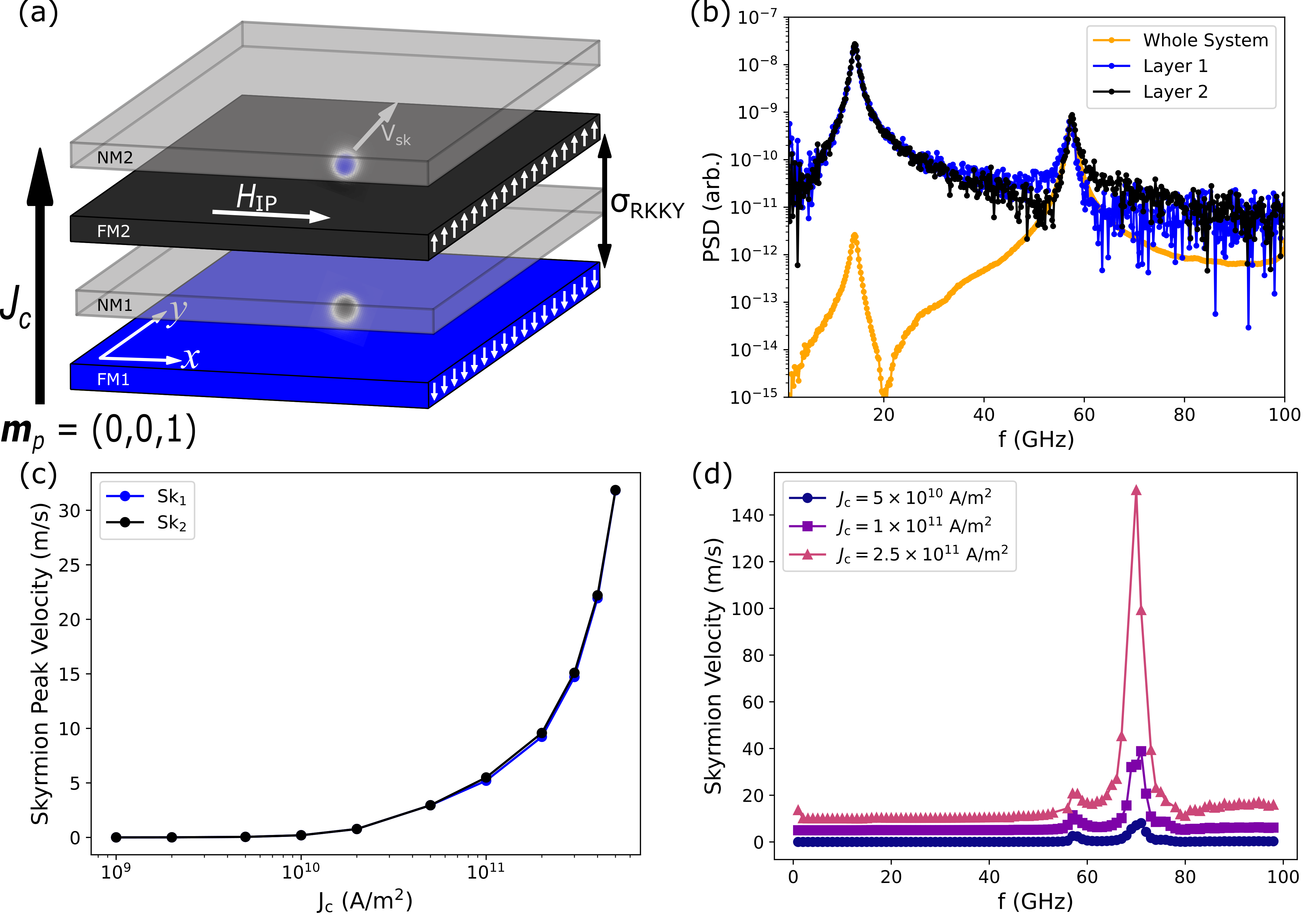}
		\caption{\highlight{(a). Illustration of the current perpendicular to plane (CPP) spin transfer torque (STT) geometry, with the direction of the current illustrated with the arrow labeled $J_c$ and the polarisation of the fixed layer illustrated with $\mathbf{m_p}$. (b). Spectral response of skyrmion system when stimulated with a CPP-STT with a current density of $1\times 10^9 \mathrm{A/m^2}$. (c). Skyrmion velocity when driven at the OOP breathing mode resonant frequeny of 57.6~GHz found from (b) as a function of current density. (d). Sweeps of the skyrmion velocity as a function of frequency for several representative current densities chosen from (c).}}
		\label{fig:CI_SWE}
	\end{figure}
	
	\highlight{In this section, we study the effects of the skyrmion breathing mode excitation by an electric current via a spin-torque mechanism (instead of a microwave magnetic field) in a fully compensated SAF. It has already been shown that a Slonczewski-type STT~\cite{Slonczewski1996, Berger1996,Xiao2004} can be used to qualitatively reproduce the effects of the microwave magnetic field when stimulating breathing modes~\cite{Lonsky2020_PRB}. Here, we extend this to considering the effects of an STT on the skyrmion velocity in the presence of a static in-plane magnetic field.\\\\
	To model a Slonczewski type STT, an additional torque term is added to the right hand side of the LLG (Equation~\ref{eq:LLG}), given by:
	\begin{equation}
		\vec{\mathbf{\tau}}_{SL} = \beta \dfrac{\epsilon - \alpha \epsilon'}{1 + \alpha^2}\Big[\vec{\mathbf{m}}\times \Big(\vec{\mathbf{m}}_P\times \vec{\mathbf{m}}\Big)\Big] - \dfrac{\epsilon' - \alpha \epsilon}{1 + \alpha^2} \Big(\vec{\mathbf{m}}\times\vec{\mathbf{m}}_P\Big).
		\end{equation}
	Here, $\vec{\mathbf{m}_P}$ is the magnetisation of the fixed layer---which is not explicitly simulated---and $\epsilon$ and  $\epsilon'$ represent the effective spin polarisation terms for the damping-like and field-like torques, respectively. We can write $\epsilon$ as
	\begin{equation}
		\epsilon = \dfrac{P\Lambda^2}{(\Lambda^2 + 1) + (\Lambda^2 - 1 ) (\vec{\mathbf{m}}\cdot\vec{\mathbf{m}_P})},
	\end{equation}
	where P is the spin polarisation and the Slonczewski $\Lambda$ parameter is a dimensionless parameter of the model. Finally, the last dimensionless parameter $\beta$ is given by
	\begin{equation}
		\beta = \dfrac{\hbar J_c}{M_{\mathrm{s}} e d},
		\end{equation}
	where $J_c$ is the current density and $d$ is the film thickness. For all subsequent results, we simulate a skyrmion in a system with an antiferromagnetic interlayer exchange coupling energy of $\sigma_{\mathrm{RKKY}} = -3\times 10^{-4}\ \mathrm{J/m^2}$, and driven by a pure damping-like torque ($\epsilon \neq 0, \epsilon' = 0$). For a Slonczewski-type STT the current density $J_c$ is applied along the out-of-plane axis, in what is described as current perpendicular to plane (CPP). The direction of the fixed layer is also set along the $+z$ direction. A schematic of the simulated system along with the direction of the current density and the polarisation of the fixed layer is shown in Figure~\ref{fig:CI_SWE}(a). For all simulations the STT is considered to be present in both ferromagnetic layers. Considering the spin diffusion length $l_{\mathrm{Ru}}$ of ruthenium, the material typically chosen as the nonmagnetic spacer layer, of 14~nm~\cite{Eid2002}, this set-up is valid~\cite{Lonsky2020_PRB}.\\\\
	Figure~\ref{fig:CI_SWE}(b) shows the spectral response of the system when excited with a sinc pulse of current density of $1\times 10^9 \ \mathrm{A/m^2}$. The frequencies of both the in-phase and out-of-phase modes reproduce those shown in Figure~\ref{fig:Skyrmion_breathing}(a). As mentioned, the excitation of skyrmion breathing modes was studied extensively in Ref.~\cite{Lonsky2020_PRB}, where they showed that for their chosen magnetic parameters the breathing modes occurred at the same frequencies regardless of whether they were excited by a magnetic field or a current density. Furthermore, they demonstrated that this was independent of the nature of the STT; demonstrating the same effects with field-like STTs ($\epsilon = 0, \epsilon'\neq 0$) or a mixture of damping-like and field-like STTs ($\epsilon \neq 0, \epsilon'\neq 0$).\\\\
	Figure~\ref{fig:CI_SWE}(c) shows the velocity of the skyrmion as a function of the current density when driven by an oscillating current density with a frequency of 57.6~GHz---which was obtained by fitting to the peak of the out-of-phase breathing mode in Figure~\ref{fig:CI_SWE}(b). The current density was varied in a quasi-logarithmic fashion to allow a sampling of a wide range of magnitudes. At current densities much above $5\times 10^{11}\ \mathrm{A/m^2}$, significant deformation of the skyrmion was observed, before eventually breaking down at current densities in the $10^{12}\ \mathrm{A/m^2}$ (not shown). We extend this to examine the effects of the skyrmion velocity as a function of frequency at current densities of $5 \times10^{10}$, $1 \times 10^{11}$ and $2\times 10^{11}\ \mathrm{A/m^2}$ in Figure~\ref{fig:CI_SWE}(d). The shape of the $v(f)$ is broadly similar to that presented in Figure~\ref{fig:velocity_fn_freq}. However, some important differences should be noted. In a similar way to the magnetic-field driven case, we observe little-to-no evidence of any motion at the resonant frequency of the in-phase breathing mode. This is because the CPP-STT is acting as an effective magnetic field on the skyrmions, and so we expect any behaviour to be consistent with the case of driving with a simple magnetic field. Likewise, the relative velocities of the skyrmions driven at the resonant frequency of the out-of-phase breathing mode and at the resonant frequency of the spin wave mode arising from the canted spins are different. While the absolute magnitude of the velocity depends on the strength of the current density or magnetic field, this relative difference is of note. It arises because a current density along the $z$-axis, with a spin polarisation also along the $z$-axis will induce an effective field that acts on the in plane spins. In the case of the skyrmions, this are only the spins in the narrow band around the skyrmion where the magnetisation rotates from up to down (and \textit{vice-versa} for the skyrmion in the other layer). When the spin wave emission is arising from the hybrid skyrmion breathing and resonance of the canted spins of the system, then we would expect this to be coupled much better to the STT, and so increase the velocity of the skyrmion, as we see for each of the current densities in Figure~\ref{fig:CI_SWE}(d).}
	 
	\section{Conclusion}
	To conclude, we have used micromagnetic simulations to demonstrate a new method for stimulating skyrmion motion in a synthetic antiferromagnet using only global magnetic fields. We have used a microwave frequency out-of-plane magnetic field to excite skyrmion breathing, which in the presence of a static in-plane field emits spin waves asymmetrically. Through characterisation of the breathing modes of the skyrmions in our systems we have understood the underlying relationship between skyrmion velocity and the frequency of the driving field, observing that the maximum skyrmion velocity occurs at the characteristic frequency of the out-of-phase skyrmion breathing mode.\\ 
	We have further measured the variation in radius of the skyrmions at different driving fields to understand that there is a threshold magnitude of radius variation, only above which will skyrmion motion be driven.\\ 
	The understanding of the coupling between the skyrmion velocity and the out-of-phase breathing mode has allowed us to properly understand the relationship between the velocity and the antiferromagnetic interlayer exchange coupling strength, where we define three different regimes of skyrmion motion depending on the magnitude of the antiferromagnetic exchange coupling.\\
	\highlight{We have also extended our study to consider synthetic antiferromagnets, which is relevant to experimental studies where a perfect compensation of moments is difficult to achieve. We show that as expected, the velocity is at a maximum when the system is fully compensated, however motion is possible over a wide range of decompensation, and this is true for low and relatively high coupling strength (in comparison to what has been reported experimentally).}\\
	Finally, we have compared our results to the case when driven by a microwave electric field \highlight{and oscillating current densities}. By doing this we have revealed both the true origin of the relationship between velocity and exchange coupling, and the reason for the lower frequency of skyrmion motion when driven by electric fields compared to magnetic fields, \highlight{as well as demonstrating the wide range of mechanisms through which this spin wave emission skyrmion motion can be driven.}\\
	These results will help to inform the design of future novel computing architectures based on the dynamics of skyrmions in synthetic antiferromagnets.

\begin{acknowledgments}
This project was funded by the UK Government Department for Science, Innovation and Technology through National Measurement System funding (Metrology of complex systems for low energy computation). The project was also supported by EPSRC Grant No. EP/T006803/1.
\end{acknowledgments}

\bibliography{bibo.bib}

\begin{thebibliography}{58}%
\makeatletter
\providecommand \@ifxundefined [1]{%
 \@ifx{#1\undefined}
}%
\providecommand \@ifnum [1]{%
 \ifnum #1\expandafter \@firstoftwo
 \else \expandafter \@secondoftwo
 \fi
}%
\providecommand \@ifx [1]{%
 \ifx #1\expandafter \@firstoftwo
 \else \expandafter \@secondoftwo
 \fi
}%
\providecommand \natexlab [1]{#1}%
\providecommand \enquote  [1]{``#1''}%
\providecommand \bibnamefont  [1]{#1}%
\providecommand \bibfnamefont [1]{#1}%
\providecommand \citenamefont [1]{#1}%
\providecommand \href@noop [0]{\@secondoftwo}%
\providecommand \href [0]{\begingroup \@sanitize@url \@href}%
\providecommand \@href[1]{\@@startlink{#1}\@@href}%
\providecommand \@@href[1]{\endgroup#1\@@endlink}%
\providecommand \@sanitize@url [0]{\catcode `\\12\catcode `\$12\catcode
  `\&12\catcode `\#12\catcode `\^12\catcode `\_12\catcode `\%12\relax}%
\providecommand \@@startlink[1]{}%
\providecommand \@@endlink[0]{}%
\providecommand \url  [0]{\begingroup\@sanitize@url \@url }%
\providecommand \@url [1]{\endgroup\@href {#1}{\urlprefix }}%
\providecommand \urlprefix  [0]{URL }%
\providecommand \Eprint [0]{\href }%
\providecommand \doibase [0]{https://doi.org/}%
\providecommand \selectlanguage [0]{\@gobble}%
\providecommand \bibinfo  [0]{\@secondoftwo}%
\providecommand \bibfield  [0]{\@secondoftwo}%
\providecommand \translation [1]{[#1]}%
\providecommand \BibitemOpen [0]{}%
\providecommand \bibitemStop [0]{}%
\providecommand \bibitemNoStop [0]{.\EOS\space}%
\providecommand \EOS [0]{\spacefactor3000\relax}%
\providecommand \BibitemShut  [1]{\csname bibitem#1\endcsname}%
\let\auto@bib@innerbib\@empty
\bibitem [{\citenamefont {Nagaosa}\ and\ \citenamefont
  {Tokura}(2013)}]{Nagaosa2013}%
  \BibitemOpen
  \bibfield  {author} {\bibinfo {author} {\bibfnamefont {N.}~\bibnamefont
  {Nagaosa}}\ and\ \bibinfo {author} {\bibfnamefont {Y.}~\bibnamefont
  {Tokura}},\ }\bibfield  {title} {\bibinfo {title} {Topological properties and
  dynamics of magnetic skyrmions},\ }\href
  {https://doi.org/10.1038/nnano.2013.243} {\bibfield  {journal} {\bibinfo
  {journal} {Nature Nanotechnology}\ }\textbf {\bibinfo {volume} {8}},\
  \bibinfo {pages} {899} (\bibinfo {year} {2013})}\BibitemShut {NoStop}%
\bibitem [{\citenamefont {Everschor-Sitte}\ \emph {et~al.}(2018)\citenamefont
  {Everschor-Sitte}, \citenamefont {Masell}, \citenamefont {Reeve},\ and\
  \citenamefont {Kläui}}]{EverschorSitte2018}%
  \BibitemOpen
  \bibfield  {author} {\bibinfo {author} {\bibfnamefont {K.}~\bibnamefont
  {Everschor-Sitte}}, \bibinfo {author} {\bibfnamefont {J.}~\bibnamefont
  {Masell}}, \bibinfo {author} {\bibfnamefont {R.~M.}\ \bibnamefont {Reeve}},\
  and\ \bibinfo {author} {\bibfnamefont {M.}~\bibnamefont {Kläui}},\
  }\bibfield  {title} {\bibinfo {title} {Perspective: Magnetic
  skyrmions—overview of recent progress in an active research field},\ }\href
  {https://doi.org/10.1063/1.5048972} {\bibfield  {journal} {\bibinfo
  {journal} {Journal of Applied Physics}\ }\textbf {\bibinfo {volume} {124}},\
  \bibinfo {pages} {240901} (\bibinfo {year} {2018})}\BibitemShut {NoStop}%
\bibitem [{\citenamefont {Marrows}\ and\ \citenamefont
  {Zeissler}(2021)}]{Marrows2021}%
  \BibitemOpen
  \bibfield  {author} {\bibinfo {author} {\bibfnamefont {C.~H.}\ \bibnamefont
  {Marrows}}\ and\ \bibinfo {author} {\bibfnamefont {K.}~\bibnamefont
  {Zeissler}},\ }\bibfield  {title} {\bibinfo {title} {Perspective on skyrmion
  spintronics},\ }\href {https://doi.org/10.1063/5.0072735} {\bibfield
  {journal} {\bibinfo  {journal} {Applied Physics Letters}\ }\textbf {\bibinfo
  {volume} {119}},\ \bibinfo {pages} {250502} (\bibinfo {year}
  {2021})}\BibitemShut {NoStop}%
\bibitem [{\citenamefont {Finocchio}\ \emph {et~al.}(2016)\citenamefont
  {Finocchio}, \citenamefont {Büttner}, \citenamefont {Tomasello},
  \citenamefont {Carpentieri},\ and\ \citenamefont {Kläui}}]{Finocchio2016}%
  \BibitemOpen
  \bibfield  {author} {\bibinfo {author} {\bibfnamefont {G.}~\bibnamefont
  {Finocchio}}, \bibinfo {author} {\bibfnamefont {F.}~\bibnamefont {Büttner}},
  \bibinfo {author} {\bibfnamefont {R.}~\bibnamefont {Tomasello}}, \bibinfo
  {author} {\bibfnamefont {M.}~\bibnamefont {Carpentieri}},\ and\ \bibinfo
  {author} {\bibfnamefont {M.}~\bibnamefont {Kläui}},\ }\bibfield  {title}
  {\bibinfo {title} {Magnetic skyrmions: from fundamental to applications},\
  }\href {https://doi.org/10.1088/0022-3727/49/42/423001} {\bibfield  {journal}
  {\bibinfo  {journal} {Journal of Physics D: Applied Physics}\ }\textbf
  {\bibinfo {volume} {49}},\ \bibinfo {pages} {423001} (\bibinfo {year}
  {2016})}\BibitemShut {NoStop}%
\bibitem [{\citenamefont {Tomasello}\ \emph {et~al.}(2014)\citenamefont
  {Tomasello}, \citenamefont {Martinez}, \citenamefont {Zivieri}, \citenamefont
  {Torres}, \citenamefont {Carpentieri},\ and\ \citenamefont
  {Finocchio}}]{Tomasello2014}%
  \BibitemOpen
  \bibfield  {author} {\bibinfo {author} {\bibfnamefont {R.}~\bibnamefont
  {Tomasello}}, \bibinfo {author} {\bibfnamefont {E.}~\bibnamefont {Martinez}},
  \bibinfo {author} {\bibfnamefont {R.}~\bibnamefont {Zivieri}}, \bibinfo
  {author} {\bibfnamefont {L.}~\bibnamefont {Torres}}, \bibinfo {author}
  {\bibfnamefont {M.}~\bibnamefont {Carpentieri}},\ and\ \bibinfo {author}
  {\bibfnamefont {G.}~\bibnamefont {Finocchio}},\ }\bibfield  {title} {\bibinfo
  {title} {A strategy for the design of skyrmion racetrack memories},\ }\href
  {https://doi.org/10.1038/srep06784} {\bibfield  {journal} {\bibinfo
  {journal} {Scientific Reports}\ }\textbf {\bibinfo {volume} {4}},\ \bibinfo
  {pages} {6784} (\bibinfo {year} {2014})}\BibitemShut {NoStop}%
\bibitem [{\citenamefont {Song}\ \emph {et~al.}(2020)\citenamefont {Song},
  \citenamefont {Jeong}, \citenamefont {Pan}, \citenamefont {Zhang},
  \citenamefont {Xia}, \citenamefont {Cha}, \citenamefont {Park}, \citenamefont
  {Kim}, \citenamefont {Finizio}, \citenamefont {Raabe}, \citenamefont {Chang},
  \citenamefont {Zhou}, \citenamefont {Zhao}, \citenamefont {Kang},
  \citenamefont {Ju},\ and\ \citenamefont {Woo}}]{Song2020}%
  \BibitemOpen
  \bibfield  {author} {\bibinfo {author} {\bibfnamefont {K.~M.}\ \bibnamefont
  {Song}}, \bibinfo {author} {\bibfnamefont {J.-S.}\ \bibnamefont {Jeong}},
  \bibinfo {author} {\bibfnamefont {B.}~\bibnamefont {Pan}}, \bibinfo {author}
  {\bibfnamefont {X.}~\bibnamefont {Zhang}}, \bibinfo {author} {\bibfnamefont
  {J.}~\bibnamefont {Xia}}, \bibinfo {author} {\bibfnamefont {S.}~\bibnamefont
  {Cha}}, \bibinfo {author} {\bibfnamefont {T.-E.}\ \bibnamefont {Park}},
  \bibinfo {author} {\bibfnamefont {K.}~\bibnamefont {Kim}}, \bibinfo {author}
  {\bibfnamefont {S.}~\bibnamefont {Finizio}}, \bibinfo {author} {\bibfnamefont
  {J.}~\bibnamefont {Raabe}}, \bibinfo {author} {\bibfnamefont
  {J.}~\bibnamefont {Chang}}, \bibinfo {author} {\bibfnamefont
  {Y.}~\bibnamefont {Zhou}}, \bibinfo {author} {\bibfnamefont {W.}~\bibnamefont
  {Zhao}}, \bibinfo {author} {\bibfnamefont {W.}~\bibnamefont {Kang}}, \bibinfo
  {author} {\bibfnamefont {H.}~\bibnamefont {Ju}},\ and\ \bibinfo {author}
  {\bibfnamefont {S.}~\bibnamefont {Woo}},\ }\bibfield  {title} {\bibinfo
  {title} {Skyrmion-based artificial synapses for neuromorphic computing},\
  }\href {https://doi.org/10.1038/s41928-020-0385-0} {\bibfield  {journal}
  {\bibinfo  {journal} {Nature Electronics}\ }\textbf {\bibinfo {volume} {3}},\
  \bibinfo {pages} {148} (\bibinfo {year} {2020})}\BibitemShut {NoStop}%
\bibitem [{\citenamefont {Marrows}\ \emph {et~al.}(2024)\citenamefont
  {Marrows}, \citenamefont {Barker}, \citenamefont {Moore},\ and\ \citenamefont
  {Moorsom}}]{Marrows2024}%
  \BibitemOpen
  \bibfield  {author} {\bibinfo {author} {\bibfnamefont {C.~H.}\ \bibnamefont
  {Marrows}}, \bibinfo {author} {\bibfnamefont {J.}~\bibnamefont {Barker}},
  \bibinfo {author} {\bibfnamefont {T.~A.}\ \bibnamefont {Moore}},\ and\
  \bibinfo {author} {\bibfnamefont {T.}~\bibnamefont {Moorsom}},\ }\bibfield
  {title} {\bibinfo {title} {Neuromorphic computing with spintronics},\ }\href
  {https://doi.org/10.1038/s44306-024-00019-2} {\bibfield  {journal} {\bibinfo
  {journal} {npj Spintronics}\ }\textbf {\bibinfo {volume} {2}},\ \bibinfo
  {pages} {12} (\bibinfo {year} {2024})}\BibitemShut {NoStop}%
\bibitem [{\citenamefont {Petrovi\'c}\ \emph {et~al.}(2024)\citenamefont
  {Petrovi\'c}, \citenamefont {Psaroudaki}, \citenamefont {Fischer},
  \citenamefont {Garst},\ and\ \citenamefont {Panagopoulos}}]{petrovic2024}%
  \BibitemOpen
  \bibfield  {author} {\bibinfo {author} {\bibfnamefont {A.~P.}\ \bibnamefont
  {Petrovi\'c}}, \bibinfo {author} {\bibfnamefont {C.}~\bibnamefont
  {Psaroudaki}}, \bibinfo {author} {\bibfnamefont {P.}~\bibnamefont {Fischer}},
  \bibinfo {author} {\bibfnamefont {M.}~\bibnamefont {Garst}},\ and\ \bibinfo
  {author} {\bibfnamefont {C.}~\bibnamefont {Panagopoulos}},\ }\href
  {https://arxiv.org/abs/2410.11427} {\bibinfo {title} {Colloquium: Quantum
  properties and functionalities of magnetic skyrmions}} (\bibinfo {year}
  {2024}),\ \Eprint {https://arxiv.org/abs/2410.11427} {arXiv:2410.11427}
  \BibitemShut {NoStop}%
\bibitem [{\citenamefont {Banerjee}\ \emph {et~al.}(2024)\citenamefont
  {Banerjee}, \citenamefont {Bell}, \citenamefont {Ciccarelli}, \citenamefont
  {Hesjedal}, \citenamefont {Johnson}, \citenamefont {Kurebayashi},
  \citenamefont {Moore}, \citenamefont {Moutafis}, \citenamefont {Stern},
  \citenamefont {Vera-Marun}, \citenamefont {Wade}, \citenamefont {Barton},
  \citenamefont {Connolly}, \citenamefont {Curson}, \citenamefont {Fallon},
  \citenamefont {Fisher}, \citenamefont {Gangloff}, \citenamefont {Griggs},
  \citenamefont {Linfield}, \citenamefont {Marrows}, \citenamefont {Rossi},
  \citenamefont {Schindler}, \citenamefont {Smith}, \citenamefont {Thomson},\
  and\ \citenamefont {Kazakova}}]{banerjee2024}%
  \BibitemOpen
  \bibfield  {author} {\bibinfo {author} {\bibfnamefont {N.}~\bibnamefont
  {Banerjee}}, \bibinfo {author} {\bibfnamefont {C.}~\bibnamefont {Bell}},
  \bibinfo {author} {\bibfnamefont {C.}~\bibnamefont {Ciccarelli}}, \bibinfo
  {author} {\bibfnamefont {T.}~\bibnamefont {Hesjedal}}, \bibinfo {author}
  {\bibfnamefont {F.}~\bibnamefont {Johnson}}, \bibinfo {author} {\bibfnamefont
  {H.}~\bibnamefont {Kurebayashi}}, \bibinfo {author} {\bibfnamefont {T.~A.}\
  \bibnamefont {Moore}}, \bibinfo {author} {\bibfnamefont {C.}~\bibnamefont
  {Moutafis}}, \bibinfo {author} {\bibfnamefont {H.~L.}\ \bibnamefont {Stern}},
  \bibinfo {author} {\bibfnamefont {I.~J.}\ \bibnamefont {Vera-Marun}},
  \bibinfo {author} {\bibfnamefont {J.}~\bibnamefont {Wade}}, \bibinfo {author}
  {\bibfnamefont {C.}~\bibnamefont {Barton}}, \bibinfo {author} {\bibfnamefont
  {M.~R.}\ \bibnamefont {Connolly}}, \bibinfo {author} {\bibfnamefont {N.~J.}\
  \bibnamefont {Curson}}, \bibinfo {author} {\bibfnamefont {K.}~\bibnamefont
  {Fallon}}, \bibinfo {author} {\bibfnamefont {A.~J.}\ \bibnamefont {Fisher}},
  \bibinfo {author} {\bibfnamefont {D.~A.}\ \bibnamefont {Gangloff}}, \bibinfo
  {author} {\bibfnamefont {W.}~\bibnamefont {Griggs}}, \bibinfo {author}
  {\bibfnamefont {E.}~\bibnamefont {Linfield}}, \bibinfo {author}
  {\bibfnamefont {C.~H.}\ \bibnamefont {Marrows}}, \bibinfo {author}
  {\bibfnamefont {A.}~\bibnamefont {Rossi}}, \bibinfo {author} {\bibfnamefont
  {F.}~\bibnamefont {Schindler}}, \bibinfo {author} {\bibfnamefont
  {J.}~\bibnamefont {Smith}}, \bibinfo {author} {\bibfnamefont
  {T.}~\bibnamefont {Thomson}},\ and\ \bibinfo {author} {\bibfnamefont
  {O.}~\bibnamefont {Kazakova}},\ }\href {https://arxiv.org/abs/2406.07720}
  {\bibinfo {title} {Materials for quantum technologies: a roadmap for spin and
  topology}} (\bibinfo {year} {2024}),\ \Eprint
  {https://arxiv.org/abs/2406.07720} {arXiv:2406.07720} \BibitemShut {NoStop}%
\bibitem [{\citenamefont {Fert}\ \emph {et~al.}(2013)\citenamefont {Fert},
  \citenamefont {Cros},\ and\ \citenamefont {Sampaio}}]{Fert2013}%
  \BibitemOpen
  \bibfield  {author} {\bibinfo {author} {\bibfnamefont {A.}~\bibnamefont
  {Fert}}, \bibinfo {author} {\bibfnamefont {V.}~\bibnamefont {Cros}},\ and\
  \bibinfo {author} {\bibfnamefont {J.}~\bibnamefont {Sampaio}},\ }\bibfield
  {title} {\bibinfo {title} {Skyrmions on the track},\ }\href
  {https://doi.org/10.1038/nnano.2013.29} {\bibfield  {journal} {\bibinfo
  {journal} {Nature Nanotechnology}\ }\textbf {\bibinfo {volume} {8}},\
  \bibinfo {pages} {152} (\bibinfo {year} {2013})}\BibitemShut {NoStop}%
\bibitem [{\citenamefont {Woo}\ \emph {et~al.}(2016)\citenamefont {Woo},
  \citenamefont {Litzius}, \citenamefont {Kr{\"u}ger}, \citenamefont {Im},
  \citenamefont {Caretta}, \citenamefont {Richter}, \citenamefont {Mann},
  \citenamefont {Krone}, \citenamefont {Reeve}, \citenamefont {Weigand},
  \citenamefont {Agrawal}, \citenamefont {Lemesh}, \citenamefont {Mawass},
  \citenamefont {Fischer}, \citenamefont {Kl{\"a}ui},\ and\ \citenamefont
  {Beach}}]{Woo2016}%
  \BibitemOpen
  \bibfield  {author} {\bibinfo {author} {\bibfnamefont {S.}~\bibnamefont
  {Woo}}, \bibinfo {author} {\bibfnamefont {K.}~\bibnamefont {Litzius}},
  \bibinfo {author} {\bibfnamefont {B.}~\bibnamefont {Kr{\"u}ger}}, \bibinfo
  {author} {\bibfnamefont {M.-Y.}\ \bibnamefont {Im}}, \bibinfo {author}
  {\bibfnamefont {L.}~\bibnamefont {Caretta}}, \bibinfo {author} {\bibfnamefont
  {K.}~\bibnamefont {Richter}}, \bibinfo {author} {\bibfnamefont
  {M.}~\bibnamefont {Mann}}, \bibinfo {author} {\bibfnamefont {A.}~\bibnamefont
  {Krone}}, \bibinfo {author} {\bibfnamefont {R.~M.}\ \bibnamefont {Reeve}},
  \bibinfo {author} {\bibfnamefont {M.}~\bibnamefont {Weigand}}, \bibinfo
  {author} {\bibfnamefont {P.}~\bibnamefont {Agrawal}}, \bibinfo {author}
  {\bibfnamefont {I.}~\bibnamefont {Lemesh}}, \bibinfo {author} {\bibfnamefont
  {M.-A.}\ \bibnamefont {Mawass}}, \bibinfo {author} {\bibfnamefont
  {P.}~\bibnamefont {Fischer}}, \bibinfo {author} {\bibfnamefont
  {M.}~\bibnamefont {Kl{\"a}ui}},\ and\ \bibinfo {author} {\bibfnamefont
  {G.~S.~D.}\ \bibnamefont {Beach}},\ }\bibfield  {title} {\bibinfo {title}
  {Observation of room-temperature magnetic skyrmions and their current-driven
  dynamics in ultrathin metallic ferromagnets},\ }\href
  {https://doi.org/10.1038/nmat4593} {\bibfield  {journal} {\bibinfo  {journal}
  {Nature Materials}\ }\textbf {\bibinfo {volume} {15}},\ \bibinfo {pages}
  {501} (\bibinfo {year} {2016})}\BibitemShut {NoStop}%
\bibitem [{\citenamefont {Juge}\ \emph {et~al.}(2019)\citenamefont {Juge},
  \citenamefont {Je}, \citenamefont {Chaves}, \citenamefont {Buda-Prejbeanu},
  \citenamefont {Pe\~na Garcia}, \citenamefont {Nath}, \citenamefont {Miron},
  \citenamefont {Rana}, \citenamefont {Aballe}, \citenamefont {Foerster},
  \citenamefont {Genuzio}, \citenamefont {Mentes}, \citenamefont {Locatelli},
  \citenamefont {Maccherozzi}, \citenamefont {Dhesi}, \citenamefont
  {Belmeguenai}, \citenamefont {Roussign\'e}, \citenamefont {Auffret},
  \citenamefont {Pizzini}, \citenamefont {Gaudin}, \citenamefont {Vogel},\ and\
  \citenamefont {Boulle}}]{Juge2019}%
  \BibitemOpen
  \bibfield  {author} {\bibinfo {author} {\bibfnamefont {R.}~\bibnamefont
  {Juge}}, \bibinfo {author} {\bibfnamefont {S.-G.}\ \bibnamefont {Je}},
  \bibinfo {author} {\bibfnamefont {D.~d.~S.}\ \bibnamefont {Chaves}}, \bibinfo
  {author} {\bibfnamefont {L.~D.}\ \bibnamefont {Buda-Prejbeanu}}, \bibinfo
  {author} {\bibfnamefont {J.}~\bibnamefont {Pe\~na Garcia}}, \bibinfo {author}
  {\bibfnamefont {J.}~\bibnamefont {Nath}}, \bibinfo {author} {\bibfnamefont
  {I.~M.}\ \bibnamefont {Miron}}, \bibinfo {author} {\bibfnamefont {K.~G.}\
  \bibnamefont {Rana}}, \bibinfo {author} {\bibfnamefont {L.}~\bibnamefont
  {Aballe}}, \bibinfo {author} {\bibfnamefont {M.}~\bibnamefont {Foerster}},
  \bibinfo {author} {\bibfnamefont {F.}~\bibnamefont {Genuzio}}, \bibinfo
  {author} {\bibfnamefont {T.~O.}\ \bibnamefont {Mentes}}, \bibinfo {author}
  {\bibfnamefont {A.}~\bibnamefont {Locatelli}}, \bibinfo {author}
  {\bibfnamefont {F.}~\bibnamefont {Maccherozzi}}, \bibinfo {author}
  {\bibfnamefont {S.~S.}\ \bibnamefont {Dhesi}}, \bibinfo {author}
  {\bibfnamefont {M.}~\bibnamefont {Belmeguenai}}, \bibinfo {author}
  {\bibfnamefont {Y.}~\bibnamefont {Roussign\'e}}, \bibinfo {author}
  {\bibfnamefont {S.}~\bibnamefont {Auffret}}, \bibinfo {author} {\bibfnamefont
  {S.}~\bibnamefont {Pizzini}}, \bibinfo {author} {\bibfnamefont
  {G.}~\bibnamefont {Gaudin}}, \bibinfo {author} {\bibfnamefont
  {J.}~\bibnamefont {Vogel}},\ and\ \bibinfo {author} {\bibfnamefont
  {O.}~\bibnamefont {Boulle}},\ }\bibfield  {title} {\bibinfo {title}
  {Current-driven skyrmion dynamics and drive-dependent skyrmion hall effect in
  an ultrathin film},\ }\href
  {https://doi.org/10.1103/PhysRevApplied.12.044007} {\bibfield  {journal}
  {\bibinfo  {journal} {Phys. Rev. Appl.}\ }\textbf {\bibinfo {volume} {12}},\
  \bibinfo {pages} {044007} (\bibinfo {year} {2019})}\BibitemShut {NoStop}%
\bibitem [{\citenamefont {Zhang}\ \emph {et~al.}(2018)\citenamefont {Zhang},
  \citenamefont {Wang}, \citenamefont {Burn}, \citenamefont {Peng},
  \citenamefont {Berger}, \citenamefont {Bauer}, \citenamefont {Pfleiderer},
  \citenamefont {van~der Laan},\ and\ \citenamefont {Hesjedal}}]{Zhang2018}%
  \BibitemOpen
  \bibfield  {author} {\bibinfo {author} {\bibfnamefont {S.~L.}\ \bibnamefont
  {Zhang}}, \bibinfo {author} {\bibfnamefont {W.~W.}\ \bibnamefont {Wang}},
  \bibinfo {author} {\bibfnamefont {D.~M.}\ \bibnamefont {Burn}}, \bibinfo
  {author} {\bibfnamefont {H.}~\bibnamefont {Peng}}, \bibinfo {author}
  {\bibfnamefont {H.}~\bibnamefont {Berger}}, \bibinfo {author} {\bibfnamefont
  {A.}~\bibnamefont {Bauer}}, \bibinfo {author} {\bibfnamefont
  {C.}~\bibnamefont {Pfleiderer}}, \bibinfo {author} {\bibfnamefont
  {G.}~\bibnamefont {van~der Laan}},\ and\ \bibinfo {author} {\bibfnamefont
  {T.}~\bibnamefont {Hesjedal}},\ }\bibfield  {title} {\bibinfo {title}
  {Manipulation of skyrmion motion by magnetic field gradients},\ }\href
  {https://doi.org/10.1038/s41467-018-04563-4} {\bibfield  {journal} {\bibinfo
  {journal} {Nature Communications}\ }\textbf {\bibinfo {volume} {9}},\
  \bibinfo {pages} {2115} (\bibinfo {year} {2018})}\BibitemShut {NoStop}%
\bibitem [{\citenamefont {Wang}\ \emph {et~al.}(2017)\citenamefont {Wang},
  \citenamefont {Xiao}, \citenamefont {Chen}, \citenamefont {Zhou},\ and\
  \citenamefont {Liu}}]{Wang_2017}%
  \BibitemOpen
  \bibfield  {author} {\bibinfo {author} {\bibfnamefont {C.}~\bibnamefont
  {Wang}}, \bibinfo {author} {\bibfnamefont {D.}~\bibnamefont {Xiao}}, \bibinfo
  {author} {\bibfnamefont {X.}~\bibnamefont {Chen}}, \bibinfo {author}
  {\bibfnamefont {Y.}~\bibnamefont {Zhou}},\ and\ \bibinfo {author}
  {\bibfnamefont {Y.}~\bibnamefont {Liu}},\ }\bibfield  {title} {\bibinfo
  {title} {Manipulating and trapping skyrmions by magnetic field gradients},\
  }\href {https://doi.org/10.1088/1367-2630/aa7812} {\bibfield  {journal}
  {\bibinfo  {journal} {New Journal of Physics}\ }\textbf {\bibinfo {volume}
  {19}},\ \bibinfo {pages} {083008} (\bibinfo {year} {2017})}\BibitemShut
  {NoStop}%
\bibitem [{\citenamefont {Moon}\ \emph {et~al.}(2016)\citenamefont {Moon},
  \citenamefont {Kim}, \citenamefont {Je}, \citenamefont {Chun}, \citenamefont
  {Kim}, \citenamefont {Qiu}, \citenamefont {Choe},\ and\ \citenamefont
  {Hwang}}]{Moon2016}%
  \BibitemOpen
  \bibfield  {author} {\bibinfo {author} {\bibfnamefont {K.-W.}\ \bibnamefont
  {Moon}}, \bibinfo {author} {\bibfnamefont {D.-H.}\ \bibnamefont {Kim}},
  \bibinfo {author} {\bibfnamefont {S.-G.}\ \bibnamefont {Je}}, \bibinfo
  {author} {\bibfnamefont {B.~S.}\ \bibnamefont {Chun}}, \bibinfo {author}
  {\bibfnamefont {W.}~\bibnamefont {Kim}}, \bibinfo {author} {\bibfnamefont
  {Z.~Q.}\ \bibnamefont {Qiu}}, \bibinfo {author} {\bibfnamefont {S.-B.}\
  \bibnamefont {Choe}},\ and\ \bibinfo {author} {\bibfnamefont
  {C.}~\bibnamefont {Hwang}},\ }\bibfield  {title} {\bibinfo {title} {Skyrmion
  motion driven by oscillating magnetic field},\ }\href
  {https://doi.org/10.1038/srep20360} {\bibfield  {journal} {\bibinfo
  {journal} {Scientific Reports}\ }\textbf {\bibinfo {volume} {6}},\ \bibinfo
  {pages} {20360} (\bibinfo {year} {2016})}\BibitemShut {NoStop}%
\bibitem [{\citenamefont {Wang}\ \emph {et~al.}(2015)\citenamefont {Wang},
  \citenamefont {Beg}, \citenamefont {Zhang}, \citenamefont {Kuch},\ and\
  \citenamefont {Fangohr}}]{Wang2015}%
  \BibitemOpen
  \bibfield  {author} {\bibinfo {author} {\bibfnamefont {W.}~\bibnamefont
  {Wang}}, \bibinfo {author} {\bibfnamefont {M.}~\bibnamefont {Beg}}, \bibinfo
  {author} {\bibfnamefont {B.}~\bibnamefont {Zhang}}, \bibinfo {author}
  {\bibfnamefont {W.}~\bibnamefont {Kuch}},\ and\ \bibinfo {author}
  {\bibfnamefont {H.}~\bibnamefont {Fangohr}},\ }\bibfield  {title} {\bibinfo
  {title} {Driving magnetic skyrmions with microwave fields},\ }\href
  {https://doi.org/10.1103/PhysRevB.92.020403} {\bibfield  {journal} {\bibinfo
  {journal} {Phys. Rev. B}\ }\textbf {\bibinfo {volume} {92}},\ \bibinfo
  {pages} {020403} (\bibinfo {year} {2015})}\BibitemShut {NoStop}%
\bibitem [{\citenamefont {Everschor-Sitte}\ and\ \citenamefont
  {Sitte}(2014)}]{EverschorSitte2014}%
  \BibitemOpen
  \bibfield  {author} {\bibinfo {author} {\bibfnamefont {K.}~\bibnamefont
  {Everschor-Sitte}}\ and\ \bibinfo {author} {\bibfnamefont {M.}~\bibnamefont
  {Sitte}},\ }\bibfield  {title} {\bibinfo {title} {Real-space berry phases:
  Skyrmion soccer (invited)},\ }\href {https://doi.org/10.1063/1.4870695}
  {\bibfield  {journal} {\bibinfo  {journal} {Journal of Applied Physics}\
  }\textbf {\bibinfo {volume} {115}},\ \bibinfo {pages} {172602} (\bibinfo
  {year} {2014})}\BibitemShut {NoStop}%
\bibitem [{\citenamefont {Jiang}\ \emph
  {et~al.}(2017{\natexlab{a}})\citenamefont {Jiang}, \citenamefont {Zhang},
  \citenamefont {Yu}, \citenamefont {Zhang}, \citenamefont {Wang},
  \citenamefont {Benjamin Jungfleisch}, \citenamefont {Pearson}, \citenamefont
  {Cheng}, \citenamefont {Heinonen}, \citenamefont {Wang}, \citenamefont
  {Zhou}, \citenamefont {Hoffmann},\ and\ \citenamefont
  {te Velthuis}}]{Jiang2017}%
  \BibitemOpen
  \bibfield  {author} {\bibinfo {author} {\bibfnamefont {W.}~\bibnamefont
  {Jiang}}, \bibinfo {author} {\bibfnamefont {X.}~\bibnamefont {Zhang}},
  \bibinfo {author} {\bibfnamefont {G.}~\bibnamefont {Yu}}, \bibinfo {author}
  {\bibfnamefont {W.}~\bibnamefont {Zhang}}, \bibinfo {author} {\bibfnamefont
  {X.}~\bibnamefont {Wang}}, \bibinfo {author} {\bibfnamefont {M.}~\bibnamefont
  {Benjamin Jungfleisch}}, \bibinfo {author} {\bibfnamefont {J.}~\bibnamefont
  {Pearson}}, \bibinfo {author} {\bibfnamefont {X.}~\bibnamefont {Cheng}},
  \bibinfo {author} {\bibfnamefont {O.}~\bibnamefont {Heinonen}}, \bibinfo
  {author} {\bibfnamefont {K.~L.}\ \bibnamefont {Wang}}, \bibinfo {author}
  {\bibfnamefont {Y.}~\bibnamefont {Zhou}}, \bibinfo {author} {\bibfnamefont
  {A.}~\bibnamefont {Hoffmann}},\ and\ \bibinfo {author} {\bibfnamefont
  {S.}~\bibnamefont {te Velthuis}},\ }\bibfield  {title} {\bibinfo {title}
  {Direct observation of the skyrmion hall effect},\ }\href
  {https://doi.org/10.1038/nphys3883} {\bibfield  {journal} {\bibinfo
  {journal} {Nature Physics}\ }\textbf {\bibinfo {volume} {13}},\ \bibinfo
  {pages} {162} (\bibinfo {year} {2017}{\natexlab{a}})}\BibitemShut {NoStop}%
\bibitem [{\citenamefont {Litzius}\ \emph {et~al.}(2017)\citenamefont
  {Litzius}, \citenamefont {Lemesh}, \citenamefont {Kr{\"u}ger}, \citenamefont
  {Bassirian}, \citenamefont {Caretta}, \citenamefont {Richter}, \citenamefont
  {B{\"u}ttner}, \citenamefont {Sato}, \citenamefont {Tretiakov}, \citenamefont
  {F{\"o}rster}, \citenamefont {Reeve}, \citenamefont {Weigand}, \citenamefont
  {Bykova}, \citenamefont {Stoll}, \citenamefont {Sch{\"u}tz}, \citenamefont
  {Beach},\ and\ \citenamefont {Kl{\"a}ui}}]{Litzius2017}%
  \BibitemOpen
  \bibfield  {author} {\bibinfo {author} {\bibfnamefont {K.}~\bibnamefont
  {Litzius}}, \bibinfo {author} {\bibfnamefont {I.}~\bibnamefont {Lemesh}},
  \bibinfo {author} {\bibfnamefont {B.}~\bibnamefont {Kr{\"u}ger}}, \bibinfo
  {author} {\bibfnamefont {P.}~\bibnamefont {Bassirian}}, \bibinfo {author}
  {\bibfnamefont {L.}~\bibnamefont {Caretta}}, \bibinfo {author} {\bibfnamefont
  {K.}~\bibnamefont {Richter}}, \bibinfo {author} {\bibfnamefont
  {F.}~\bibnamefont {B{\"u}ttner}}, \bibinfo {author} {\bibfnamefont
  {K.}~\bibnamefont {Sato}}, \bibinfo {author} {\bibfnamefont {O.~A.}\
  \bibnamefont {Tretiakov}}, \bibinfo {author} {\bibfnamefont {J.}~\bibnamefont
  {F{\"o}rster}}, \bibinfo {author} {\bibfnamefont {R.~M.}\ \bibnamefont
  {Reeve}}, \bibinfo {author} {\bibfnamefont {M.}~\bibnamefont {Weigand}},
  \bibinfo {author} {\bibfnamefont {I.}~\bibnamefont {Bykova}}, \bibinfo
  {author} {\bibfnamefont {H.}~\bibnamefont {Stoll}}, \bibinfo {author}
  {\bibfnamefont {G.}~\bibnamefont {Sch{\"u}tz}}, \bibinfo {author}
  {\bibfnamefont {G.~S.~D.}\ \bibnamefont {Beach}},\ and\ \bibinfo {author}
  {\bibfnamefont {M.}~\bibnamefont {Kl{\"a}ui}},\ }\bibfield  {title} {\bibinfo
  {title} {Skyrmion hall effect revealed by direct time-resolved x-ray
  microscopy},\ }\href {https://doi.org/10.1038/nphys4000} {\bibfield
  {journal} {\bibinfo  {journal} {Nature Physics}\ }\textbf {\bibinfo {volume}
  {13}},\ \bibinfo {pages} {170} (\bibinfo {year} {2017})}\BibitemShut
  {NoStop}%
\bibitem [{\citenamefont {Zeissler}\ \emph {et~al.}(2020)\citenamefont
  {Zeissler}, \citenamefont {Finizio}, \citenamefont {Barton}, \citenamefont
  {Huxtable}, \citenamefont {Massey}, \citenamefont {Raabe}, \citenamefont
  {Sadovnikov}, \citenamefont {Nikitov}, \citenamefont {Brearton},
  \citenamefont {Hesjedal}, \citenamefont {van~der Laan}, \citenamefont
  {Rosamond}, \citenamefont {Linfield}, \citenamefont {Burnell},\ and\
  \citenamefont {Marrows}}]{Zeissler2020}%
  \BibitemOpen
  \bibfield  {author} {\bibinfo {author} {\bibfnamefont {K.}~\bibnamefont
  {Zeissler}}, \bibinfo {author} {\bibfnamefont {S.}~\bibnamefont {Finizio}},
  \bibinfo {author} {\bibfnamefont {C.}~\bibnamefont {Barton}}, \bibinfo
  {author} {\bibfnamefont {A.~J.}\ \bibnamefont {Huxtable}}, \bibinfo {author}
  {\bibfnamefont {J.}~\bibnamefont {Massey}}, \bibinfo {author} {\bibfnamefont
  {J.}~\bibnamefont {Raabe}}, \bibinfo {author} {\bibfnamefont {A.~V.}\
  \bibnamefont {Sadovnikov}}, \bibinfo {author} {\bibfnamefont {S.~A.}\
  \bibnamefont {Nikitov}}, \bibinfo {author} {\bibfnamefont {R.}~\bibnamefont
  {Brearton}}, \bibinfo {author} {\bibfnamefont {T.}~\bibnamefont {Hesjedal}},
  \bibinfo {author} {\bibfnamefont {G.}~\bibnamefont {van~der Laan}}, \bibinfo
  {author} {\bibfnamefont {M.~C.}\ \bibnamefont {Rosamond}}, \bibinfo {author}
  {\bibfnamefont {E.~H.}\ \bibnamefont {Linfield}}, \bibinfo {author}
  {\bibfnamefont {G.}~\bibnamefont {Burnell}},\ and\ \bibinfo {author}
  {\bibfnamefont {C.~H.}\ \bibnamefont {Marrows}},\ }\bibfield  {title}
  {\bibinfo {title} {Diameter-independent skyrmion hall angle observed in
  chiral magnetic multilayers},\ }\href
  {https://doi.org/10.1038/s41467-019-14232-9} {\bibfield  {journal} {\bibinfo
  {journal} {Nature Communications}\ }\textbf {\bibinfo {volume} {11}},\
  \bibinfo {pages} {428} (\bibinfo {year} {2020})}\BibitemShut {NoStop}%
\bibitem [{\citenamefont {Zhang}\ \emph {et~al.}(2016)\citenamefont {Zhang},
  \citenamefont {Zhou},\ and\ \citenamefont {Ezawa}}]{Zhang2016}%
  \BibitemOpen
  \bibfield  {author} {\bibinfo {author} {\bibfnamefont {X.}~\bibnamefont
  {Zhang}}, \bibinfo {author} {\bibfnamefont {Y.}~\bibnamefont {Zhou}},\ and\
  \bibinfo {author} {\bibfnamefont {M.}~\bibnamefont {Ezawa}},\ }\bibfield
  {title} {\bibinfo {title} {Magnetic bilayer-skyrmions without skyrmion hall
  effect},\ }\href {https://doi.org/10.1038/ncomms10293} {\bibfield  {journal}
  {\bibinfo  {journal} {Nature Communications}\ }\textbf {\bibinfo {volume}
  {7}},\ \bibinfo {pages} {10293} (\bibinfo {year} {2016})}\BibitemShut
  {NoStop}%
\bibitem [{\citenamefont {Duine}\ \emph {et~al.}(2018)\citenamefont {Duine},
  \citenamefont {Lee}, \citenamefont {Parkin},\ and\ \citenamefont
  {Stiles}}]{Duine2018}%
  \BibitemOpen
  \bibfield  {author} {\bibinfo {author} {\bibfnamefont {R.~A.}\ \bibnamefont
  {Duine}}, \bibinfo {author} {\bibfnamefont {K.-J.}\ \bibnamefont {Lee}},
  \bibinfo {author} {\bibfnamefont {S.~S.~P.}\ \bibnamefont {Parkin}},\ and\
  \bibinfo {author} {\bibfnamefont {M.~D.}\ \bibnamefont {Stiles}},\ }\bibfield
   {title} {\bibinfo {title} {Synthetic antiferromagnetic spintronics},\ }\href
  {https://doi.org/10.1038/s41567-018-0050-y} {\bibfield  {journal} {\bibinfo
  {journal} {Nature Physics}\ }\textbf {\bibinfo {volume} {14}},\ \bibinfo
  {pages} {217} (\bibinfo {year} {2018})}\BibitemShut {NoStop}%
\bibitem [{\citenamefont {Barker}\ and\ \citenamefont
  {Tretiakov}(2016)}]{Barker2016}%
  \BibitemOpen
  \bibfield  {author} {\bibinfo {author} {\bibfnamefont {J.}~\bibnamefont
  {Barker}}\ and\ \bibinfo {author} {\bibfnamefont {O.~A.}\ \bibnamefont
  {Tretiakov}},\ }\bibfield  {title} {\bibinfo {title} {Static and dynamical
  properties of antiferromagnetic skyrmions in the presence of applied current
  and temperature},\ }\href {https://doi.org/10.1103/PhysRevLett.116.147203}
  {\bibfield  {journal} {\bibinfo  {journal} {Phys. Rev. Lett.}\ }\textbf
  {\bibinfo {volume} {116}},\ \bibinfo {pages} {147203} (\bibinfo {year}
  {2016})}\BibitemShut {NoStop}%
\bibitem [{\citenamefont {B{\"u}ttner}\ \emph {et~al.}(2018)\citenamefont
  {B{\"u}ttner}, \citenamefont {Lemesh},\ and\ \citenamefont
  {Beach}}]{Buttner2018}%
  \BibitemOpen
  \bibfield  {author} {\bibinfo {author} {\bibfnamefont {F.}~\bibnamefont
  {B{\"u}ttner}}, \bibinfo {author} {\bibfnamefont {I.}~\bibnamefont
  {Lemesh}},\ and\ \bibinfo {author} {\bibfnamefont {G.~S.~D.}\ \bibnamefont
  {Beach}},\ }\bibfield  {title} {\bibinfo {title} {Theory of isolated magnetic
  skyrmions: From fundamentals to room temperature applications},\ }\href
  {https://doi.org/10.1038/s41598-018-22242-8} {\bibfield  {journal} {\bibinfo
  {journal} {Scientific Reports}\ }\textbf {\bibinfo {volume} {8}},\ \bibinfo
  {pages} {4464} (\bibinfo {year} {2018})}\BibitemShut {NoStop}%
\bibitem [{\citenamefont {Legrand}\ \emph {et~al.}(2020)\citenamefont
  {Legrand}, \citenamefont {Maccariello}, \citenamefont {Ajejas}, \citenamefont
  {Collin}, \citenamefont {Vecchiola}, \citenamefont {Bouzehouane},
  \citenamefont {Reyren}, \citenamefont {Cros},\ and\ \citenamefont
  {Fert}}]{Legrand2020}%
  \BibitemOpen
  \bibfield  {author} {\bibinfo {author} {\bibfnamefont {W.}~\bibnamefont
  {Legrand}}, \bibinfo {author} {\bibfnamefont {D.}~\bibnamefont
  {Maccariello}}, \bibinfo {author} {\bibfnamefont {F.}~\bibnamefont {Ajejas}},
  \bibinfo {author} {\bibfnamefont {S.}~\bibnamefont {Collin}}, \bibinfo
  {author} {\bibfnamefont {A.}~\bibnamefont {Vecchiola}}, \bibinfo {author}
  {\bibfnamefont {K.}~\bibnamefont {Bouzehouane}}, \bibinfo {author}
  {\bibfnamefont {N.}~\bibnamefont {Reyren}}, \bibinfo {author} {\bibfnamefont
  {V.}~\bibnamefont {Cros}},\ and\ \bibinfo {author} {\bibfnamefont
  {A.}~\bibnamefont {Fert}},\ }\bibfield  {title} {\bibinfo {title}
  {Room-temperature stabilization of antiferromagnetic skyrmions in synthetic
  antiferromagnets},\ }\href {https://doi.org/10.1038/s41563-019-0468-3}
  {\bibfield  {journal} {\bibinfo  {journal} {Nature Materials}\ }\textbf
  {\bibinfo {volume} {19}},\ \bibinfo {pages} {34} (\bibinfo {year}
  {2020})}\BibitemShut {NoStop}%
\bibitem [{\citenamefont {Juge}\ \emph {et~al.}(2022)\citenamefont {Juge},
  \citenamefont {Sisodia}, \citenamefont {Larra{\~{n}}aga}, \citenamefont
  {Zhang}, \citenamefont {Pham}, \citenamefont {Rana}, \citenamefont {Sarpi},
  \citenamefont {Mille}, \citenamefont {Stanescu}, \citenamefont {Belkhou},
  \citenamefont {Mawass}, \citenamefont {Novakovic-Marinkovic}, \citenamefont
  {Kronast}, \citenamefont {Weigand}, \citenamefont {Gr{\"a}fe}, \citenamefont
  {Wintz}, \citenamefont {Finizio}, \citenamefont {Raabe}, \citenamefont
  {Aballe}, \citenamefont {Foerster}, \citenamefont {Belmeguenai},
  \citenamefont {Buda-Prejbeanu}, \citenamefont {Pelloux-Prayer}, \citenamefont
  {Shaw}, \citenamefont {Nembach}, \citenamefont {Ranno}, \citenamefont
  {Gaudin},\ and\ \citenamefont {Boulle}}]{Juge2022}%
  \BibitemOpen
  \bibfield  {author} {\bibinfo {author} {\bibfnamefont {R.}~\bibnamefont
  {Juge}}, \bibinfo {author} {\bibfnamefont {N.}~\bibnamefont {Sisodia}},
  \bibinfo {author} {\bibfnamefont {J.~U.}\ \bibnamefont {Larra{\~{n}}aga}},
  \bibinfo {author} {\bibfnamefont {Q.}~\bibnamefont {Zhang}}, \bibinfo
  {author} {\bibfnamefont {V.~T.}\ \bibnamefont {Pham}}, \bibinfo {author}
  {\bibfnamefont {K.~G.}\ \bibnamefont {Rana}}, \bibinfo {author}
  {\bibfnamefont {B.}~\bibnamefont {Sarpi}}, \bibinfo {author} {\bibfnamefont
  {N.}~\bibnamefont {Mille}}, \bibinfo {author} {\bibfnamefont
  {S.}~\bibnamefont {Stanescu}}, \bibinfo {author} {\bibfnamefont
  {R.}~\bibnamefont {Belkhou}}, \bibinfo {author} {\bibfnamefont {M.-A.}\
  \bibnamefont {Mawass}}, \bibinfo {author} {\bibfnamefont {N.}~\bibnamefont
  {Novakovic-Marinkovic}}, \bibinfo {author} {\bibfnamefont {F.}~\bibnamefont
  {Kronast}}, \bibinfo {author} {\bibfnamefont {M.}~\bibnamefont {Weigand}},
  \bibinfo {author} {\bibfnamefont {J.}~\bibnamefont {Gr{\"a}fe}}, \bibinfo
  {author} {\bibfnamefont {S.}~\bibnamefont {Wintz}}, \bibinfo {author}
  {\bibfnamefont {S.}~\bibnamefont {Finizio}}, \bibinfo {author} {\bibfnamefont
  {J.}~\bibnamefont {Raabe}}, \bibinfo {author} {\bibfnamefont
  {L.}~\bibnamefont {Aballe}}, \bibinfo {author} {\bibfnamefont
  {M.}~\bibnamefont {Foerster}}, \bibinfo {author} {\bibfnamefont
  {M.}~\bibnamefont {Belmeguenai}}, \bibinfo {author} {\bibfnamefont {L.~D.}\
  \bibnamefont {Buda-Prejbeanu}}, \bibinfo {author} {\bibfnamefont
  {J.}~\bibnamefont {Pelloux-Prayer}}, \bibinfo {author} {\bibfnamefont
  {J.~M.}\ \bibnamefont {Shaw}}, \bibinfo {author} {\bibfnamefont {H.~T.}\
  \bibnamefont {Nembach}}, \bibinfo {author} {\bibfnamefont {L.}~\bibnamefont
  {Ranno}}, \bibinfo {author} {\bibfnamefont {G.}~\bibnamefont {Gaudin}},\ and\
  \bibinfo {author} {\bibfnamefont {O.}~\bibnamefont {Boulle}},\ }\bibfield
  {title} {\bibinfo {title} {Skyrmions in synthetic antiferromagnets and their
  nucleation via electrical current and ultra-fast laser illumination},\ }\href
  {https://doi.org/10.1038/s41467-022-32525-4} {\bibfield  {journal} {\bibinfo
  {journal} {Nature Communications}\ }\textbf {\bibinfo {volume} {13}},\
  \bibinfo {pages} {4807} (\bibinfo {year} {2022})}\BibitemShut {NoStop}%
\bibitem [{\citenamefont {Pham}\ \emph {et~al.}(2024)\citenamefont {Pham},
  \citenamefont {Sisodia}, \citenamefont {Manici}, \citenamefont
  {Urrestarazu-Larra{\~n}aga}, \citenamefont {Bairagi}, \citenamefont
  {Pelloux-Prayer}, \citenamefont {Guedas}, \citenamefont {Buda-Prejbeanu},
  \citenamefont {Auffret}, \citenamefont {Locatelli}, \citenamefont {Menteş},
  \citenamefont {Pizzini}, \citenamefont {Kumar}, \citenamefont {Finco},
  \citenamefont {Jacques}, \citenamefont {Gaudin},\ and\ \citenamefont
  {Boulle}}]{Pham2024}%
  \BibitemOpen
  \bibfield  {author} {\bibinfo {author} {\bibfnamefont {V.~T.}\ \bibnamefont
  {Pham}}, \bibinfo {author} {\bibfnamefont {N.}~\bibnamefont {Sisodia}},
  \bibinfo {author} {\bibfnamefont {I.~D.}\ \bibnamefont {Manici}}, \bibinfo
  {author} {\bibfnamefont {J.}~\bibnamefont {Urrestarazu-Larra{\~n}aga}},
  \bibinfo {author} {\bibfnamefont {K.}~\bibnamefont {Bairagi}}, \bibinfo
  {author} {\bibfnamefont {J.}~\bibnamefont {Pelloux-Prayer}}, \bibinfo
  {author} {\bibfnamefont {R.}~\bibnamefont {Guedas}}, \bibinfo {author}
  {\bibfnamefont {L.~D.}\ \bibnamefont {Buda-Prejbeanu}}, \bibinfo {author}
  {\bibfnamefont {S.}~\bibnamefont {Auffret}}, \bibinfo {author} {\bibfnamefont
  {A.}~\bibnamefont {Locatelli}}, \bibinfo {author} {\bibfnamefont {T.~O.}\
  \bibnamefont {Menteş}}, \bibinfo {author} {\bibfnamefont {S.}~\bibnamefont
  {Pizzini}}, \bibinfo {author} {\bibfnamefont {P.}~\bibnamefont {Kumar}},
  \bibinfo {author} {\bibfnamefont {A.}~\bibnamefont {Finco}}, \bibinfo
  {author} {\bibfnamefont {V.}~\bibnamefont {Jacques}}, \bibinfo {author}
  {\bibfnamefont {G.}~\bibnamefont {Gaudin}},\ and\ \bibinfo {author}
  {\bibfnamefont {O.}~\bibnamefont {Boulle}},\ }\bibfield  {title} {\bibinfo
  {title} {Fast current-induced skyrmion motion in synthetic
  antiferromagnets},\ }\href {https://doi.org/10.1126/science.add5751}
  {\bibfield  {journal} {\bibinfo  {journal} {Science}\ }\textbf {\bibinfo
  {volume} {384}},\ \bibinfo {pages} {307} (\bibinfo {year}
  {2024})}\BibitemShut {NoStop}%
\bibitem [{\citenamefont {Yang}\ \emph {et~al.}(2015)\citenamefont {Yang},
  \citenamefont {Ryu},\ and\ \citenamefont {Parkin}}]{Yang2015}%
  \BibitemOpen
  \bibfield  {author} {\bibinfo {author} {\bibfnamefont {S.-H.}\ \bibnamefont
  {Yang}}, \bibinfo {author} {\bibfnamefont {K.-S.}\ \bibnamefont {Ryu}},\ and\
  \bibinfo {author} {\bibfnamefont {S.}~\bibnamefont {Parkin}},\ }\bibfield
  {title} {\bibinfo {title} {Domain-wall velocities of up to 750 m s-1 driven
  by exchange coupling torque in synthetic antiferromagnets},\ }\href
  {https://doi.org/10.1038/nnano.2014.324} {\bibfield  {journal} {\bibinfo
  {journal} {Nature Nanotechnology}\ }\textbf {\bibinfo {volume} {10}},\
  \bibinfo {pages} {221} (\bibinfo {year} {2015})}\BibitemShut {NoStop}%
\bibitem [{\citenamefont {Lepadatu}\ \emph {et~al.}(2017)\citenamefont
  {Lepadatu}, \citenamefont {Saarikoski}, \citenamefont {Beacham},
  \citenamefont {Benitez}, \citenamefont {Moore}, \citenamefont {Burnell},
  \citenamefont {Sugimoto}, \citenamefont {Yesudas}, \citenamefont {Wheeler},
  \citenamefont {Miguel}, \citenamefont {Dhesi}, \citenamefont {McGrouther},
  \citenamefont {McVitie}, \citenamefont {Tatara},\ and\ \citenamefont
  {Marrows}}]{Lepadatu2017}%
  \BibitemOpen
  \bibfield  {author} {\bibinfo {author} {\bibfnamefont {S.}~\bibnamefont
  {Lepadatu}}, \bibinfo {author} {\bibfnamefont {H.}~\bibnamefont
  {Saarikoski}}, \bibinfo {author} {\bibfnamefont {R.}~\bibnamefont {Beacham}},
  \bibinfo {author} {\bibfnamefont {M.~J.}\ \bibnamefont {Benitez}}, \bibinfo
  {author} {\bibfnamefont {T.~A.}\ \bibnamefont {Moore}}, \bibinfo {author}
  {\bibfnamefont {G.}~\bibnamefont {Burnell}}, \bibinfo {author} {\bibfnamefont
  {S.}~\bibnamefont {Sugimoto}}, \bibinfo {author} {\bibfnamefont
  {D.}~\bibnamefont {Yesudas}}, \bibinfo {author} {\bibfnamefont {M.~C.}\
  \bibnamefont {Wheeler}}, \bibinfo {author} {\bibfnamefont {J.}~\bibnamefont
  {Miguel}}, \bibinfo {author} {\bibfnamefont {S.~S.}\ \bibnamefont {Dhesi}},
  \bibinfo {author} {\bibfnamefont {D.}~\bibnamefont {McGrouther}}, \bibinfo
  {author} {\bibfnamefont {S.}~\bibnamefont {McVitie}}, \bibinfo {author}
  {\bibfnamefont {G.}~\bibnamefont {Tatara}},\ and\ \bibinfo {author}
  {\bibfnamefont {C.~H.}\ \bibnamefont {Marrows}},\ }\bibfield  {title}
  {\bibinfo {title} {Synthetic ferrimagnet nanowires with very low critical
  current density for coupled domain wall motion},\ }\href
  {https://doi.org/10.1038/s41598-017-01748-7} {\bibfield  {journal} {\bibinfo
  {journal} {Scientific Reports}\ }\textbf {\bibinfo {volume} {7}},\ \bibinfo
  {pages} {1640} (\bibinfo {year} {2017})}\BibitemShut {NoStop}%
\bibitem [{\citenamefont {Barker}\ \emph
  {et~al.}(2023{\natexlab{a}})\citenamefont {Barker}, \citenamefont {Finizio},
  \citenamefont {Haltz}, \citenamefont {Mayr}, \citenamefont {Shepley},
  \citenamefont {Moore}, \citenamefont {Burnell}, \citenamefont {Raabe},\ and\
  \citenamefont {Marrows}}]{Barker2023_JPhysD}%
  \BibitemOpen
  \bibfield  {author} {\bibinfo {author} {\bibfnamefont {C.~E.~A.}\
  \bibnamefont {Barker}}, \bibinfo {author} {\bibfnamefont {S.}~\bibnamefont
  {Finizio}}, \bibinfo {author} {\bibfnamefont {E.}~\bibnamefont {Haltz}},
  \bibinfo {author} {\bibfnamefont {S.}~\bibnamefont {Mayr}}, \bibinfo {author}
  {\bibfnamefont {P.~M.}\ \bibnamefont {Shepley}}, \bibinfo {author}
  {\bibfnamefont {T.~A.}\ \bibnamefont {Moore}}, \bibinfo {author}
  {\bibfnamefont {G.}~\bibnamefont {Burnell}}, \bibinfo {author} {\bibfnamefont
  {J.}~\bibnamefont {Raabe}},\ and\ \bibinfo {author} {\bibfnamefont {C.~H.}\
  \bibnamefont {Marrows}},\ }\bibfield  {title} {\bibinfo {title} {Domain wall
  motion at low current density in a synthetic antiferromagnet nanowire},\
  }\href {https://doi.org/10.1088/1361-6463/ace6b4} {\bibfield  {journal}
  {\bibinfo  {journal} {Journal of Physics D: Applied Physics}\ }\textbf
  {\bibinfo {volume} {56}},\ \bibinfo {pages} {425002} (\bibinfo {year}
  {2023}{\natexlab{a}})}\BibitemShut {NoStop}%
\bibitem [{\citenamefont {Wang}\ \emph {et~al.}(2023)\citenamefont {Wang},
  \citenamefont {Bheemarasetty},\ and\ \citenamefont {Xiao}}]{Wang2023}%
  \BibitemOpen
  \bibfield  {author} {\bibinfo {author} {\bibfnamefont {K.}~\bibnamefont
  {Wang}}, \bibinfo {author} {\bibfnamefont {V.}~\bibnamefont
  {Bheemarasetty}},\ and\ \bibinfo {author} {\bibfnamefont {G.}~\bibnamefont
  {Xiao}},\ }\bibfield  {title} {\bibinfo {title} {{Spin textures in synthetic
  antiferromagnets: Challenges, opportunities, and future directions}},\ }\href
  {https://doi.org/10.1063/5.0153349} {\bibfield  {journal} {\bibinfo
  {journal} {APL Materials}\ }\textbf {\bibinfo {volume} {11}},\ \bibinfo
  {pages} {070902} (\bibinfo {year} {2023})}\BibitemShut {NoStop}%
\bibitem [{\citenamefont {Dohi}\ \emph {et~al.}(2019)\citenamefont {Dohi},
  \citenamefont {DuttaGupta}, \citenamefont {Fukami},\ and\ \citenamefont
  {Ohno}}]{Dohi2019}%
  \BibitemOpen
  \bibfield  {author} {\bibinfo {author} {\bibfnamefont {T.}~\bibnamefont
  {Dohi}}, \bibinfo {author} {\bibfnamefont {S.}~\bibnamefont {DuttaGupta}},
  \bibinfo {author} {\bibfnamefont {S.}~\bibnamefont {Fukami}},\ and\ \bibinfo
  {author} {\bibfnamefont {H.}~\bibnamefont {Ohno}},\ }\bibfield  {title}
  {\bibinfo {title} {Formation and current-induced motion of synthetic
  antiferromagnetic skyrmion bubbles},\ }\href
  {https://doi.org/10.1038/s41467-019-13182-6} {\bibfield  {journal} {\bibinfo
  {journal} {Nature Communications}\ }\textbf {\bibinfo {volume} {10}},\
  \bibinfo {pages} {5153} (\bibinfo {year} {2019})}\BibitemShut {NoStop}%
\bibitem [{\citenamefont {Lonsky}\ and\ \citenamefont
  {Hoffmann}(2020{\natexlab{a}})}]{Lonsky2020_APLMat}%
  \BibitemOpen
  \bibfield  {author} {\bibinfo {author} {\bibfnamefont {M.}~\bibnamefont
  {Lonsky}}\ and\ \bibinfo {author} {\bibfnamefont {A.}~\bibnamefont
  {Hoffmann}},\ }\bibfield  {title} {\bibinfo {title} {{Dynamic excitations of
  chiral magnetic textures}},\ }\href {https://doi.org/10.1063/5.0027042}
  {\bibfield  {journal} {\bibinfo  {journal} {APL Materials}\ }\textbf
  {\bibinfo {volume} {8}},\ \bibinfo {pages} {100903} (\bibinfo {year}
  {2020}{\natexlab{a}})}\BibitemShut {NoStop}%
\bibitem [{\citenamefont {Mochizuki}(2012)}]{Mochizuki2012}%
  \BibitemOpen
  \bibfield  {author} {\bibinfo {author} {\bibfnamefont {M.}~\bibnamefont
  {Mochizuki}},\ }\bibfield  {title} {\bibinfo {title} {Spin-wave modes and
  their intense excitation effects in skyrmion crystals},\ }\href
  {https://doi.org/10.1103/PhysRevLett.108.017601} {\bibfield  {journal}
  {\bibinfo  {journal} {Phys. Rev. Lett.}\ }\textbf {\bibinfo {volume} {108}},\
  \bibinfo {pages} {017601} (\bibinfo {year} {2012})}\BibitemShut {NoStop}%
\bibitem [{\citenamefont {Onose}\ \emph {et~al.}(2012)\citenamefont {Onose},
  \citenamefont {Okamura}, \citenamefont {Seki}, \citenamefont {Ishiwata},\
  and\ \citenamefont {Tokura}}]{Onose2012}%
  \BibitemOpen
  \bibfield  {author} {\bibinfo {author} {\bibfnamefont {Y.}~\bibnamefont
  {Onose}}, \bibinfo {author} {\bibfnamefont {Y.}~\bibnamefont {Okamura}},
  \bibinfo {author} {\bibfnamefont {S.}~\bibnamefont {Seki}}, \bibinfo {author}
  {\bibfnamefont {S.}~\bibnamefont {Ishiwata}},\ and\ \bibinfo {author}
  {\bibfnamefont {Y.}~\bibnamefont {Tokura}},\ }\bibfield  {title} {\bibinfo
  {title} {Observation of magnetic excitations of skyrmion crystal in a
  helimagnetic insulator ${\mathrm{cu}}_{2}{\mathrm{oseo}}_{3}$},\ }\href
  {https://doi.org/10.1103/PhysRevLett.109.037603} {\bibfield  {journal}
  {\bibinfo  {journal} {Phys. Rev. Lett.}\ }\textbf {\bibinfo {volume} {109}},\
  \bibinfo {pages} {037603} (\bibinfo {year} {2012})}\BibitemShut {NoStop}%
\bibitem [{\citenamefont {Kim}\ \emph {et~al.}(2014)\citenamefont {Kim},
  \citenamefont {Garcia-Sanchez}, \citenamefont {Sampaio}, \citenamefont
  {Moreau-Luchaire}, \citenamefont {Cros},\ and\ \citenamefont
  {Fert}}]{Kim2014}%
  \BibitemOpen
  \bibfield  {author} {\bibinfo {author} {\bibfnamefont {J.-V.}\ \bibnamefont
  {Kim}}, \bibinfo {author} {\bibfnamefont {F.}~\bibnamefont {Garcia-Sanchez}},
  \bibinfo {author} {\bibfnamefont {J.~a.}\ \bibnamefont {Sampaio}}, \bibinfo
  {author} {\bibfnamefont {C.}~\bibnamefont {Moreau-Luchaire}}, \bibinfo
  {author} {\bibfnamefont {V.}~\bibnamefont {Cros}},\ and\ \bibinfo {author}
  {\bibfnamefont {A.}~\bibnamefont {Fert}},\ }\bibfield  {title} {\bibinfo
  {title} {Breathing modes of confined skyrmions in ultrathin magnetic dots},\
  }\href {https://doi.org/10.1103/PhysRevB.90.064410} {\bibfield  {journal}
  {\bibinfo  {journal} {Phys. Rev. B}\ }\textbf {\bibinfo {volume} {90}},\
  \bibinfo {pages} {064410} (\bibinfo {year} {2014})}\BibitemShut {NoStop}%
\bibitem [{\citenamefont {Satywali}\ \emph {et~al.}(2021)\citenamefont
  {Satywali}, \citenamefont {Kravchuk}, \citenamefont {Pan}, \citenamefont
  {Raju}, \citenamefont {He}, \citenamefont {Ma}, \citenamefont
  {Petrovi{\'{c}}}, \citenamefont {Garst},\ and\ \citenamefont
  {Panagopoulos}}]{Satywali2021}%
  \BibitemOpen
  \bibfield  {author} {\bibinfo {author} {\bibfnamefont {B.}~\bibnamefont
  {Satywali}}, \bibinfo {author} {\bibfnamefont {V.~P.}\ \bibnamefont
  {Kravchuk}}, \bibinfo {author} {\bibfnamefont {L.}~\bibnamefont {Pan}},
  \bibinfo {author} {\bibfnamefont {M.}~\bibnamefont {Raju}}, \bibinfo {author}
  {\bibfnamefont {S.}~\bibnamefont {He}}, \bibinfo {author} {\bibfnamefont
  {F.}~\bibnamefont {Ma}}, \bibinfo {author} {\bibfnamefont {A.~P.}\
  \bibnamefont {Petrovi{\'{c}}}}, \bibinfo {author} {\bibfnamefont
  {M.}~\bibnamefont {Garst}},\ and\ \bibinfo {author} {\bibfnamefont
  {C.}~\bibnamefont {Panagopoulos}},\ }\bibfield  {title} {\bibinfo {title}
  {Microwave resonances of magnetic skyrmions in thin film multilayers},\
  }\href {https://doi.org/10.1038/s41467-021-22220-1} {\bibfield  {journal}
  {\bibinfo  {journal} {Nature Communications}\ }\textbf {\bibinfo {volume}
  {12}},\ \bibinfo {pages} {1909} (\bibinfo {year} {2021})}\BibitemShut
  {NoStop}%
\bibitem [{\citenamefont {Srivastava}\ \emph {et~al.}(2023)\citenamefont
  {Srivastava}, \citenamefont {Sassi}, \citenamefont {Ajejas}, \citenamefont
  {Vecchiola}, \citenamefont {Ngouagnia~Yemeli}, \citenamefont {Hurdequint},
  \citenamefont {Bouzehouane}, \citenamefont {Reyren}, \citenamefont {Cros},
  \citenamefont {Devolder}, \citenamefont {Kim},\ and\ \citenamefont
  {de~Loubens}}]{Srivastava2023}%
  \BibitemOpen
  \bibfield  {author} {\bibinfo {author} {\bibfnamefont {T.}~\bibnamefont
  {Srivastava}}, \bibinfo {author} {\bibfnamefont {Y.}~\bibnamefont {Sassi}},
  \bibinfo {author} {\bibfnamefont {F.}~\bibnamefont {Ajejas}}, \bibinfo
  {author} {\bibfnamefont {A.}~\bibnamefont {Vecchiola}}, \bibinfo {author}
  {\bibfnamefont {I.}~\bibnamefont {Ngouagnia~Yemeli}}, \bibinfo {author}
  {\bibfnamefont {H.}~\bibnamefont {Hurdequint}}, \bibinfo {author}
  {\bibfnamefont {K.}~\bibnamefont {Bouzehouane}}, \bibinfo {author}
  {\bibfnamefont {N.}~\bibnamefont {Reyren}}, \bibinfo {author} {\bibfnamefont
  {V.}~\bibnamefont {Cros}}, \bibinfo {author} {\bibfnamefont {T.}~\bibnamefont
  {Devolder}}, \bibinfo {author} {\bibfnamefont {J.-V.}\ \bibnamefont {Kim}},\
  and\ \bibinfo {author} {\bibfnamefont {G.}~\bibnamefont {de~Loubens}},\
  }\bibfield  {title} {\bibinfo {title} {{Resonant dynamics of
  three-dimensional skyrmionic textures in thin film multilayers}},\ }\href
  {https://doi.org/10.1063/5.0150265} {\bibfield  {journal} {\bibinfo
  {journal} {APL Materials}\ }\textbf {\bibinfo {volume} {11}},\ \bibinfo
  {pages} {061110} (\bibinfo {year} {2023})}\BibitemShut {NoStop}%
\bibitem [{\citenamefont {Lonsky}\ and\ \citenamefont
  {Hoffmann}(2020{\natexlab{b}})}]{Lonsky2020_PRB}%
  \BibitemOpen
  \bibfield  {author} {\bibinfo {author} {\bibfnamefont {M.}~\bibnamefont
  {Lonsky}}\ and\ \bibinfo {author} {\bibfnamefont {A.}~\bibnamefont
  {Hoffmann}},\ }\bibfield  {title} {\bibinfo {title} {Coupled skyrmion
  breathing modes in synthetic ferri- and antiferromagnets},\ }\href
  {https://doi.org/10.1103/PhysRevB.102.104403} {\bibfield  {journal} {\bibinfo
   {journal} {Phys. Rev. B}\ }\textbf {\bibinfo {volume} {102}},\ \bibinfo
  {pages} {104403} (\bibinfo {year} {2020}{\natexlab{b}})}\BibitemShut
  {NoStop}%
\bibitem [{\citenamefont {Barker}\ \emph
  {et~al.}(2023{\natexlab{b}})\citenamefont {Barker}, \citenamefont {Haltz},
  \citenamefont {Moore},\ and\ \citenamefont {Marrows}}]{Barker2023_JAP}%
  \BibitemOpen
  \bibfield  {author} {\bibinfo {author} {\bibfnamefont {C.~E.~A.}\
  \bibnamefont {Barker}}, \bibinfo {author} {\bibfnamefont {E.}~\bibnamefont
  {Haltz}}, \bibinfo {author} {\bibfnamefont {T.~A.}\ \bibnamefont {Moore}},\
  and\ \bibinfo {author} {\bibfnamefont {C.~H.}\ \bibnamefont {Marrows}},\
  }\bibfield  {title} {\bibinfo {title} {{Breathing modes of skyrmion strings
  in a synthetic antiferromagnet multilayer}},\ }\href
  {https://doi.org/10.1063/5.0142772} {\bibfield  {journal} {\bibinfo
  {journal} {Journal of Applied Physics}\ }\textbf {\bibinfo {volume} {133}},\
  \bibinfo {pages} {113901} (\bibinfo {year} {2023}{\natexlab{b}})}\BibitemShut
  {NoStop}%
\bibitem [{\citenamefont {Xing}\ \emph {et~al.}(2018)\citenamefont {Xing},
  \citenamefont {Hua},\ and\ \citenamefont {Wang}}]{Xing2018}%
  \BibitemOpen
  \bibfield  {author} {\bibinfo {author} {\bibfnamefont {L.}~\bibnamefont
  {Xing}}, \bibinfo {author} {\bibfnamefont {D.}~\bibnamefont {Hua}},\ and\
  \bibinfo {author} {\bibfnamefont {W.}~\bibnamefont {Wang}},\ }\bibfield
  {title} {\bibinfo {title} {{Magnetic excitations of skyrmions in
  antiferromagnetic-exchange coupled disks}},\ }\href
  {https://doi.org/10.1063/1.5042794} {\bibfield  {journal} {\bibinfo
  {journal} {Journal of Applied Physics}\ }\textbf {\bibinfo {volume} {124}},\
  \bibinfo {pages} {123904} (\bibinfo {year} {2018})}\BibitemShut {NoStop}%
\bibitem [{\citenamefont {Chen}\ \emph {et~al.}(2021)\citenamefont {Chen},
  \citenamefont {Hu},\ and\ \citenamefont {Yu}}]{Chen2021}%
  \BibitemOpen
  \bibfield  {author} {\bibinfo {author} {\bibfnamefont {J.}~\bibnamefont
  {Chen}}, \bibinfo {author} {\bibfnamefont {J.}~\bibnamefont {Hu}},\ and\
  \bibinfo {author} {\bibfnamefont {H.}~\bibnamefont {Yu}},\ }\bibfield
  {title} {\bibinfo {title} {Chiral emission of exchange spin waves by magnetic
  skyrmions},\ }\href {https://doi.org/10.1021/acsnano.0c07805} {\bibfield
  {journal} {\bibinfo  {journal} {ACS Nano}\ }\textbf {\bibinfo {volume}
  {15}},\ \bibinfo {pages} {4372} (\bibinfo {year} {2021})}\BibitemShut
  {NoStop}%
\bibitem [{\citenamefont {D{\'i}az}\ \emph {et~al.}(2020)\citenamefont
  {D{\'i}az}, \citenamefont {Hirosawa}, \citenamefont {Loss},\ and\
  \citenamefont {Psaroudaki}}]{Diaz2020}%
  \BibitemOpen
  \bibfield  {author} {\bibinfo {author} {\bibfnamefont {S.~A.}\ \bibnamefont
  {D{\'i}az}}, \bibinfo {author} {\bibfnamefont {T.}~\bibnamefont {Hirosawa}},
  \bibinfo {author} {\bibfnamefont {D.}~\bibnamefont {Loss}},\ and\ \bibinfo
  {author} {\bibfnamefont {C.}~\bibnamefont {Psaroudaki}},\ }\bibfield  {title}
  {\bibinfo {title} {Spin wave radiation by a topological charge dipole},\
  }\href {https://doi.org/10.1021/acs.nanolett.0c02192} {\bibfield  {journal}
  {\bibinfo  {journal} {Nano Letters}\ }\textbf {\bibinfo {volume} {20}},\
  \bibinfo {pages} {6556} (\bibinfo {year} {2020})}\BibitemShut {NoStop}%
\bibitem [{\citenamefont {Tang}\ \emph {et~al.}(2023)\citenamefont {Tang},
  \citenamefont {Liyanage}, \citenamefont {Montoya}, \citenamefont {Patel},
  \citenamefont {Quigley}, \citenamefont {Grutter}, \citenamefont
  {Fitzsimmons}, \citenamefont {Sinha}, \citenamefont {Borchers}, \citenamefont
  {Fullerton}, \citenamefont {DeBeer-Schmitt},\ and\ \citenamefont
  {Gilbert}}]{Tang2023}%
  \BibitemOpen
  \bibfield  {author} {\bibinfo {author} {\bibfnamefont {N.}~\bibnamefont
  {Tang}}, \bibinfo {author} {\bibfnamefont {W.~L. N.~C.}\ \bibnamefont
  {Liyanage}}, \bibinfo {author} {\bibfnamefont {S.~A.}\ \bibnamefont
  {Montoya}}, \bibinfo {author} {\bibfnamefont {S.}~\bibnamefont {Patel}},
  \bibinfo {author} {\bibfnamefont {L.~J.}\ \bibnamefont {Quigley}}, \bibinfo
  {author} {\bibfnamefont {A.~J.}\ \bibnamefont {Grutter}}, \bibinfo {author}
  {\bibfnamefont {M.~R.}\ \bibnamefont {Fitzsimmons}}, \bibinfo {author}
  {\bibfnamefont {S.}~\bibnamefont {Sinha}}, \bibinfo {author} {\bibfnamefont
  {J.~A.}\ \bibnamefont {Borchers}}, \bibinfo {author} {\bibfnamefont {E.~E.}\
  \bibnamefont {Fullerton}}, \bibinfo {author} {\bibfnamefont {L.}~\bibnamefont
  {DeBeer-Schmitt}},\ and\ \bibinfo {author} {\bibfnamefont {D.~A.}\
  \bibnamefont {Gilbert}},\ }\bibfield  {title} {\bibinfo {title}
  {Skyrmion-excited spin-wave fractal networks},\ }\href
  {https://doi.org/https://doi.org/10.1002/adma.202300416} {\bibfield
  {journal} {\bibinfo  {journal} {Advanced Materials}\ }\textbf {\bibinfo
  {volume} {35}},\ \bibinfo {pages} {2300416} (\bibinfo {year} {2023})},\
  \Eprint
  {https://arxiv.org/abs/https://advanced.onlinelibrary.wiley.com/doi/pdf/10.1002/adma.202300416}
  {https://advanced.onlinelibrary.wiley.com/doi/pdf/10.1002/adma.202300416}
  \BibitemShut {NoStop}%
\bibitem [{\citenamefont {Yuan}\ \emph {et~al.}(2019)\citenamefont {Yuan},
  \citenamefont {Wang}, \citenamefont {Yung},\ and\ \citenamefont
  {Wang}}]{Yuan2019}%
  \BibitemOpen
  \bibfield  {author} {\bibinfo {author} {\bibfnamefont {H.~Y.}\ \bibnamefont
  {Yuan}}, \bibinfo {author} {\bibfnamefont {X.~S.}\ \bibnamefont {Wang}},
  \bibinfo {author} {\bibfnamefont {M.-H.}\ \bibnamefont {Yung}},\ and\
  \bibinfo {author} {\bibfnamefont {X.~R.}\ \bibnamefont {Wang}},\ }\bibfield
  {title} {\bibinfo {title} {Wiggling skyrmion propagation under parametric
  pumping},\ }\href {https://doi.org/10.1103/PhysRevB.99.014428} {\bibfield
  {journal} {\bibinfo  {journal} {Phys. Rev. B}\ }\textbf {\bibinfo {volume}
  {99}},\ \bibinfo {pages} {014428} (\bibinfo {year} {2019})}\BibitemShut
  {NoStop}%
\bibitem [{\citenamefont {Qiu}\ \emph {et~al.}(2021)\citenamefont {Qiu},
  \citenamefont {Shen}, \citenamefont {Zhang}, \citenamefont {Zhou},
  \citenamefont {Zhao}, \citenamefont {Xia}, \citenamefont {Luo},\ and\
  \citenamefont {Liu}}]{Qiu2021}%
  \BibitemOpen
  \bibfield  {author} {\bibinfo {author} {\bibfnamefont {L.}~\bibnamefont
  {Qiu}}, \bibinfo {author} {\bibfnamefont {L.}~\bibnamefont {Shen}}, \bibinfo
  {author} {\bibfnamefont {X.}~\bibnamefont {Zhang}}, \bibinfo {author}
  {\bibfnamefont {Y.}~\bibnamefont {Zhou}}, \bibinfo {author} {\bibfnamefont
  {G.}~\bibnamefont {Zhao}}, \bibinfo {author} {\bibfnamefont {W.}~\bibnamefont
  {Xia}}, \bibinfo {author} {\bibfnamefont {H.-B.}\ \bibnamefont {Luo}},\ and\
  \bibinfo {author} {\bibfnamefont {J.~P.}\ \bibnamefont {Liu}},\ }\bibfield
  {title} {\bibinfo {title} {{Interlayer coupling effect on skyrmion dynamics
  in synthetic antiferromagnets}},\ }\href {https://doi.org/10.1063/5.0039470}
  {\bibfield  {journal} {\bibinfo  {journal} {Applied Physics Letters}\
  }\textbf {\bibinfo {volume} {118}},\ \bibinfo {pages} {082403} (\bibinfo
  {year} {2021})}\BibitemShut {NoStop}%
\bibitem [{\citenamefont {Göbel}\ \emph {et~al.}(2021)\citenamefont {Göbel},
  \citenamefont {Mertig},\ and\ \citenamefont {Tretiakov}}]{GOBEL20211}%
  \BibitemOpen
  \bibfield  {author} {\bibinfo {author} {\bibfnamefont {B.}~\bibnamefont
  {Göbel}}, \bibinfo {author} {\bibfnamefont {I.}~\bibnamefont {Mertig}},\
  and\ \bibinfo {author} {\bibfnamefont {O.~A.}\ \bibnamefont {Tretiakov}},\
  }\bibfield  {title} {\bibinfo {title} {Beyond skyrmions: Review and
  perspectives of alternative magnetic quasiparticles},\ }\href
  {https://doi.org/https://doi.org/10.1016/j.physrep.2020.10.001} {\bibfield
  {journal} {\bibinfo  {journal} {Physics Reports}\ }\textbf {\bibinfo {volume}
  {895}},\ \bibinfo {pages} {1} (\bibinfo {year} {2021})},\ \bibinfo {note}
  {beyond skyrmions: Review and perspectives of alternative magnetic
  quasiparticles}\BibitemShut {NoStop}%
\bibitem [{\citenamefont {Liu}\ \emph {et~al.}(2025)\citenamefont {Liu},
  \citenamefont {Ai}, \citenamefont {Reisbick}, \citenamefont {Zong},
  \citenamefont {Pofelski}, \citenamefont {Han}, \citenamefont {Camino},
  \citenamefont {Jing}, \citenamefont {Lomakin},\ and\ \citenamefont
  {Zhu}}]{Liu2025}%
  \BibitemOpen
  \bibfield  {author} {\bibinfo {author} {\bibfnamefont {C.}~\bibnamefont
  {Liu}}, \bibinfo {author} {\bibfnamefont {F.}~\bibnamefont {Ai}}, \bibinfo
  {author} {\bibfnamefont {S.}~\bibnamefont {Reisbick}}, \bibinfo {author}
  {\bibfnamefont {A.}~\bibnamefont {Zong}}, \bibinfo {author} {\bibfnamefont
  {A.}~\bibnamefont {Pofelski}}, \bibinfo {author} {\bibfnamefont {M.-G.}\
  \bibnamefont {Han}}, \bibinfo {author} {\bibfnamefont {F.}~\bibnamefont
  {Camino}}, \bibinfo {author} {\bibfnamefont {C.}~\bibnamefont {Jing}},
  \bibinfo {author} {\bibfnamefont {V.}~\bibnamefont {Lomakin}},\ and\ \bibinfo
  {author} {\bibfnamefont {Y.}~\bibnamefont {Zhu}},\ }\bibfield  {title}
  {\bibinfo {title} {Correlated spin-wave generation and domain-wall
  oscillation in a topologically textured magnetic film},\ }\bibfield
  {journal} {\bibinfo  {journal} {Nature Materials}\ }\href
  {https://doi.org/10.1038/s41563-024-02085-7} {10.1038/s41563-024-02085-7}
  (\bibinfo {year} {2025})\BibitemShut {NoStop}%
\bibitem [{\citenamefont {Girardi}\ \emph {et~al.}(2024)\citenamefont
  {Girardi}, \citenamefont {Finizio}, \citenamefont {Donnelly}, \citenamefont
  {Rubini}, \citenamefont {Mayr}, \citenamefont {Levati}, \citenamefont
  {Cuccurullo}, \citenamefont {Maspero}, \citenamefont {Raabe}, \citenamefont
  {Petti},\ and\ \citenamefont {Albisetti}}]{Girardi2024}%
  \BibitemOpen
  \bibfield  {author} {\bibinfo {author} {\bibfnamefont {D.}~\bibnamefont
  {Girardi}}, \bibinfo {author} {\bibfnamefont {S.}~\bibnamefont {Finizio}},
  \bibinfo {author} {\bibfnamefont {C.}~\bibnamefont {Donnelly}}, \bibinfo
  {author} {\bibfnamefont {G.}~\bibnamefont {Rubini}}, \bibinfo {author}
  {\bibfnamefont {S.}~\bibnamefont {Mayr}}, \bibinfo {author} {\bibfnamefont
  {V.}~\bibnamefont {Levati}}, \bibinfo {author} {\bibfnamefont
  {S.}~\bibnamefont {Cuccurullo}}, \bibinfo {author} {\bibfnamefont
  {F.}~\bibnamefont {Maspero}}, \bibinfo {author} {\bibfnamefont
  {J.}~\bibnamefont {Raabe}}, \bibinfo {author} {\bibfnamefont
  {D.}~\bibnamefont {Petti}},\ and\ \bibinfo {author} {\bibfnamefont
  {E.}~\bibnamefont {Albisetti}},\ }\bibfield  {title} {\bibinfo {title}
  {Three-dimensional spin-wave dynamics, localization and interference in a
  synthetic antiferromagnet},\ }\href
  {https://doi.org/10.1038/s41467-024-47339-9} {\bibfield  {journal} {\bibinfo
  {journal} {Nature Communications}\ }\textbf {\bibinfo {volume} {15}},\
  \bibinfo {pages} {3057} (\bibinfo {year} {2024})}\BibitemShut {NoStop}%
\bibitem [{\citenamefont {Vansteenkiste}\ \emph {et~al.}(2014)\citenamefont
  {Vansteenkiste}, \citenamefont {Leliaert}, \citenamefont {Dvornik},
  \citenamefont {Helsen}, \citenamefont {Garcia-Sanchez},\ and\ \citenamefont
  {Van~Waeyenberge}}]{Vansteenkiste2014}%
  \BibitemOpen
  \bibfield  {author} {\bibinfo {author} {\bibfnamefont {A.}~\bibnamefont
  {Vansteenkiste}}, \bibinfo {author} {\bibfnamefont {J.}~\bibnamefont
  {Leliaert}}, \bibinfo {author} {\bibfnamefont {M.}~\bibnamefont {Dvornik}},
  \bibinfo {author} {\bibfnamefont {M.}~\bibnamefont {Helsen}}, \bibinfo
  {author} {\bibfnamefont {F.}~\bibnamefont {Garcia-Sanchez}},\ and\ \bibinfo
  {author} {\bibfnamefont {B.}~\bibnamefont {Van~Waeyenberge}},\ }\bibfield
  {title} {\bibinfo {title} {{The design and verification of MuMax3}},\ }\href
  {https://doi.org/10.1063/1.4899186} {\bibfield  {journal} {\bibinfo
  {journal} {AIP Advances}\ }\textbf {\bibinfo {volume} {4}},\ \bibinfo {pages}
  {107133} (\bibinfo {year} {2014})}\BibitemShut {NoStop}%
\bibitem [{\citenamefont {Jiang}\ \emph
  {et~al.}(2017{\natexlab{b}})\citenamefont {Jiang}, \citenamefont {Chen},
  \citenamefont {Liu}, \citenamefont {Zang}, \citenamefont {{te Velthuis}},\
  and\ \citenamefont {Hoffmann}}]{Jiang_2017}%
  \BibitemOpen
  \bibfield  {author} {\bibinfo {author} {\bibfnamefont {W.}~\bibnamefont
  {Jiang}}, \bibinfo {author} {\bibfnamefont {G.}~\bibnamefont {Chen}},
  \bibinfo {author} {\bibfnamefont {K.}~\bibnamefont {Liu}}, \bibinfo {author}
  {\bibfnamefont {J.}~\bibnamefont {Zang}}, \bibinfo {author} {\bibfnamefont
  {S.~G.}\ \bibnamefont {{te Velthuis}}},\ and\ \bibinfo {author}
  {\bibfnamefont {A.}~\bibnamefont {Hoffmann}},\ }\bibfield  {title} {\bibinfo
  {title} {Skyrmions in magnetic multilayers},\ }\href
  {https://doi.org/https://doi.org/10.1016/j.physrep.2017.08.001} {\bibfield
  {journal} {\bibinfo  {journal} {Physics Reports}\ }\textbf {\bibinfo {volume}
  {704}},\ \bibinfo {pages} {1} (\bibinfo {year}
  {2017}{\natexlab{b}})}\BibitemShut {NoStop}%
\bibitem [{\citenamefont {Barker}\ \emph {et~al.}(2024)\citenamefont {Barker},
  \citenamefont {Fallon}, \citenamefont {Barton}, \citenamefont {Haltz},
  \citenamefont {Almeida}, \citenamefont {Villa}, \citenamefont {Kirkbride},
  \citenamefont {Maccherozzi}, \citenamefont {Sarpi}, \citenamefont {Dhesi},
  \citenamefont {McGrouther}, \citenamefont {McVitie}, \citenamefont {Moore},
  \citenamefont {Kazakova},\ and\ \citenamefont {Marrows}}]{Barker2024_PRB}%
  \BibitemOpen
  \bibfield  {author} {\bibinfo {author} {\bibfnamefont {C.~E.~A.}\
  \bibnamefont {Barker}}, \bibinfo {author} {\bibfnamefont {K.}~\bibnamefont
  {Fallon}}, \bibinfo {author} {\bibfnamefont {C.}~\bibnamefont {Barton}},
  \bibinfo {author} {\bibfnamefont {E.}~\bibnamefont {Haltz}}, \bibinfo
  {author} {\bibfnamefont {T.~P.}\ \bibnamefont {Almeida}}, \bibinfo {author}
  {\bibfnamefont {S.}~\bibnamefont {Villa}}, \bibinfo {author} {\bibfnamefont
  {C.}~\bibnamefont {Kirkbride}}, \bibinfo {author} {\bibfnamefont
  {F.}~\bibnamefont {Maccherozzi}}, \bibinfo {author} {\bibfnamefont
  {B.}~\bibnamefont {Sarpi}}, \bibinfo {author} {\bibfnamefont {S.~S.}\
  \bibnamefont {Dhesi}}, \bibinfo {author} {\bibfnamefont {D.}~\bibnamefont
  {McGrouther}}, \bibinfo {author} {\bibfnamefont {S.}~\bibnamefont {McVitie}},
  \bibinfo {author} {\bibfnamefont {T.~A.}\ \bibnamefont {Moore}}, \bibinfo
  {author} {\bibfnamefont {O.}~\bibnamefont {Kazakova}},\ and\ \bibinfo
  {author} {\bibfnamefont {C.~H.}\ \bibnamefont {Marrows}},\ }\bibfield
  {title} {\bibinfo {title} {Phase coexistence and transitions between
  antiferromagnetic and ferromagnetic states in a synthetic antiferromagnet},\
  }\href {https://doi.org/10.1103/PhysRevB.109.134437} {\bibfield  {journal}
  {\bibinfo  {journal} {Phys. Rev. B}\ }\textbf {\bibinfo {volume} {109}},\
  \bibinfo {pages} {134437} (\bibinfo {year} {2024})}\BibitemShut {NoStop}%
\bibitem [{\citenamefont {Haltz}\ \emph {et~al.}(2023)\citenamefont {Haltz},
  \citenamefont {Barker},\ and\ \citenamefont {Marrows}}]{haltz2023_arxiv}%
  \BibitemOpen
  \bibfield  {author} {\bibinfo {author} {\bibfnamefont {E.}~\bibnamefont
  {Haltz}}, \bibinfo {author} {\bibfnamefont {C.~E.~A.}\ \bibnamefont
  {Barker}},\ and\ \bibinfo {author} {\bibfnamefont {C.~H.}\ \bibnamefont
  {Marrows}},\ }\href {https://arxiv.org/abs/2309.03697} {\bibinfo {title}
  {Statics and dynamics of skyrmions in balanced and unbalanced synthetic
  antiferromagnets}} (\bibinfo {year} {2023}),\ \Eprint
  {https://arxiv.org/abs/2309.03697} {arXiv:2309.03697 [cond-mat.mes-hall]}
  \BibitemShut {NoStop}%
\bibitem [{\citenamefont {Mishra}\ \emph {et~al.}(2024)\citenamefont {Mishra},
  \citenamefont {Sravani}, \citenamefont {Bose},\ and\ \citenamefont
  {Bhuktare}}]{Mishra2024}%
  \BibitemOpen
  \bibfield  {author} {\bibinfo {author} {\bibfnamefont {P.~K.}\ \bibnamefont
  {Mishra}}, \bibinfo {author} {\bibfnamefont {M.}~\bibnamefont {Sravani}},
  \bibinfo {author} {\bibfnamefont {A.}~\bibnamefont {Bose}},\ and\ \bibinfo
  {author} {\bibfnamefont {S.}~\bibnamefont {Bhuktare}},\ }\bibfield  {title}
  {\bibinfo {title} {Voltage-controlled magnetic anisotropy-based spintronic
  devices for magnetic memory applications: Challenges and perspectives},\
  }\href {https://doi.org/10.1063/5.0201648} {\bibfield  {journal} {\bibinfo
  {journal} {Journal of Applied Physics}\ }\textbf {\bibinfo {volume} {135}},\
  \bibinfo {pages} {220701} (\bibinfo {year} {2024})}\BibitemShut {NoStop}%
\bibitem [{\citenamefont {Slonczewski}(1996)}]{Slonczewski1996}%
  \BibitemOpen
  \bibfield  {author} {\bibinfo {author} {\bibfnamefont {J.}~\bibnamefont
  {Slonczewski}},\ }\bibfield  {title} {\bibinfo {title} {Current-driven
  excitation of magnetic multilayers},\ }\href
  {https://doi.org/https://doi.org/10.1016/0304-8853(96)00062-5} {\bibfield
  {journal} {\bibinfo  {journal} {Journal of Magnetism and Magnetic Materials}\
  }\textbf {\bibinfo {volume} {159}},\ \bibinfo {pages} {L1} (\bibinfo {year}
  {1996})}\BibitemShut {NoStop}%
\bibitem [{\citenamefont {Berger}(1996)}]{Berger1996}%
  \BibitemOpen
  \bibfield  {author} {\bibinfo {author} {\bibfnamefont {L.}~\bibnamefont
  {Berger}},\ }\bibfield  {title} {\bibinfo {title} {Emission of spin waves by
  a magnetic multilayer traversed by a current},\ }\href
  {https://doi.org/10.1103/PhysRevB.54.9353} {\bibfield  {journal} {\bibinfo
  {journal} {Phys. Rev. B}\ }\textbf {\bibinfo {volume} {54}},\ \bibinfo
  {pages} {9353} (\bibinfo {year} {1996})}\BibitemShut {NoStop}%
\bibitem [{\citenamefont {Xiao}\ \emph {et~al.}(2004)\citenamefont {Xiao},
  \citenamefont {Zangwill},\ and\ \citenamefont {Stiles}}]{Xiao2004}%
  \BibitemOpen
  \bibfield  {author} {\bibinfo {author} {\bibfnamefont {J.}~\bibnamefont
  {Xiao}}, \bibinfo {author} {\bibfnamefont {A.}~\bibnamefont {Zangwill}},\
  and\ \bibinfo {author} {\bibfnamefont {M.~D.}\ \bibnamefont {Stiles}},\
  }\bibfield  {title} {\bibinfo {title} {Boltzmann test of slonczewski's theory
  of spin-transfer torque},\ }\href
  {https://doi.org/10.1103/PhysRevB.70.172405} {\bibfield  {journal} {\bibinfo
  {journal} {Phys. Rev. B}\ }\textbf {\bibinfo {volume} {70}},\ \bibinfo
  {pages} {172405} (\bibinfo {year} {2004})}\BibitemShut {NoStop}%
\bibitem [{\citenamefont {Eid}\ \emph {et~al.}(2002)\citenamefont {Eid},
  \citenamefont {Fonck}, \citenamefont {Darwish}, \citenamefont {Pratt},\ and\
  \citenamefont {Bass}}]{Eid2002}%
  \BibitemOpen
  \bibfield  {author} {\bibinfo {author} {\bibfnamefont {K.}~\bibnamefont
  {Eid}}, \bibinfo {author} {\bibfnamefont {R.}~\bibnamefont {Fonck}}, \bibinfo
  {author} {\bibfnamefont {M.~A.}\ \bibnamefont {Darwish}}, \bibinfo {author}
  {\bibfnamefont {J.}~\bibnamefont {Pratt}, \bibfnamefont {W.~P.}},\ and\
  \bibinfo {author} {\bibfnamefont {J.}~\bibnamefont {Bass}},\ }\bibfield
  {title} {\bibinfo {title} {Current-perpendicular-to-plane-magnetoresistance
  properties of ru and co/ru interfaces},\ }\href
  {https://doi.org/10.1063/1.1447294} {\bibfield  {journal} {\bibinfo
  {journal} {Journal of Applied Physics}\ }\textbf {\bibinfo {volume} {91}},\
  \bibinfo {pages} {8102} (\bibinfo {year} {2002})},\ \Eprint
  {https://arxiv.org/abs/https://pubs.aip.org/aip/jap/article-pdf/91/10/8102/19074680/8102\_1\_online.pdf}
  {https://pubs.aip.org/aip/jap/article-pdf/91/10/8102/19074680/8102\_1\_online.pdf}
  \BibitemShut {NoStop}%
\end{thebibliography}%

\end{document}


\title{Skyrmion motion in a synthetic antiferromagnet driven by asymmetric spin wave emission\\
	Supplementary information}
	
	\author{Christopher E. A. Barker}
	\email[]{christopher.barker@npl.co.uk}
	\affiliation{National Physical Laboratory, Hampton Road, Teddington, TW11 0LW, United Kingdom}
	
	\author{Charles Parton-Barr}
	\affiliation{School of Physics and Astronomy, University of Leeds, Leeds, LS2 9JT, United Kingdom}
	
	\author{Christopher H. Marrows}
	\email[]{c.h.marrows@leeds.ac.uk}
	\affiliation{School of Physics and Astronomy, University of Leeds, Leeds, LS2 9JT, United Kingdom}
	
	\author{Olga Kazakova}
	\affiliation{National Physical Laboratory, Hampton Road, Teddington, TW11 0LW, United Kingdom}
	\affiliation{Department of Electrical and Electronic Engineering, University of Manchester, Manchester, M13 9PL, United Kingdom}
	
	\author{Craig Barton}
	\email[Author to whom correspondence should be addressed: ]{craig.barton@npl.co.uk}
	\affiliation{National Physical Laboratory, Hampton Road, Teddington, TW11 0LW, United Kingdom}
	
	\date{\today}
	
	\maketitle
	\newpage
	
	\section{Measurement of hysteresis loops}\label{supp:note:hyst_loops}
\begin{figure}[h!]
		\centering
		\includegraphics[width=0.68\linewidth]{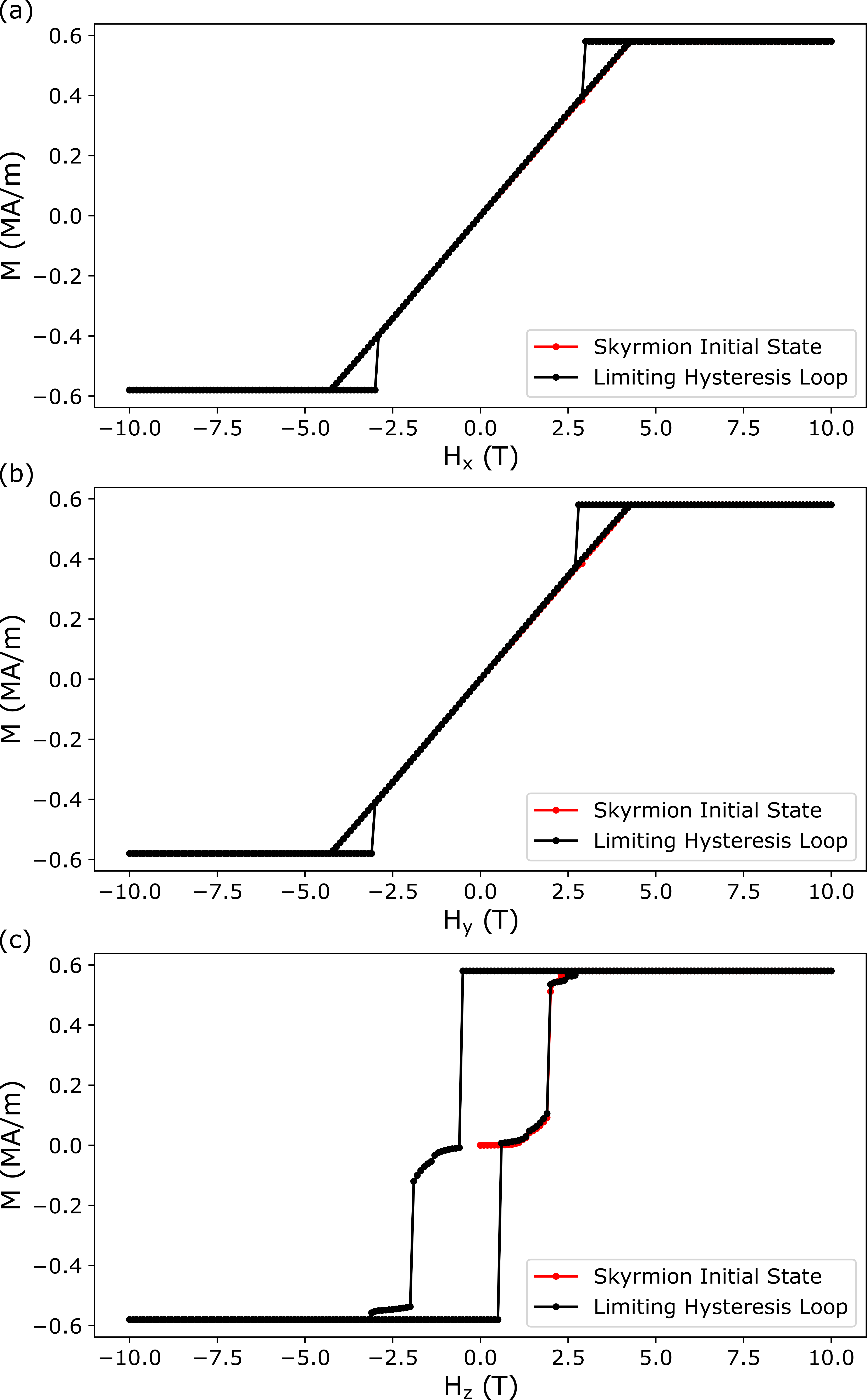}
		\caption{Hysteresis loops of the simulated system for field along the three principal axes. (a) Field along the x axis. (b) Field along the y axis. (c) Field along the z axis. All three cases start with a relaxed skyrmion state, and the data for the field sweep from this to positive saturation is in red. However, with the exception of a small section of panel (c), this always lies at the same values as the limiting hyseresis loop in black.}
		\label{supp:fig:hyst_loops}
	\end{figure}

	To give a comparison of the fields used in the main paper to the saturation fields of the system, hysteresis loops were simulated. These are shown in Fig.~\ref{supp:fig:hyst_loops}. This was done for magnetic fields along all three principal axes: Fig~\ref{supp:fig:hyst_loops}(a) shows the field along the x-axis, Fig~\ref{supp:fig:hyst_loops}(b) shows the field along the y-axis, and Fig.~\ref{supp:fig:hyst_loops}(c) shows the field along the z-axis. In all cases, the system was first initialised with a skyrmion in the centre and allowed to relax, in the same way as for all simulations in the main paper. After this, the field was first swept to positive saturation, before a full sweep was performed between positive and negative saturation and back. The initial data for the sweep of the field from the skyrmion state to saturation is shown in red, while the remaining data---which forms the limiting hysteresis loop---is shown in black for all three datasets. From Figs.~\ref{supp:fig:hyst_loops}(a) and~\ref{supp:fig:hyst_loops}(b), we can determine that the in-plane saturation field for the sample is 4.3~T. This means that our chosen static in-plane field of 0.5~T is only $\sim$10~\% of the saturation field.
	\newpage
	
	\newpage
	
	\section{Identification of higher frequency spin-wave mode driven by in-plane field}\label{supp:sec:high_freq_mode}
		\begin{figure}[h!]
			\centering
			\includegraphics[width=\linewidth]{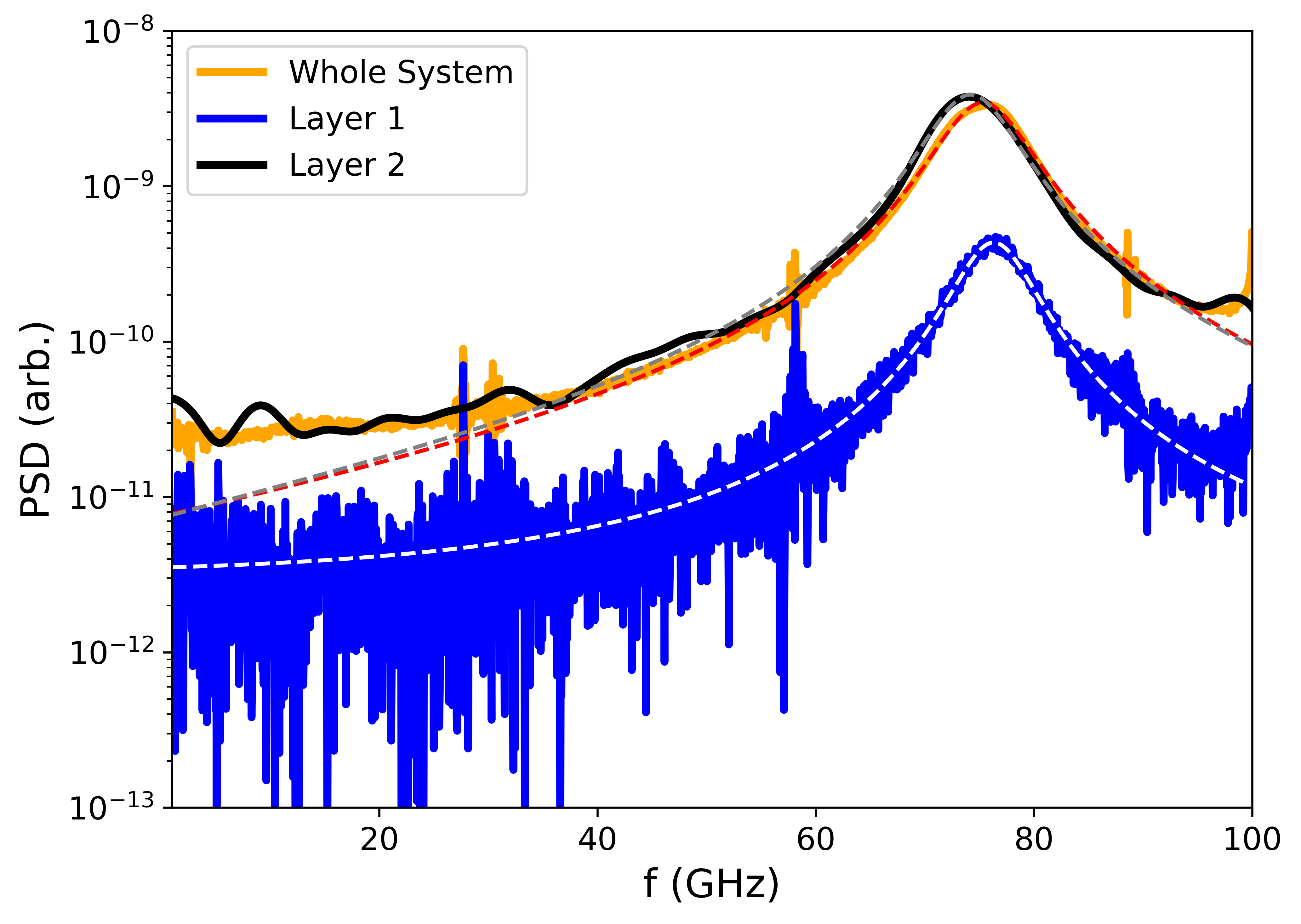}
			\caption{PSD of the simulation when initialised with a uniform magnetization, and then excited with an out of plane sinc field of 0.5~mT in a static in-plane field of 0.5~T. Shown are the response of the entire simulated system, as well as the two individual magnetic layers, along with fits to Lorentzian functions for all three datasets.}
			\label{supp:fig:IP_SW_Mode}
		\end{figure}
		
		In Fig.~\ref{fig:velocity_fn_freq}(a) of the main paper, a peak in the skyrmion velocity is observed that is well coupled to the frequency of the out-of-phase breathing mode in Fig.~\ref{fig:Skyrmion_breathing} of the main paper. However, there is an additional higher frequency peak in velocity that does not correspond to a skyrmion breathing mode. In the main paper we suggested this is likely due to a spin wave mode that arises from the presence of the in-plane magnetic field. To confirm whether this is the case we perform a simulation of the system when excited with an out-of-plane sinc field of 0.5~mT in the presence of a static in-plane field of 0.5~T. If we were to perform the simulation in these conditions in the presence of a skyrmion, this would induce motion which would confuse the interpretation of the results. For this reason, we perform the simulation with an initial magnetisation that is uniform. The results of this are shown in Fig.~\ref{supp:fig:IP_SW_Mode} for the entire simulated volume as well as the two individual magnetic layers. Lorentzian fits to each response are shown with dashed lines. The centre frequencies extracted from these are $75.321 \pm 0.005$~GHz for the entire system and $76.32 \pm 0.01$~GHz and $74.212 \pm 0.004$~GHz for the first and second layers respectively. These values align well with the higher-frequency peak in velocity observed in Fig.~\ref{fig:velocity_fn_freq}(a) in the main paper.\\\\
		\begin{figure}[h!]
			\centering
			\includegraphics[width=\linewidth]{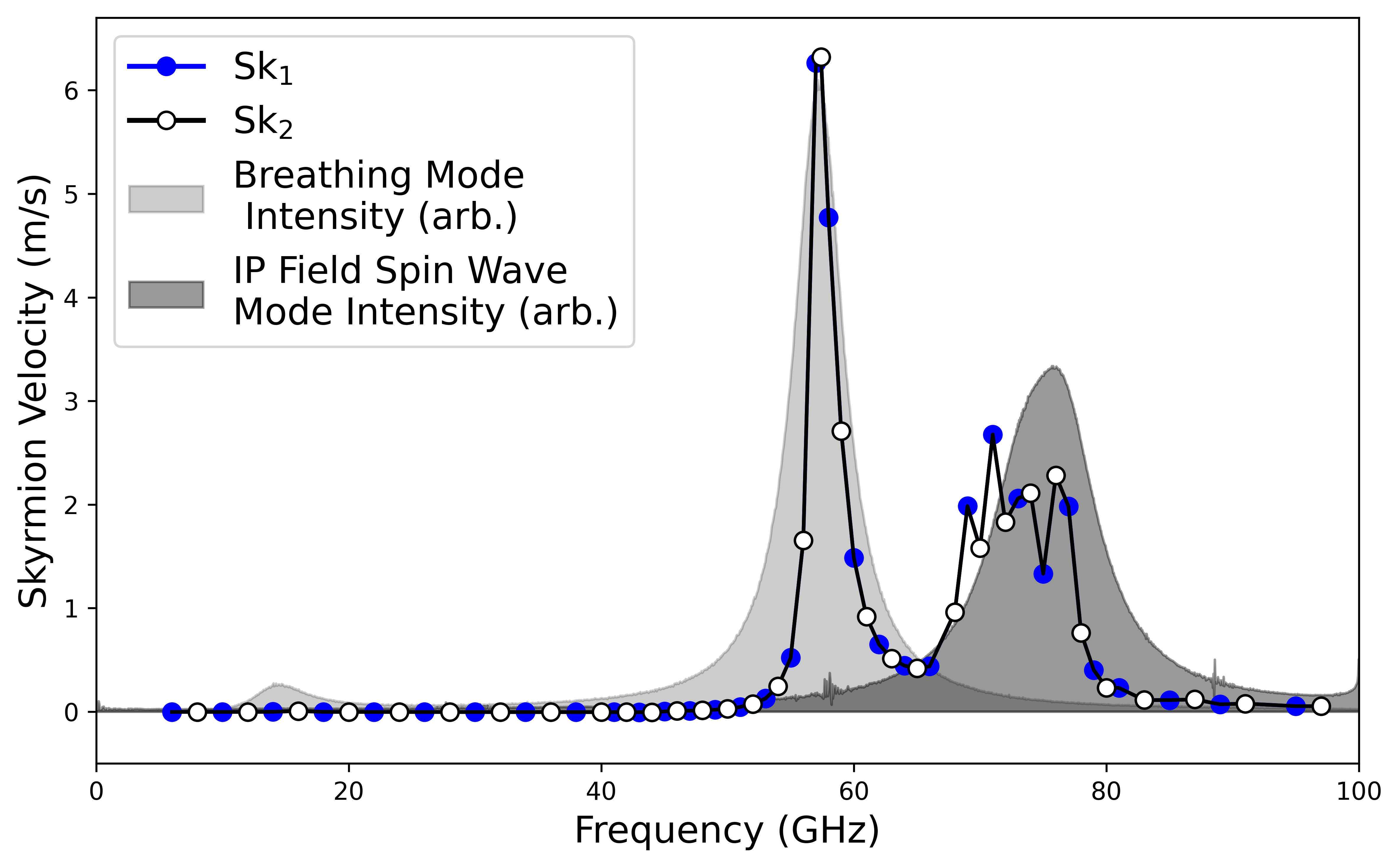}
			\caption{The replotted data of Fig.~\ref{fig:velocity_fn_freq}(a) from the main text, along with an additional darker shaded region that has been rescaled from Fig.~\ref{supp:fig:IP_SW_Mode} to highlight the overlap in frequency of the higher frequency peak in velocity with the spin-wave mode arising from the in-plane magnetic field.}
			\label{supp:fig:vel_fn_freq}
		\end{figure}
		
		To highlight this, we also replot Fig.~\ref{fig:velocity_fn_freq}(a) from the main paper here with an additional shaded region drawn from Fig.~\ref{supp:fig:IP_SW_Mode} in Fig.~\ref{supp:fig:vel_fn_freq}. Any small differences in frequency between the measured frequency of the spin-wave mode and the frequency of the peak in skyrmion velocity are likely due to a slight modification in the resonant frequency caused by hybridisation with the skyrmion breathing mode, an effect which is well understood~\cite{Kim2014}.
		\newpage
		
	\section{A description of supplementary videos 1-6}\label{supp:sec:videos}
		This note briefly details the supplementary videos that are included with this paper. Each video consists of two panels, on the left is the skyrmion in the bottom magnetic layer, and on the right is the skyrmion in the top magnetic layer. Each video also links to a pair of panels in Fig.~\ref{fig:Velocity_Amplitude_Radius_Compare} from the main paper, expanding the snapshot presented there at 0.2~ns into the simulation into a video covering the whole first 1~ns of the simulation. We list them here:
		\begin{itemize}
			\item \textbf{Supplementary Video S1:} A video of the colormap of the y-component of the magnetisation. The driving magnetic field is 20~mT at a frequency of 57.4~GHz, with a static in-plane magnetic field of 0.5~T. This video is linked to Fig.~\ref{fig:Velocity_Amplitude_Radius_Compare}(a,b) in the main paper.  
			
			\item \textbf{Supplementary Video S2:} A video of the colormap of the y-component of the magnetisation. The driving magnetic field is 20~mT at a frequency of 14.3~GHz, with a static in-plane magnetic field of 0.5~T. This video is linked to Fig.~\ref{fig:Velocity_Amplitude_Radius_Compare}(g,h) in the main paper.
			
			\item \textbf{Supplementary Video S3:} A video of the vertical slice through the centre of the y-component of the skyrmions magnetisation. The driving magnetic field is 20~mT at a frequency of 57.4~GHz, with a static in-plane magnetic field of 0.5~T. This video is linked to Fig.~\ref{fig:Velocity_Amplitude_Radius_Compare}(c,d) in the main paper.
			
			\item \textbf{Supplementary Video S4:} A video of the vertical slice through the centre of the y-component of the skyrmions magnetisation. The driving magnetic field is 20~mT at a frequency of 14.3~GHz, with a static in-plane magnetic field of 0.5~T. This video is linked to Fig.~\ref{fig:Velocity_Amplitude_Radius_Compare}(i,j) in the main paper.
			
			\item \textbf{Supplementary Video S5:} A video of the vertical slice through the centre of the y-component of the skyrmions magnetisation. The driving magnetic field is 20~mT at a frequency of 57.4~GHz, and there is no static in-plane magnetic field. This video is linked to Fig.~\ref{fig:Velocity_Amplitude_Radius_Compare}(e,f) in the main paper.
			
			\item \textbf{Supplementary Video S6:} A video of the vertical slice through the centre of the y-component of the skyrmions magnetisation. The driving magnetic field is 20~mT at a frequency of 14.3~GHz, and there is no static in-plane magnetic field. This video is linked to Fig.~\ref{fig:Velocity_Amplitude_Radius_Compare}(k,l) in the main paper.
		\end{itemize}
		
		\newpage
		
	\section{Analysis of the variation in skyrmion radius}\label{supp:sec:rad_var}
\begin{figure}
			\centering
			\includegraphics[width=0.8\linewidth]{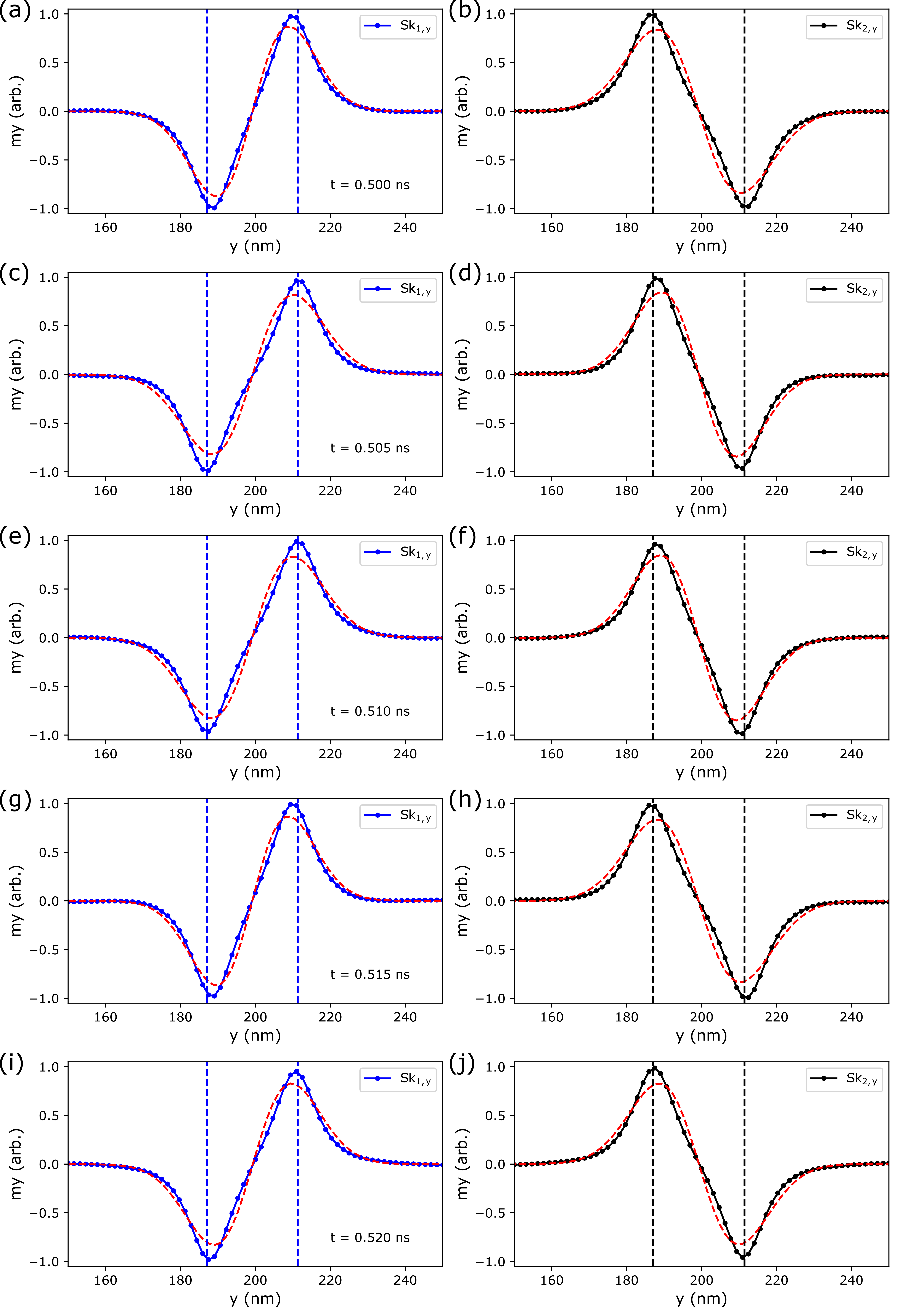}
			\caption{Snapshots of slices through the centre of the skyrmion in the $y$-component of the magnetisation. (a,c,e,g,i) show the skyrmion in layer 1 and (b,d,f,h,j) show the skyrmion in layer 2. The time of each row of images is shown in the left hand panel, and the range spans over 0.02~ns, approximately one period of the excitation field. Dashes lines show the skyrmion radius before the excitation field was switched on.}
			\label{supp:fig:rad_var_examples}
		\end{figure}

\begin{figure}
			\centering
			\includegraphics[width = \linewidth]{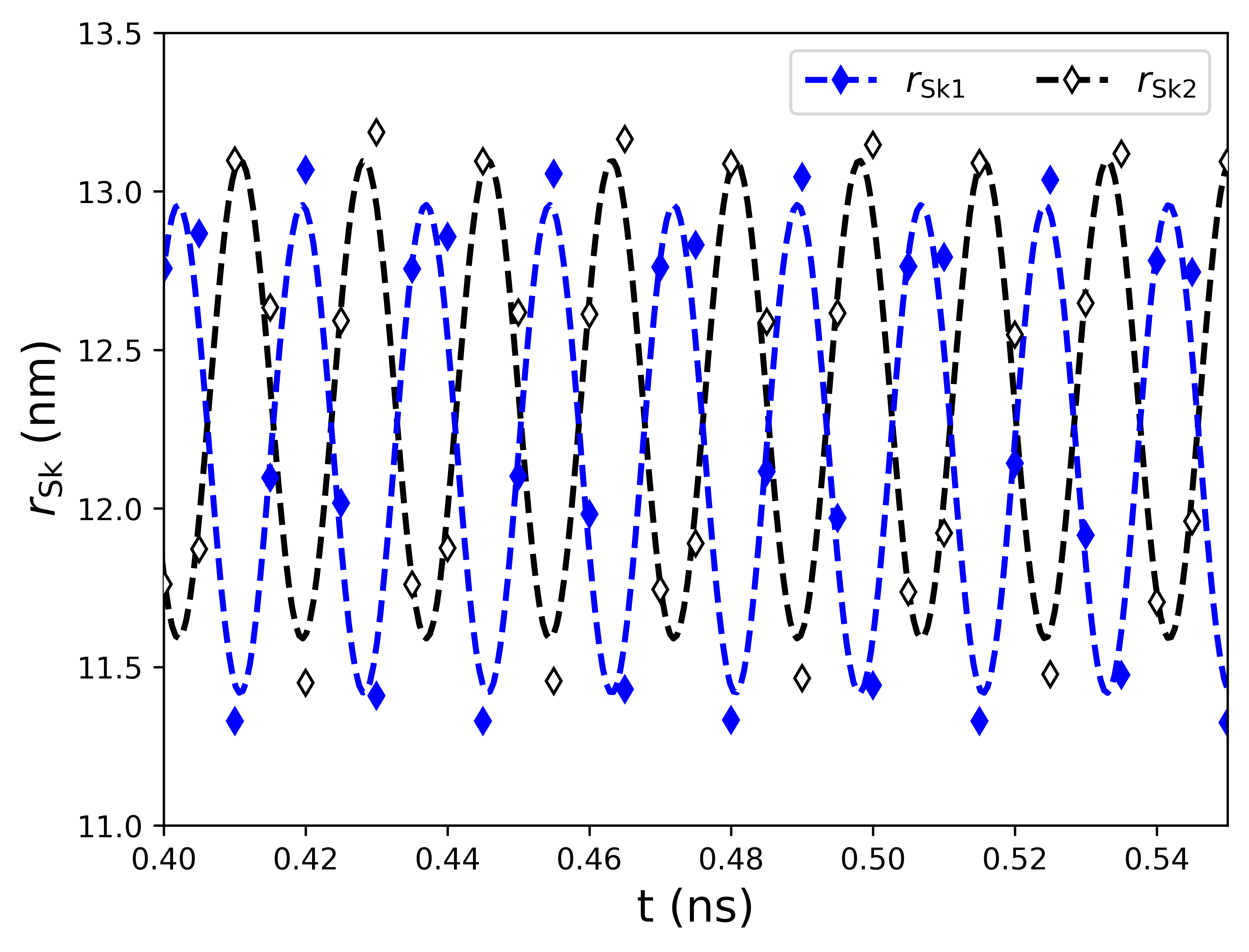}
			\caption{Extracted values of the skyrmion radius from data of the form presented in Fig.~\ref{supp:fig:rad_var_examples} together with fits to Eq.~\ref{supp:eq:sk_rad_var} in order to extract the average variation in radius. The driving field in this example is 20~mT, and radius data has been extracted from the $\mathbf{m}_{\mathrm{y}}$ profile.}
			\label{supp:fig:rad_var_fit}
		\end{figure}

		In order to calculate the variation of the skyrmion radius that is presented in Fig.~\ref{fig:Velocity_Amplitude_Radius_Compare}, a accurate and fast method of establishing the skyrmion radius was required. This also needed to be sensitive enough to small changes in the radius, and not be limited by the cell size. In order to accomplish this, a derivative of a Gaussian function was chosen for its good match to a slice through the $\mathbf{m}_{\mathrm{x}}$ or $\mathbf{m}_{\mathrm{y}}$ component of the skyrmion's magnetisation. This is of the form
		\begin{equation}
			-A\dfrac{(x - x_0)}{\sigma^2}\exp\Big(-\dfrac{(x - x_0)^2}{2\sigma^2}\Big) + C,
		\end{equation}
		where $A$ is the amplitude, $x_0$ is the centre position, $\sigma$ is the standard deviation, and $C$ is an arbitrary offset factor. We can then extract the radius which will be related to $\sigma$ through $r_{\mathrm{Sk}} = \sqrt{2\ln2} \cdot \sigma$. Snapshots of the $\mathbf{m}_{\mathrm{y}}$ component of the slice through the centre of the skyrmion are shown in Fig.~\ref{supp:fig:rad_var_examples} for a 20~mT excitation field. These show the data for the skyrmions in both layer 1 and layer 2 respectively, over 0.02~ns, approximately one period of the excitation field. The dashed lines come from $x_0 \pm \sqrt{2\ln 2}\cdot \sigma $ of the skyrmion before switching on the excitation field. They serve as a guide to the eye to illustrate the small variations in skyrmion radius during a cycle of the field.\\\\
		Once the radii of the two skyrmions have been collected over the whole simulation time, we calculate the average variation in radius. To do this we plot the radius as a function of time, and fit this data to a sinusoidal curve of the form
		\begin{equation}
			r_{\mathrm{Sk}} = A\sin(2\pi f t + \phi) + r_{\mathrm{Sk}}(0).
			\label{supp:eq:sk_rad_var}
		\end{equation}
		Here, $A$ is a prefactor determining the amplitude of the sine wave, $f$ is the excitation frequency, $\phi$ is an arbitrary phase factor, and $r_{\mathrm{Sk}}(0)$ is the radius of the skyrmion at $t=0$ which acts as an offset factor. An example fit of this data is shown in Fig.~\ref{supp:fig:rad_var_fit} for a driving field of 20~mT and radii calculated from slices through $\mathbf{m}_{\mathrm{y}}$. The fitted value of $A$ from Eq.~\ref{supp:eq:sk_rad_var} can then be multiplied by 2 to give the average variation in radius expressed as $\langle \Delta r_{\mathrm{Sk}} \rangle$ shown in Fig.~\ref{fig:Velocity_Amplitude_Radius_Compare}(b) in the main text.
		\newpage
		\newpage

	\section{Effect of unbalancing the $M_{\mathrm{s}}$ for $\sigma_{\mathrm{RKKY}} = 3\times 10^{-4} \mathrm{J/m^2}$}		
		\begin{figure}[h]
			\centering
			\includegraphics[width=\linewidth]{Breathing_Modes_Ms1_RKKY3Eneg4}
			\caption{Skyrmion breathing mode frequency as a function of the $M_{\mathrm{s}}$ of layer one of the SAF for: (a) the whole system, (b) layer 1 only and (c) layer 2 only.}
			\label{supp:fig:breathing_SyFM}
		\end{figure}
		
\begin{figure}[h]
			\centering
			\includegraphics[width=\linewidth]{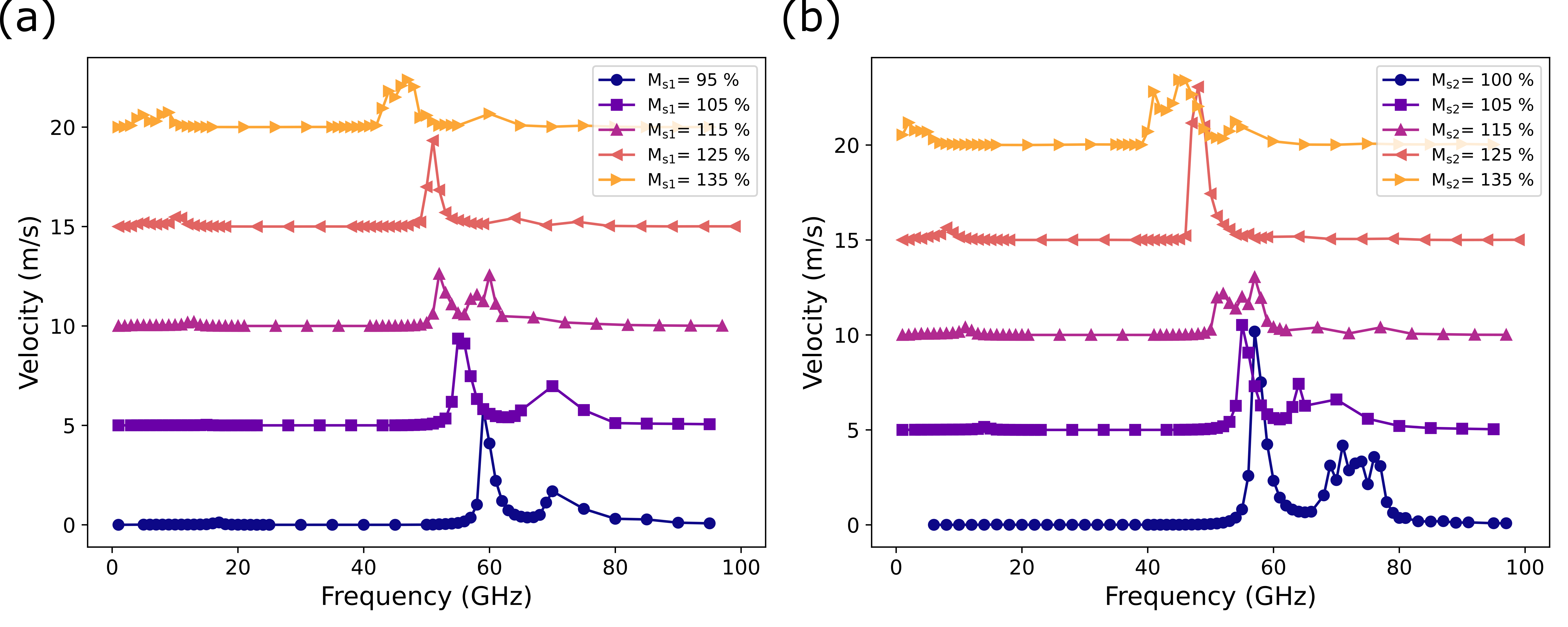}
			\caption{Skyrmion velocity as a function of frequency, for a range of: (a) layer 1 $M_{\mathrm{s}}$ and (b) Layer 2 $M_{\mathrm{s}}$. Each velocity as a function of frequency dataset is offset by a fixed amount to more clearly present the data.}
			\label{supp:fig:skyrmion_motion_freq_ms}
		\end{figure}

		In Figure~\ref{fig:SyFM} of the main text, results are presented for unbalancing the $M_{\mathrm{s}}$ for two characteristic values of $\sigma_{\mathrm{RKKY}}$. The frequencies of the breathing modes as a function of the variation of one layer's $M_{\mathrm{s}}$, and the velocity of the skyrmion as a function of frequency for several values of one layer's $M_{\mathrm{s}}$ were both presented in the figure for $\sigma_{\mathrm{RKKY}} = 1\times 10^{-4}\ \mathrm{J/m^2}$ in Figure~\ref{fig:SyFM}(a,b) of the main text. Here we summarise these same results for $\sigma_{\mathrm{RKKY}} = 3\times10^{-4}\ \mathrm{J/m^2}$, which contribute to the data presented in Figure~\ref{fig:SyFM}(d) of the main text.\\\\
		The frequencies of the breathing modes as a function of the variation of a single layer's $M_{s}$ are shown in Figure~\ref{supp:fig:breathing_SyFM}, with the spectrum measured for the whole system as well as each of the individual layers. This data follows the same broad trend as that for $\sigma_{\mathrm{RKKY}} = 1\times10^{-4}\ \mathrm{J/m^2}$ presented in Figure~\ref{fig:SyFM}(a) of the main manuscript. The frequency is of course higher for this antiferromagnetic interlayer coupling strength, with the data at $M_{s1} = 100\%$ matching that in Figure~\ref{fig:Skyrmion_breathing}(b-d) of the main text for $\sigma_{\mathrm{RKKY}} = 3\times10^{-4}\ \mathrm{J/m^2}$.\\\\
		With the understanding of the frequencies of the resonant modes gained from Figure~\ref{supp:fig:breathing_SyFM}, we can study the velocity of the skyrmions as a function of frequency. The skyrmion velocity as a function of frequency is shown in Figure~\ref{supp:fig:skyrmion_motion_freq_ms}(a) and (b). Panel (a) shows the data when the $M_{\mathrm{s}}$ of layer 1 is varied about the original value, while panel (b) shows the same for the $M_{\mathrm{s}}$ of layer 2. Each $M_{\mathrm{s}}$ dataset is offset by a fixed amount to more clearly show the data. As discussed in the main text for the $\sigma_{\mathrm{RKKY}} = 1\times 10^{-4}\ \mathrm{J/m^2}$ case, the peak in velocity follows the breathing mode as we would expect. Likewise, the higher frequency peak in velocity (associated with the resonant frequency of the canted spins from the in plane field) also decreases, albeit faster than the velocity peak associated with the skyrmion breathing mode frequency. This results---at the highest unbalance factors---in the overlap between the resonant frequency of the spin wave mode from the canted spins and that of the out-of-phase skyrmion breathing mode. This is the driver for the more chaotic behaviour of the data in Figure~\ref{fig:SyFM}(d) of the main text for values of $M_{\mathrm{s,i}} \geq 120\% $. Because the two resonances are combining, there is an unexpected increase in the spin-wave emission amplitude, which causes the peak velocities to increase. while there are some small differences between the values of $M_{\mathrm{s}}$ that this effect occurs when varying the $M_{\mathrm{s}}$ of layer 1 vs layer 2, the effects are broadly similar, and occur at the same high decompensation of the SAF.\\\\ 
		Additionally, at the highest levels of decompensation, there is some evidence of the peak in velocity at the in-phase skyrmion resonant frequency splitting into two. This is likely due to the extremely large decompensation. The variation in breathing mode frequency with $M_{\mathrm{s}}$ has been well studied in SAFs~\cite{Lonsky2020_PRB}, and likely here the divergence of the intrinsic resonant frequencies of each layer is large enough to split into two different peaks when driven with the strong microwave field required for motion. Any similar effect at the out-of-phase resonant frequency is obscured by the overlap with the resonant frequency of the canted spins, as discussed above. 
